\documentclass[a4paper,11pt]{article}
\usepackage{jinstpub} 
\usepackage{lineno}
\usepackage{mathtools}
\usepackage{xcolor}
\usepackage{enumitem}
\usepackage{booktabs}
\usepackage{tabularx}
\usepackage{siunitx}
\DeclareSIUnit{\gate}{gate}
\usepackage{silence}
\RequirePackage{fix-cm}
\title{Commissioning and Performance of the Time-of-Flight Detector for the T2K Neutrino Oscillation Experiment}

\author[h]{C.~Alt}
\author[a]{L.~Amziane}
\author[l]{N.~Baudis}
\author[a]{A.~Blanchet}
\author[a,1]{S.~Bordoni}
\author[b]{T.~H.~Bui}
\author[a]{F.~Cadoux}
\author[a,j,4]{P.~Collard} 
\author[a]{M.~El Baz}
\author[a]{Y.~Favre}
\author[a]{L.~Giannessi}
\author[b]{G.~Ha}
\author[d]{C.~Jes\'us-Valls}
\author[a]{V.~S.~Kasturi}
\author[k]{A.~Klustová}
\author[a,2]{A.~Korzenev}
\author[e]{T.~A.~Le}
\author[f]{T.~Lux}
\author[c]{A.~D.~Nguyen}
\author[b]{D.~T.~Nguyen}
\author[b]{H.~Nguyen}
\author[a,1]{S.~Samani}
\author[a]{F.~S\'anchez}
\author[b]{M.~Ta}
\author[i,5]{T.~Thaiduc} 
\author[g,a,1,3]{E.~Villa}

\note{Corresponding author.}
\note{Now at Joint Institute for Nuclear Research, Dubna, Russia.}
\note{Now at Swiss Federal Institute of Technology Zurich (ETH Zurich), Zurich, Switzerland.}
\note{Now at Institut de Fisica d'Altes Energies (IFAE), The Barcelona Institute of Science and Technology (BIST), Campus UAB, 08193 Bellaterra, Barcelona, Spain.}
\note{Now at Ecole Polytechnique Federale de Lausanne (EPFL), Lausanne, Switzerland.}

\affiliation[a]{D\'epartement de Physique Nucl\'eaire et Corpusculaire, Universit\'e de Gen\`eve, Geneva, Switzerland}

\affiliation[b]{Faculty of Physics, VNU University of Science,
Hanoi, Vietnam}

\affiliation[c]{Institute of Physics, Vietnam Academy of Science and Technology, Hanoi, Vietnam}

\affiliation[d]{ Kavli Institute for the Physics and Mathematics of the Universe (WPI), University of Tokyo Institutes for Advanced Study, University of Tokyo, Kashiwa, Japan } 

\affiliation[e]{INST, Vietnam Atomic Energy Institute,
Hanoi, Vietnam}

\affiliation[f]{Institut de Física d’Altes Energies (IFAE), The Barcelona Institute of Science and Technology (BIST), Campus UAB, E-08193 Bellaterra, Barcelona, Spain}

\affiliation[g]{CERN, European Organization for Nuclear Research, CH-1211 Geneva 23, Switzerland}

\affiliation[h]{Institute for Particle Physics and Astrophysics, ETH Zurich, 8093 Z\"urich, Switzerland}

\affiliation[i]{Boston University, Department of Physics, Boston, MA 02215, USA}

\affiliation[j]{Universit\'{e} Claude Bernard Lyon 1, Facult\'{e} des Sciences, D\'{e}partement de Physique, Villeurbanne, France}

\affiliation[k]{Imperial College London, Department of Physics, Blackett Laboratory, SW7 2BW London, United Kingdom}

\affiliation[l]{University of Oxford, Department of Physics, Oxford, United Kingdom}
\emailAdd{soniya.samani@unige.ch} 
\emailAdd{stefania.bordoni@unige.ch} 
\emailAdd{emanuele.villa@cern.ch}

\abstract{The T2K ND280 Upgrade aims to reduce systematic uncertainties in measurements of neutrino oscillation parameters and improve sensitivity to the charge-parity (CP)-violating phase, $\delta_{\mathrm{CP}}$. A key component is the Time-of-Flight (ToF) detector, comprising six panels with 118 EJ-200 plastic-scintillator bars surrounding the Super Fine-Grained Detector (SuperFGD) and two High-Angle Time Projection Chambers (HA-TPCs). Each bar is read out at both ends by silicon photomultiplier arrays and digitised using SAMPIC waveform electronics. The ToF provides precise timing and particle-direction information for particle identification and rejection of backgrounds entering the tracker from outside. This article presents the detector design, construction, signal reconstruction, integration, commissioning, and performance. Dedicated single-bar measurements achieve a time resolution of approximately 130~ps and a longitudinal position resolution of 2.6~cm. After installation in ND280, cosmic-ray calibration yields an in situ single-bar time resolution of $169 \pm 1$~ps for Top--Bottom crossing events. Beam data clearly resolve the eight-bunch T2K spill structure, confirming synchronisation with the ND280 trigger and data-acquisition systems. The ToF has been successfully commissioned and operates stably within the upgraded ND280 detector.}

\keywords{Time-of-Flight detector, detector commissioning, plastic scintillators, silicon photomultipliers, waveform digitisation, T2K ND280 Upgrade.}

\begin{document}
\maketitle
\flushbottom
\section{Introduction}
\label{sec:intro}
The increasing precision required by modern long-baseline accelerator neutrino oscillation experiments has pushed the capabilities of existing near detectors to their limits. In these experiments, the near detector plays a crucial role in constraining the neutrino flux and reducing uncertainties associated with neutrino--nucleus interaction cross sections. To meet these requirements, the Tokai-to-Kamioka~(T2K) experiment in Japan~\cite{T2K:2011qtm} initiated a major upgrade programme in 2016 for its off-axis near detector ND280~\cite{T2K:2019bbb}. The upgrade addresses two key limitations of the original detector design: non-uniform angular acceptance and limited granularity for the detection of low-momentum particles. In addition, the upgraded detector improves neutron detection capabilities and particle-direction reconstruction, vital for suppressing backgrounds originating outside the fiducial volume. Together with upgrades to the Japan Proton Accelerator Research Complex
(J-PARC) Main Ring and the T2K beamline~\cite{Igarashi:2021npv}, these improvements will enable the experiment to collect larger data samples while reducing systematic uncertainties in oscillation measurements from approximately 6\% to 4\%~\cite{Blondel:2299599,T2K:2019bbb}. This reduction is crucial for enhancing the sensitivity to the charge-parity-violating phase $\delta_{\mathrm{CP}}$, with the experiment aiming to reach approximately a $3\sigma$ confidence level~\cite{T2K:2019bbb,T2K:2019bcf}.

\begin{figure}[htbp]
    \centering
    \includegraphics[width=\textwidth]{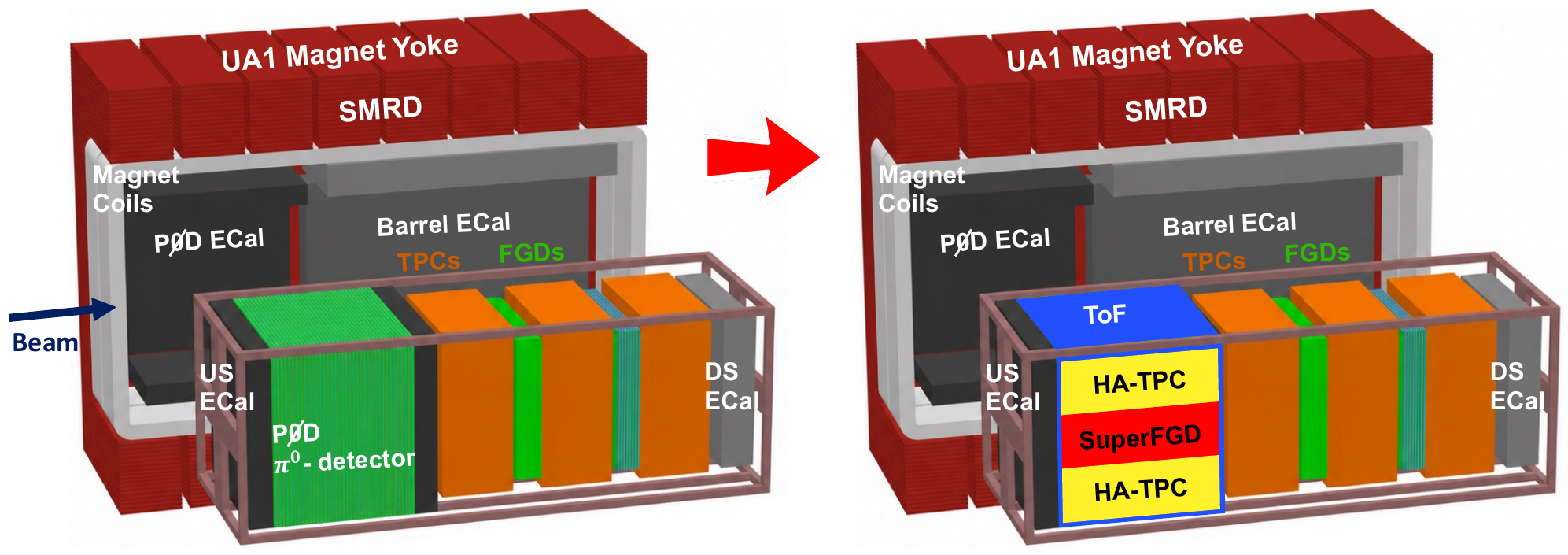}
    \caption{Schematic view of the ND280 detector before and after the upgrade. \textit{Left}: The original ND280 configuration. \textit{Right}: The ND280 Upgrade, with the P\O D replaced by the SuperFGD, two HA-TPCs, and the surrounding ToF detector. The near-side ToF panel, facing the reader, is not shown in order to display the inner detectors.}
    \label{fig:ND280up}
\end{figure}

As part of the ND280 Upgrade, the Pi-Zero Detector (P\O D)~\cite{Assylbekov:2011sh}
from the original configuration was removed and replaced by four new detector
components: the Super Fine-Grained Detector (SuperFGD)~\cite{Blondel:2020hml,Abe:2026elv},
two High-Angle Time Projection Chambers (HA-TPCs)~\cite{Attie:2021yeh}, and a
surrounding Time-of-Flight (ToF) detector~\cite{Korzenev:2021mny}, as illustrated
in Figure~\ref{fig:ND280up}. The remaining components of the original ND280
detector were retained, including the UA1 magnet~\cite{T2K:2011qtm}, Time Projection Chambers (TPCs)~\cite{ABGRALL201125}, Fine-Grained Detectors (FGDs)~\cite{AMAUDRUZ20121}, electromagnetic
calorimeter (ECal)~\cite{D_Allan_2013}, and Side Muon Range Detector
(SMRD)~\cite{AOKI2013135}.

The ToF detector precisely measures the flight time of charged particles, enabling particle-direction reconstruction and rejection of backgrounds entering from outside the fiducial volume. Combined with the reconstructed momentum and path length, the time-of-flight measurement also contributes to particle identification.

This paper presents the design, construction, installation, integration, commissioning, and performance of the ToF detector. Section~\ref{sec:tof} describes the detector design, readout and trigger systems, slow control, mechanical support, and integration within ND280. Section~\ref{sec:reconstruction} presents the signal-reconstruction methods and dedicated single-bar characterisation studies. Section~\ref{sec:commissioning} covers online monitoring, data quality, cosmic-ray response, and timing calibration. Finally, Section~\ref{sec:performance} presents the main detector-performance studies, including beam timing, delayed activity, readout dead-time, unpaired hits, and the in situ time resolution.

\section{The Time-of-Flight Detector}
\label{sec:tof}\label{sec:detector}
\begin{figure}[htbp]
    \centering
    \includegraphics[width=0.79\textwidth]{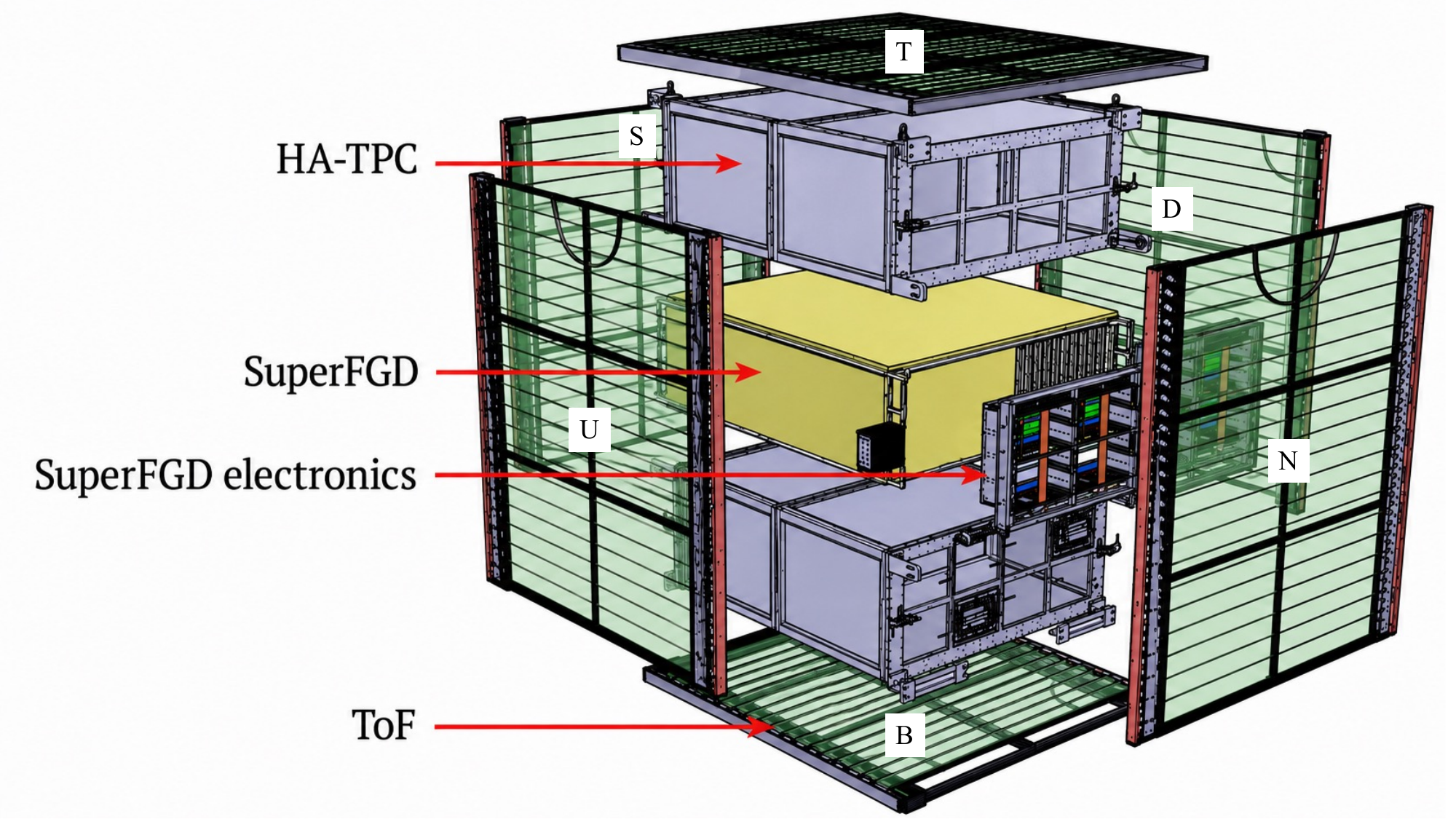}
    \caption{Exploded view of the ND280 Upgrade detector. The SuperFGD is positioned between the two HA-TPCs and is surrounded by six ToF panels, forming a nearly cubic enclosure around the active detector volume. The ToF panels are labelled according to their positions as North (N), South (S), Top (T), Bottom (B), Upstream (U), and Downstream (D). The SuperFGD readout electronics are mounted on its lateral sides, beneath the North and South ToF panels.}

    \label{fig:upgrade_exploded}
\end{figure}

The ToF detector surrounds the SuperFGD and the HA-TPCs of the ND280 Upgrade, providing nearly $4\pi$ coverage around the active detector volume, as shown in Figure~\ref{fig:upgrade_exploded}. In addition to measuring particle flight times, the ToF provides an external reference time, $t_0$, for tracks reconstructed in the HA-TPCs, allowing the measured electron drift time to be related to the absolute position along the drift direction. For tracks reconstructed in the SuperFGD, the ToF provides an independent timing measurement that complements the fine-grained spatial information and contributes to particle-direction determination and track matching.

The detector concept was first developed and prototyped for the SHiP experiment~\cite{SHiP:2015vad}, before being adopted for the ND280 Upgrade. Bench-top characterisation studies demonstrated a single-bar time resolution of approximately 130~ps, as discussed in Section~\ref{sec:singlebar}. This is small compared with the several-nanosecond flight times expected across the metre-scale distances between ToF panels, enabling reliable particle-direction determination.

\subsection{Detector Design and Geometry}
The ToF detector comprises six panels of plastic-scintillator bars that surround the SuperFGD and two HA-TPCs. The panels are positioned upstream, downstream, north, south, above, and below the tracker volume relative to the beam direction. This arrangement provides broad active coverage while limiting the required number of readout channels. Five panels contain twenty scintillator bars, while the Bottom panel contains eighteen to accommodate services and cabling for the inner detectors, giving a total of 118 bars. The active area of each 20-bar panel is approximately $240 \times 220$~cm$^2$~\cite{Korzenev:2021mny}, resulting in a total active area of approximately 30~m$^2$.

\begin{figure}[htbp]
    \centering

    \resizebox{\textwidth}{!}{%
        \includegraphics[
            height=0.30\textheight,
            keepaspectratio,
            trim=0 8 0 8,
            clip
        ]{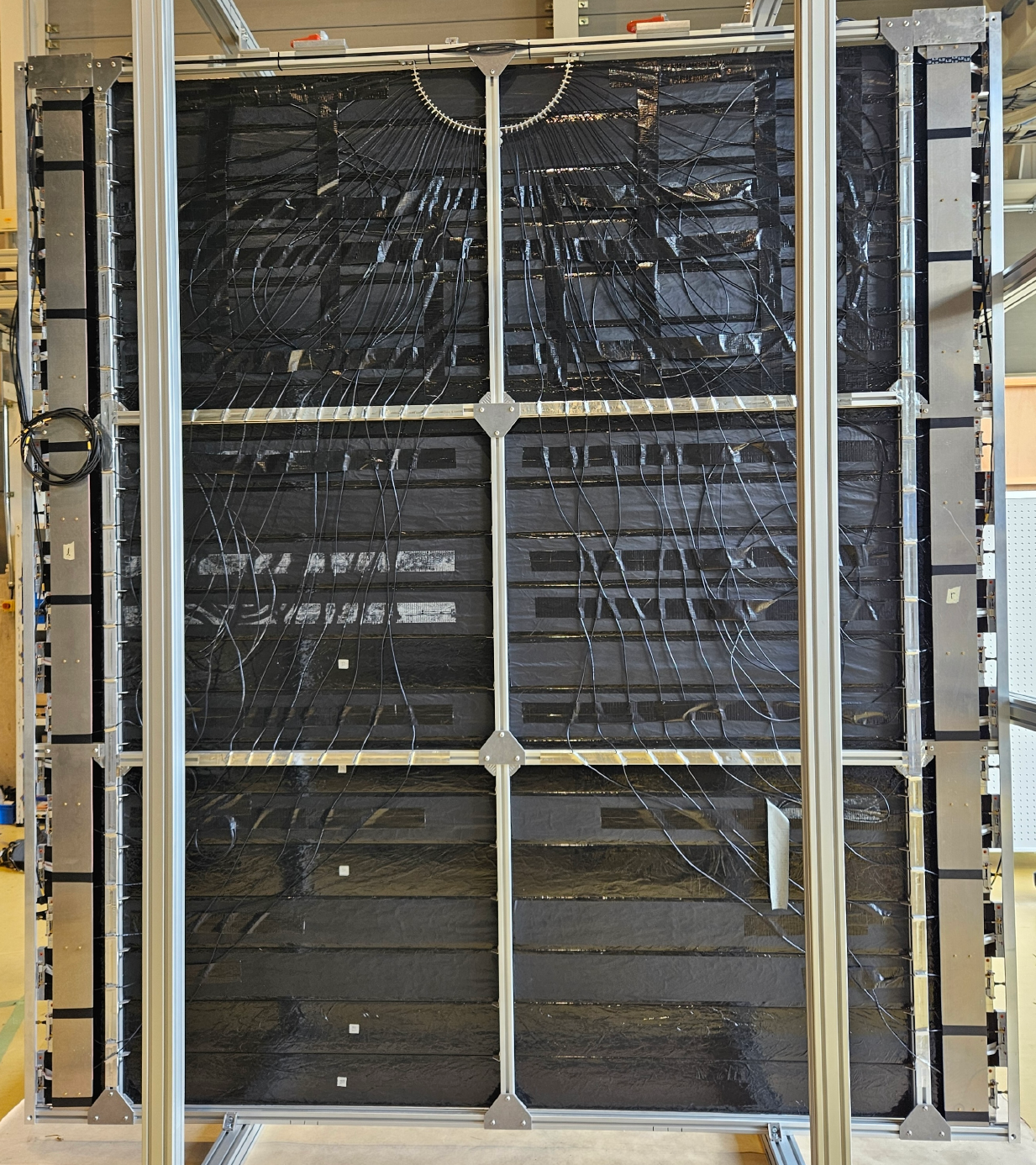}
        \hspace{0.025\textwidth}
        \includegraphics[
            height=0.30\textheight,
            keepaspectratio,
            clip
        ]{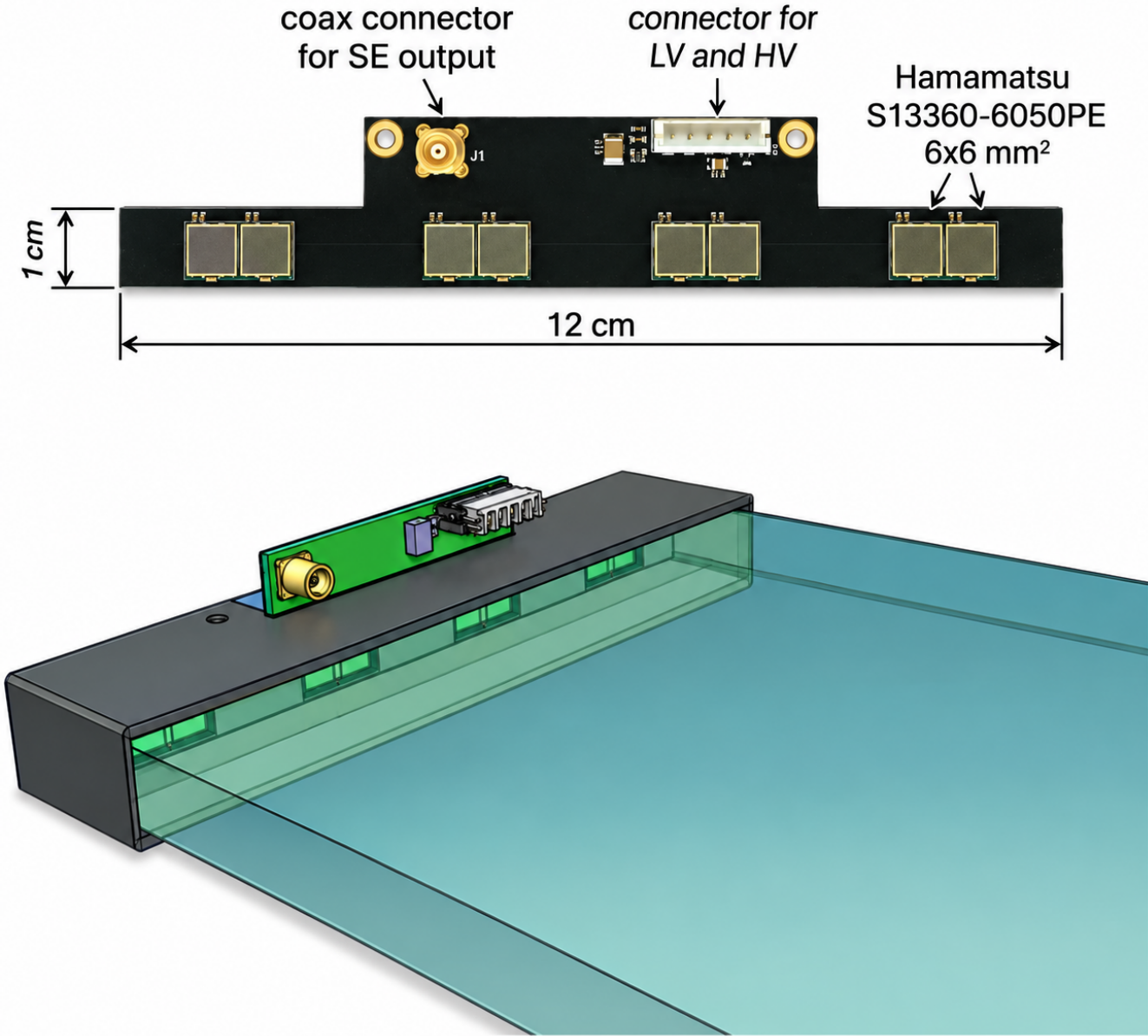}
    }

    \caption{\textit{Left}: assembled ToF panel showing the scintillator bars and signal cabling. \textit{Right}: SiPM-array readout board and its optical coupling to the scintillator bar~\cite{Korzenev:2021mny}.}
    \label{fig:sipm_readout}
\end{figure}

The scintillator bars are made of EJ-200 plastic scintillator and measure $220 \times 12 \times 1$~cm$^3$~\cite{Korzenev:2021mny}. EJ-200 has a peak emission wavelength of 425~nm and produces approximately $10^4$ photons per MeV of deposited energy~\cite{EljenScintillators}. The material has an attenuation length of about 380~cm, a rise time of 0.9~ns, and a decay time of 2.1~ns. The scintillation light produced in each bar is detected at both ends by silicon photomultipliers~(SiPMs). Each bar is wrapped in aluminium foil and additionally covered with opaque black film to prevent optical cross-talk between neighbouring bars. The bars are mounted on a rigid support frame to form each panel, ensuring stable alignment and minimising inactive regions. Adjacent bars are slightly offset in the direction normal to the panel and arranged such that their edges overlap, avoiding gaps between neighbouring scintillator bars. The assembly of the scintillator bars and their readout components is illustrated in Figure~\ref{fig:sipm_readout}. The sensor configuration and readout electronics are described in Section~\ref{sec:photosensors}. 

\subsection{Photosensors and Readout}
\label{sec:photosensors}
The scintillation light from each bar is read out at both ends by an array of eight Hamamatsu S13360-6050PE SiPMs~\cite{Hamamatsu2025}. Each SiPM has an active area of $6\times6$~mm$^2$, providing high gain and high photon-detection efficiency, with a spectral response well matched to the emission spectrum of the EJ-200 scintillator. The eight sensors at each bar end cover approximately 24\% of the scintillator end surface, as illustrated in Figure~\ref{fig:sipm_readout}. They are mounted on front-end printed circuit boards~(PCBs), which provide the bias voltage, sum the SiPM signals, and perform the first stage of amplification. The summed signal from each array forms one readout channel per bar end, increasing the effective photosensitive area while keeping the total number of channels manageable~\cite{Korzenev:2021mny}.

The SiPMs are grouped according to their breakdown voltage, allowing all sensors within an array to be operated at a common bias voltage without tuning. This simplifies detector operation and promotes a uniform response across the readout channels. The amplified signals are routed via coaxial cables to the SAMPIC electronics through two intermediate patch panels, PP1 and PP2, with cable lengths that differ between ToF panels according to their position within ND280.

\subsection{Data Acquisition System and Digitisation}
\label{Sec:DAQ}
The ToF detector readout is based on the SAMpler for PICosecond time (SAMPIC) chip~\cite{Delagnes:2014SAMPIC,Breton:2020kva}. The SAMPIC was developed at the Laboratoire de l'Accélérateur Linéaire (LAL, Orsay, France) to provide precise timing measurements with simultaneous waveform sampling. The chip combines a delay-locked loop (DLL)-based time-to-digital converter (TDC) with a fast analogue memory, allowing the extraction of both precise timing information and pulse shape characteristics.

Each SAMPIC channel contains a circular analogue memory composed of 64 sampling cells that continuously sample the incoming signal at rates up to 6.4~GHz. When a trigger occurs, the sampling buffer is frozen and the timestamp corresponding to the first sample of the waveform is recorded as the so-called \textit{Cell0Time}. The timestamp associated with this cell is measured by the DLL-based TDC with a precision of approximately 5~ps. The waveform is subsequently digitised via an integrated Wilkinson analogue-to-digital converter (ADC) and read out for offline processing.

\begin{figure}[htbp]
    \centering
    \includegraphics[
        height=0.75\textwidth,
        angle=270,
        keepaspectratio,
        trim=0 45 0 45,
        clip
    ]{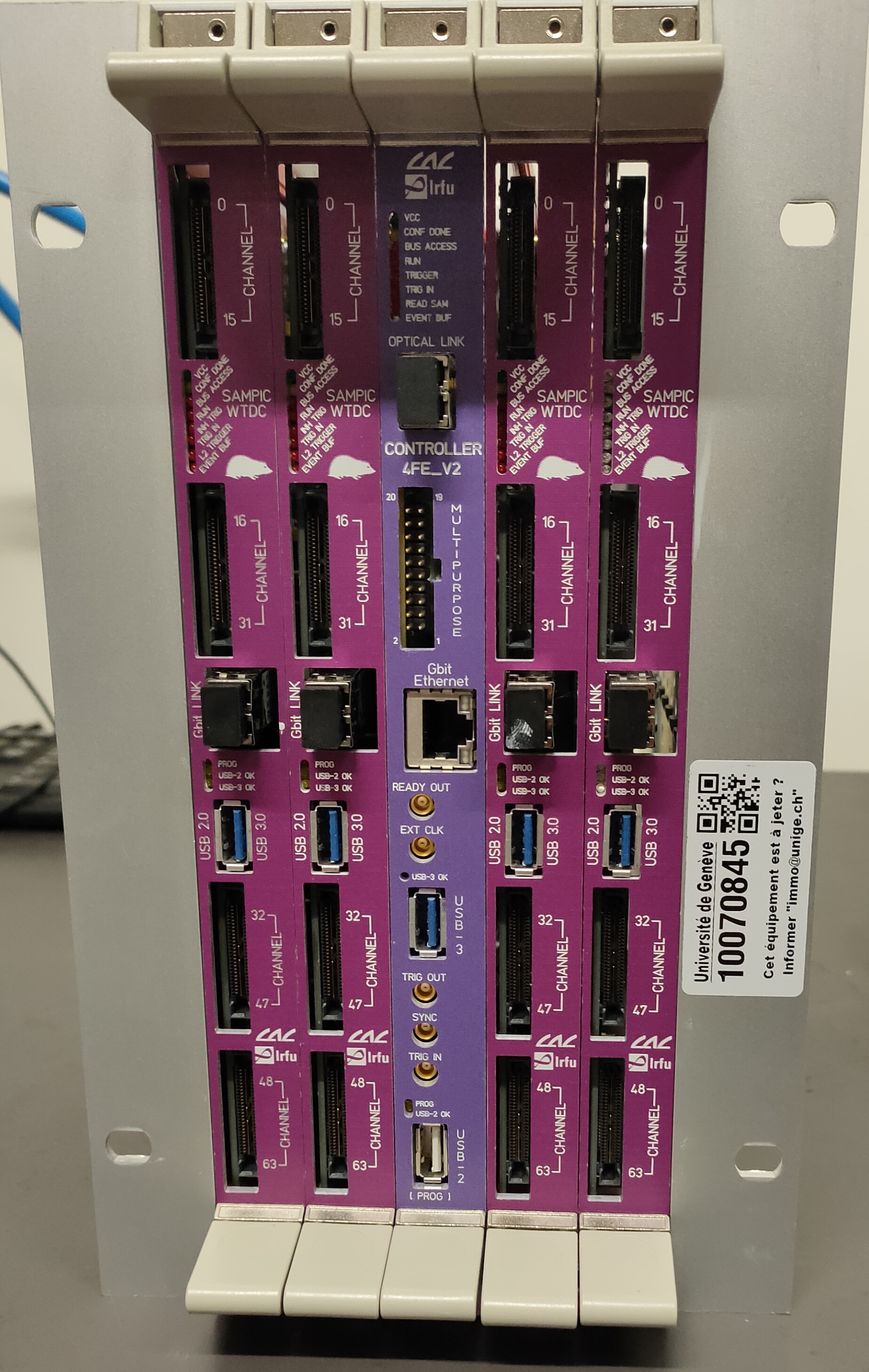}

    \caption{SAMPIC waveform-digitiser assembly showing four FEBs and a central controller board, which coordinates trigger distribution and data readout.}
    \label{fig:Sampic}
\end{figure}

The digitisation electronics are housed in a single crate comprising four front-end boards~(FEBs) connected to a central controller board, as shown in Figure~\ref{fig:Sampic}. Each FEB hosts four 16-channel SAMPIC chips, resulting in 64 readout channels per board and 256 channels in total. The controller board distributes the system clock and trigger signals to the FEBs and collects the digitised data for transmission to the data acquisition (DAQ) computer via optical link. The ToF DAQ system is integrated into the global ND280 DAQ framework through a dedicated software application based on Maximum Integrated Data Acquisition System (MIDAS)~\cite{MIDAS}. This application receives the digitised data from the SAMPIC controller board and publishes the events within the ND280 DAQ.

\subsection{Trigger Logic and Readout}
\label{Sec:Trigger}
The digitised ToF data are matched to the trigger information associated with the selected operating mode. The ToF system can operate as a stand-alone trigger endpoint, respond to triggers distributed by the global ND280 DAQ, or provide trigger signals to the global trigger chain. The SAMPIC electronics implement a hierarchical trigger logic organised in three levels:
\begin{itemize}[itemsep=0pt]
    \item \textbf{Channel self-trigger (LT1):}
    Each channel contains a programmable discriminator. A signal exceeding the configured threshold generates an LT1 trigger.

    \item \textbf{Board-level coincidence (HLT2):}
    HLT2 requires time-coincident LT1 signals from channels connected to different SAMPIC chips on the same FEB. This coincidence identifies signals detected at both ends of a bar and forms the basic ToF bar-level trigger condition.

    \item \textbf{Crate-level coincidence (HLT3):}
    HLT3 combines signals from different SAMPIC boards within the single ToF readout crate. It can form coincidences across the full ToF detector, allowing the selection of multi-panel event topologies.
\end{itemize}

We exploit these capabilities with the following configuration. First, every signal must satisfy a channel self-trigger (LT1) within a threshold set to $20$~mV. Second, we require a board-level coincidence (HLT2) with a time window of $50$~ns. Since the HLT2 logic forms coincidences only between LT1 signals originating from different SAMPIC chips on the same FEB, the two readout ends of each scintillator bar are deliberately connected to different chips. A particle crossing a bar can therefore generate the required HLT2 coincidence through the near-synchronous signals observed at its two ends. Consequently, the HLT2 condition preferentially selects synchronous signals originating from a single bar, significantly suppressing electronic noise while maintaining high signal efficiency. All hits satisfying the board-level coincidence condition and LT1 are recorded by the DAQ, ensuring that valid signals are retained regardless of which trigger mode below initiates the readout.

Crate-level coincidences (HLT3) are used to select events spanning multiple panels, requiring HLT2 triggers from different boards to occur within a 70~ns coincidence window. In stand-alone ToF operation, the local readout can be triggered using either HLT2 or HLT3, depending on the desired event selection. The HLT3 configuration is primarily used for cosmic-ray data, where the multi-panel coincidence efficiently selects through-going tracks for calibration and alignment.

An accepted coincidence can either initiate local ToF readout or be transmitted to the global ND280 trigger chain to initiate the readout of other subdetectors. For an accepted event, only channels satisfying the LT1 condition are digitised.

\begin{figure}[t]
    \centering
    \includegraphics[width=\textwidth]{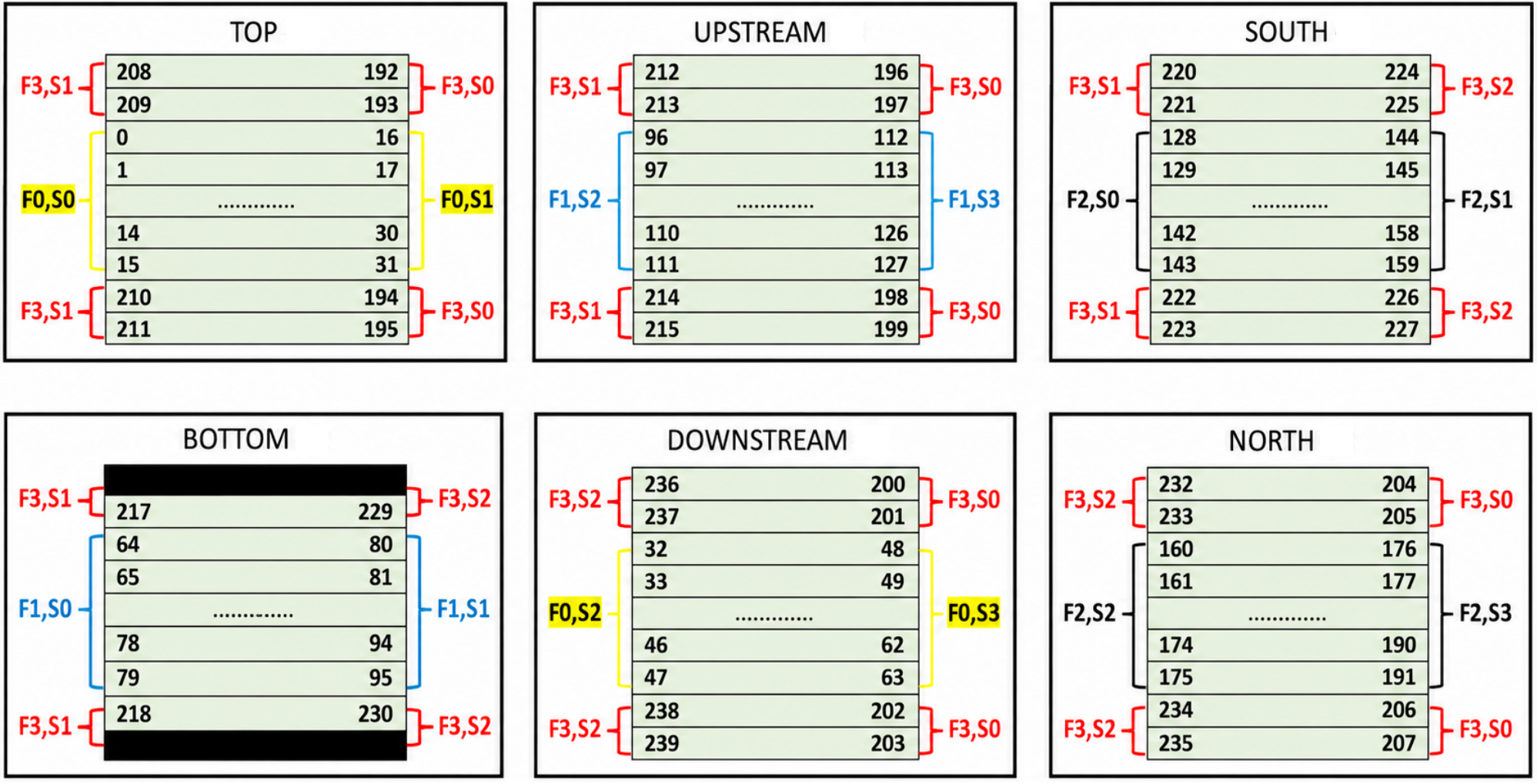}
    \vspace{1 mm}

    \footnotesize
    \setlength{\tabcolsep}{3.5pt}
    \renewcommand{\arraystretch}{0.92}

    \begin{tabularx}{\textwidth}{
        @{}
        c c >{\raggedright\arraybackslash}X c
        @{\hspace{0.4cm}}
        c c >{\raggedright\arraybackslash}X c
        @{}
    }
        \toprule
        \multicolumn{4}{c}{\textbf{Central bars}} &
        \multicolumn{4}{c}{\textbf{Edge bars}} \\
        \cmidrule(r){1-4}
        \cmidrule(l){5-8}

        FEB [$F$] & SAMPIC [$S$] & Panel / end & Bars &
        FEB [$F$] & SAMPIC [$S$] & Panel / end & Bars \\
        \midrule

        0 & 0 & Top US            & 2--17 & 3 & 0 & Upstream North    & 0, 1, 18, 19 \\
          & 1 & Top DS            & 2--17 &   & 0 & Downstream South  & 0, 1, 18, 19 \\
          & 2 & Downstream North  & 2--17 &   & 0 & Top DS            & 0, 1, 18, 19 \\
          & 3 & Downstream South  & 2--17 &   & 0 & North DS          & 0, 1, 18, 19 \\

        \addlinespace[1mm]

        1 & 0 & Bottom US         & 2--17 &   & 1 & Upstream South    & 0, 1, 18, 19 \\
          & 1 & Bottom DS         & 2--17 &   & 1 & Top US            & 0, 1, 18, 19 \\
          & 2 & Upstream South    & 2--17 &   & 1 & Bottom US         & 1, 18 \\
          & 3 & Upstream North    & 2--17 &   & 1 & South DS          & 0, 1, 18, 19 \\

        \addlinespace[1mm]

        2 & 0 & South DS          & 2--17 &   & 2 & Bottom DS         & 1, 18 \\
          & 1 & South US          & 2--17 &   & 2 & South US          & 0, 1, 18, 19 \\
          & 2 & North US          & 2--17 &   & 2 & Downstream North  & 0, 1, 18, 19 \\
          & 3 & North DS          & 2--17 &   & 2 & North US          & 0, 1, 18, 19 \\
          &   &                   &       &   & 3 & Spare             & -- \\

        \bottomrule
    \end{tabularx}

    \caption{Mapping between the ToF scintillator bars and the SAMPIC readout electronics.
    \textit{Top}: schematic of the six ToF panels to the 256 SAMPIC readout channels, with the numbers at each bar end indicating the corresponding global DAQ channel.
    \textit{Bottom}: FEB/SAMPIC assignments for the central and edge bars, using the
    $F$ and $S$ coding shown in the schematic. In the panel/end labels, the first term
    identifies the ToF panel and the second the bar-readout end; US and DS denote the
    upstream and downstream ends, respectively. Central bars are grouped on FEBs~$F0$--$F2$,
    while edge bars are assigned to FEB~$F3$.}
    \label{fig:tof_mapping}
\end{figure}

The implementation of the HLT2 and HLT3 coincidence logic places specific requirements on the channel mapping. To maximise the number of bar coincidences formed within a single board, most channels from each ToF panel are grouped on the same FEB, with the two ends of a bar connected to different SAMPIC chips. Because the number of channels per panel does not map exactly onto the available board capacity, a small number of edge bars are assigned to the fourth board, FEB3. The resulting mapping between the scintillator bars and SAMPIC readout channels is shown in Figure~\ref{fig:tof_mapping}, with the corresponding FEB and SAMPIC assignments summarised in the lower panel. This arrangement preserves efficient two-ended bar coincidences for most of the detector while accommodating all 118 scintillator bars within the four available FEBs.

\subsection{Integration with the ND280 Trigger}
Unlike the other ND280 Upgrade subdetectors, which continuously buffer data for retrieval by the DAQ using a \textit{pull} model, the ToF uses a gate-based \textit{push} architecture without continuous buffering. An acquisition gate defines the interval in which signals can be recorded; once the gate closes, the recorded data are digitised and actively transmitted to the DAQ.

\begin{figure}[t]
    \centering
    \includegraphics[width=0.87\textwidth]{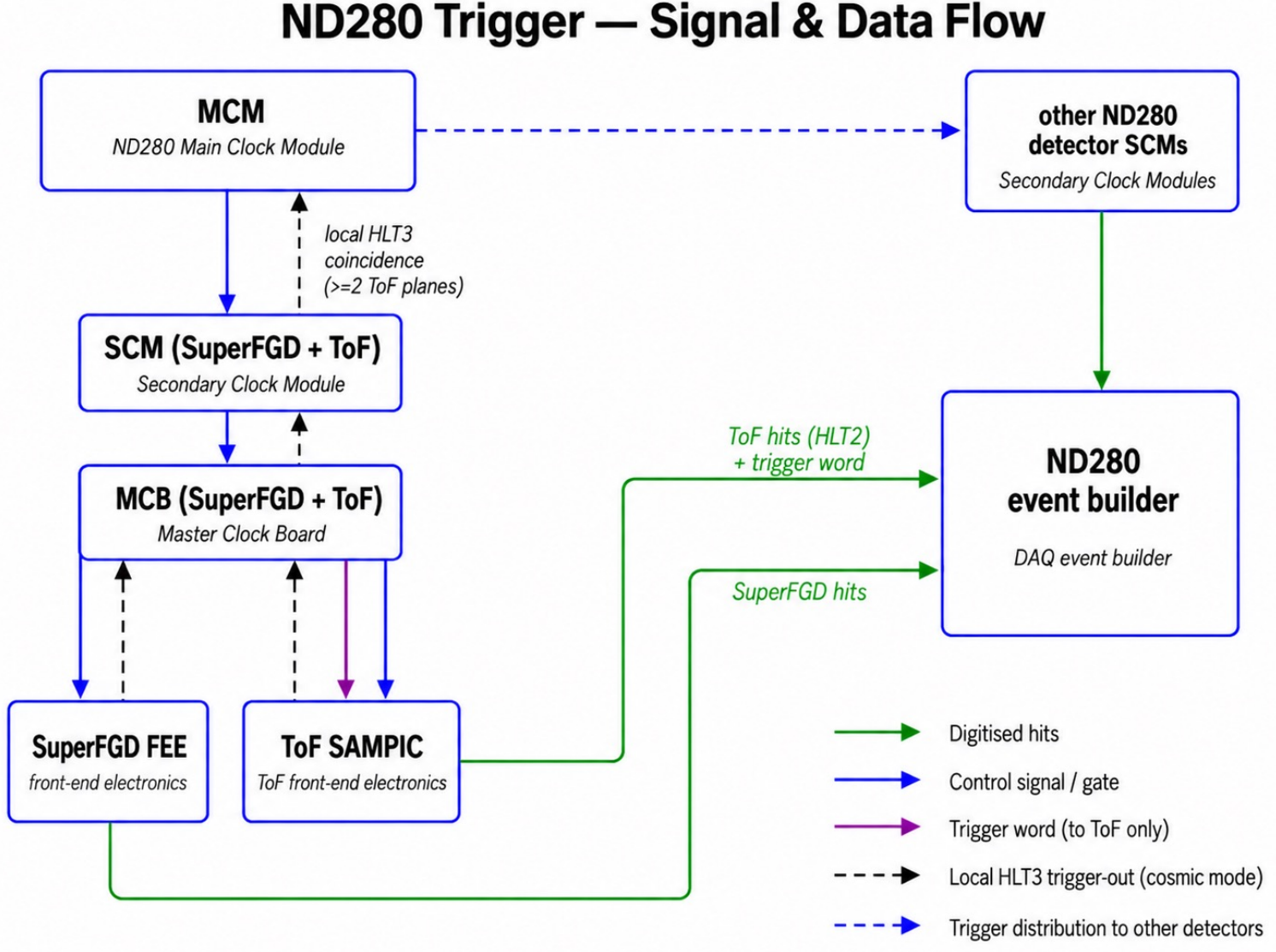}
    \caption{Integration of the ToF detector into the ND280 trigger and DAQ system, showing the trigger, gate, control, and data-flow paths relevant to ToF operation.}
    \label{fig:tof_trigger_flow}
\end{figure}

Beyond stand-alone operation, the ToF participates in three configurations within the global ND280 trigger system, as illustrated in Figure~\ref{fig:tof_trigger_flow}. The established ND280 back-end architecture uses the Main Clock Module (MCM) to coordinate clock and trigger distribution, with Secondary Clock Modules (SCMs) distributing these signals to the individual subdetector systems~\cite{Thorpe:2011T2KDAQ}. For the ND280 Upgrade, the Master Clock Board (MCB) provides the interface between this common timing system and the upgraded detector electronics~\cite{T2K:2019bbb}. In the ToF configuration, the acquisition gate and trigger word are transmitted through the SCM and MCB to the ToF electronics.

This interface also provides the timing reference needed to synchronise the ToF with the other subdetectors. The MCM, SCM, MCB, and the SuperFGD front-end electronics operate with a 100~MHz clock, whereas the ToF SAMPIC electronics use a 50~MHz clock; the corresponding clock-domain handling is included in the ToF trigger and timestamp reconstruction. The three configurations differ primarily in the origin of the trigger request: the beam trigger, an external cosmic-ray trigger from another subdetector, or a locally generated ToF trigger:
\begin{enumerate}
    
    \item \textbf{Beam trigger:} During standard beam data taking, the full ND280 trigger chain is used. The MCM distributes a trigger signal and a serial trigger word containing the event metadata, including the spill number, event number, and trigger type, through the SCMs. The ToF and SuperFGD share an SCM, and the MCB connected to this SCM forwards the trigger signal to both subdetectors, while the trigger word is sent only to the ToF. This common gate reference enables precise timing comparisons between the ToF and other subdetectors, particularly the SuperFGD.

    For the ToF, the trigger signal defines the acquisition gate, during which the SAMPIC boards apply the HLT2 coincidence logic. After the gate closes, the accepted signals are digitised and the reconstructed ToF hits are sent to the ND280 event builder together with the trigger word. A readout-complete acknowledgement is also issued by the ToF DAQ. The event builder combines the ToF data with the corresponding data from the other subdetectors to form a complete ND280 event.

    \item \textbf{ND280 cosmic trigger:} Outside beam spills, the ToF can be read out in response to a cosmic-ray trigger generated by another subdetector, either from the FGDs~\cite{AMAUDRUZ20121} or from subdetectors using the Trip-T-based readout and trigger system~\cite{Thorpe:2011T2KDAQ}. As in the beam-trigger configuration, the MCM opens a ToF acquisition gate through the MCB, during which the SAMPIC boards operate in HLT2 mode.

    The trigger word is issued only if the external subdetector identifies a valid cosmic-ray candidate. Because this decision requires additional processing, the trigger word generally arrives with a longer latency than in beam-triggered operation. If it is received within the allowed time, the ToF data recorded during the gate are associated with the trigger word and sent to the ND280 event builder. Otherwise, the local ToF data for that gate are discarded.

    \item \textbf{ToF-triggered readout:} The ToF provides the trigger for the readout of other ND280 Upgrade subdetectors, including the HA-TPCs and SuperFGD. In this mode, a locally generated HLT3 coincidence, requiring hits in at least two ToF panels, is sent to the MCM. Once accepted, the trigger is distributed through the same global trigger path as in the other configurations, and the participating subdetectors follow their standard readout sequence. The ToF hits satisfying the HLT2 condition are recorded simultaneously and included in the resulting ND280 event. This configuration is primarily used for cosmic-ray and calibration runs.
\end{enumerate}

\subsection{Slow Control System}
The ToF slow control (SC) system provides centralised monitoring and operational control of the detector subsystems, including SiPM bias voltages, front-end amplifier power, SAMPIC power supplies, and environmental conditions. The system was developed at the University of Geneva and is an integral part of the ND280 control infrastructure.

\begin{figure}[htbp]
    \centering
    \includegraphics[
        width=0.7\textwidth
    ]{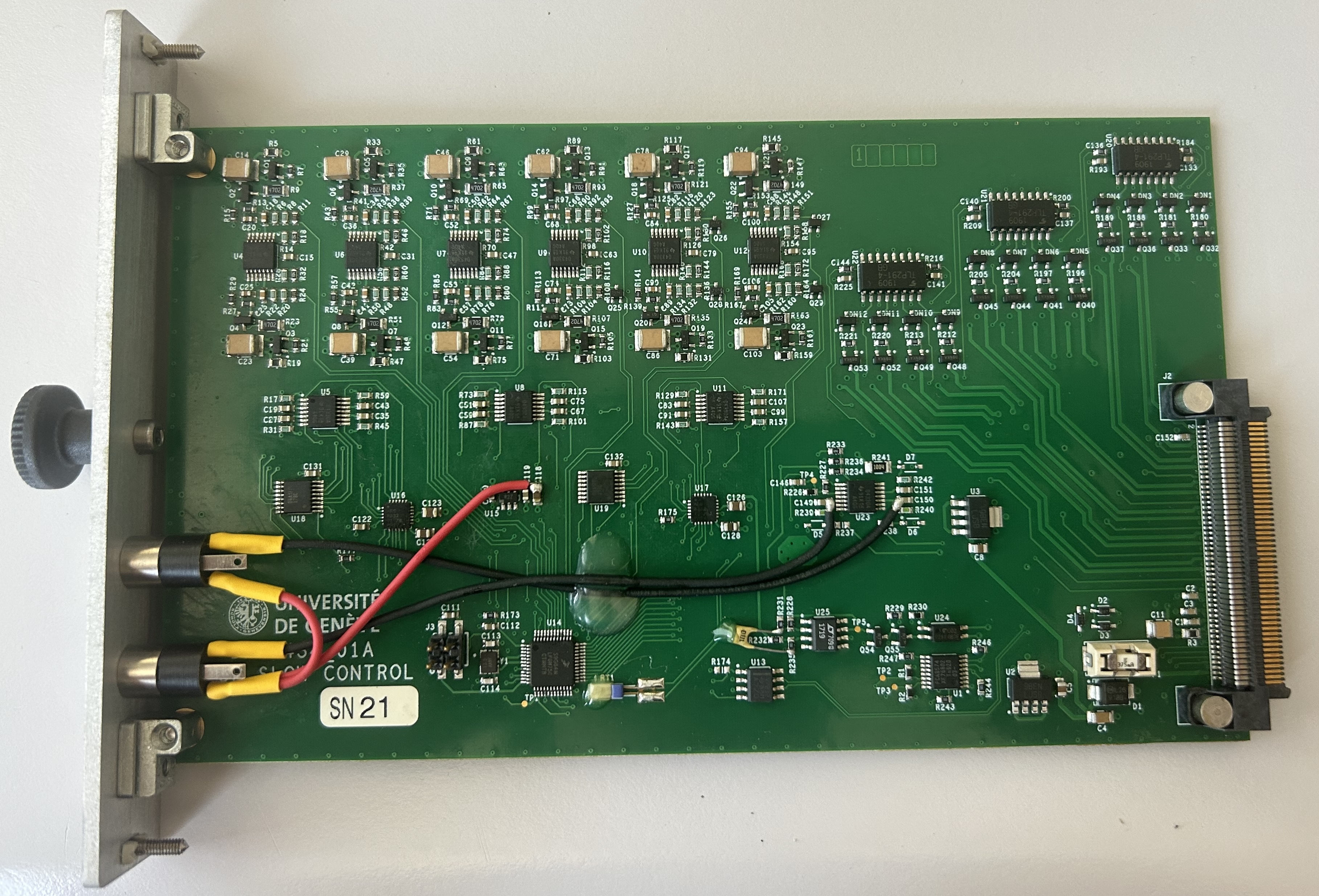}

    \caption{Example slow-control board providing SiPM bias regulation, front-end amplifier power, and temperature-sensor connections.\vspace{-2 ex}}
    \label{fig:sc_midas}
\end{figure}

The SC hardware consists of 20 custom electronic boards housed in two crates located on opposite sides of the detector. An example board is shown in Figure~\ref{fig:sc_midas}. Each board services 12 readout channels, providing the SiPM bias voltage and the low-voltage supply required by the front-end amplifiers. The complete system therefore provides 240 regulated channels, sufficient for the 236 instrumented ends of the 118 scintillator bars, with four spare channels.
Power is provided by a Wiener multi-channel power supply (MPOD) through three nominal supply rails: 111~V for the per-channel SiPM bias regulators, 9~V for the front-end amplifiers, and 4~V for the slow-control electronics. Distributing these common supplies through the SC boards reduces the number of external power connections and allows the different detector components to be powered and monitored independently.

The SiPM bias voltage is stabilised using a closed-loop regulation system. For each channel, the delivered voltage is continuously measured and compared with its programmed set point, allowing the regulator output to compensate for variations in the input supply and electrical load. This maintains a stable applied bias voltage and limits the corresponding variations in SiPM gain. Channel voltages and currents are also monitored against configurable operating limits, allowing abnormal conditions to generate alarms and, where configured, disable the affected output.

Each SC board reads out two negative-temperature-coefficient (NTC) thermistors installed on the associated ToF panels, providing measurements of the local detector temperature. A~PT100 sensor mounted on each board independently monitors the temperature of the slow-control electronics. The temperature measurements are used for detector monitoring, since the SiPM breakdown voltage and operating gain are temperature dependent. However, no automatic temperature-dependent adjustment of the SiPM bias voltage is applied.

The SC crates are connected to the slow-control computer through a Controller Area Network~(CAN) bus interface. The control and monitoring software is implemented within the MIDAS data-acquisition framework~\cite{MIDAS}. A dedicated MIDAS front-end provides a graphical interface through which operators can control the SC and SAMPIC power supplies, switch detector subsystems on or off, and monitor voltages, currents, and temperatures in real time. These quantities are stored in an online database and displayed in periodically updated history plots. Configurable warning and alarm thresholds identify conditions such as overcurrent, voltage excursions, excessive temperatures, or communication errors, allowing the system to take appropriate protective action. The ToF SC system was integrated into the global ND280 slow-control infrastructure before the start of data taking in October~2023 and is connected to the experiment interlock system for detector safety.

\subsection{Mechanical Support and Rotation System}
\label{ssc:mechanicalSupport}
\begin{figure}[htbp]
    \centering

    \includegraphics[
        width=0.85\textwidth
    ]{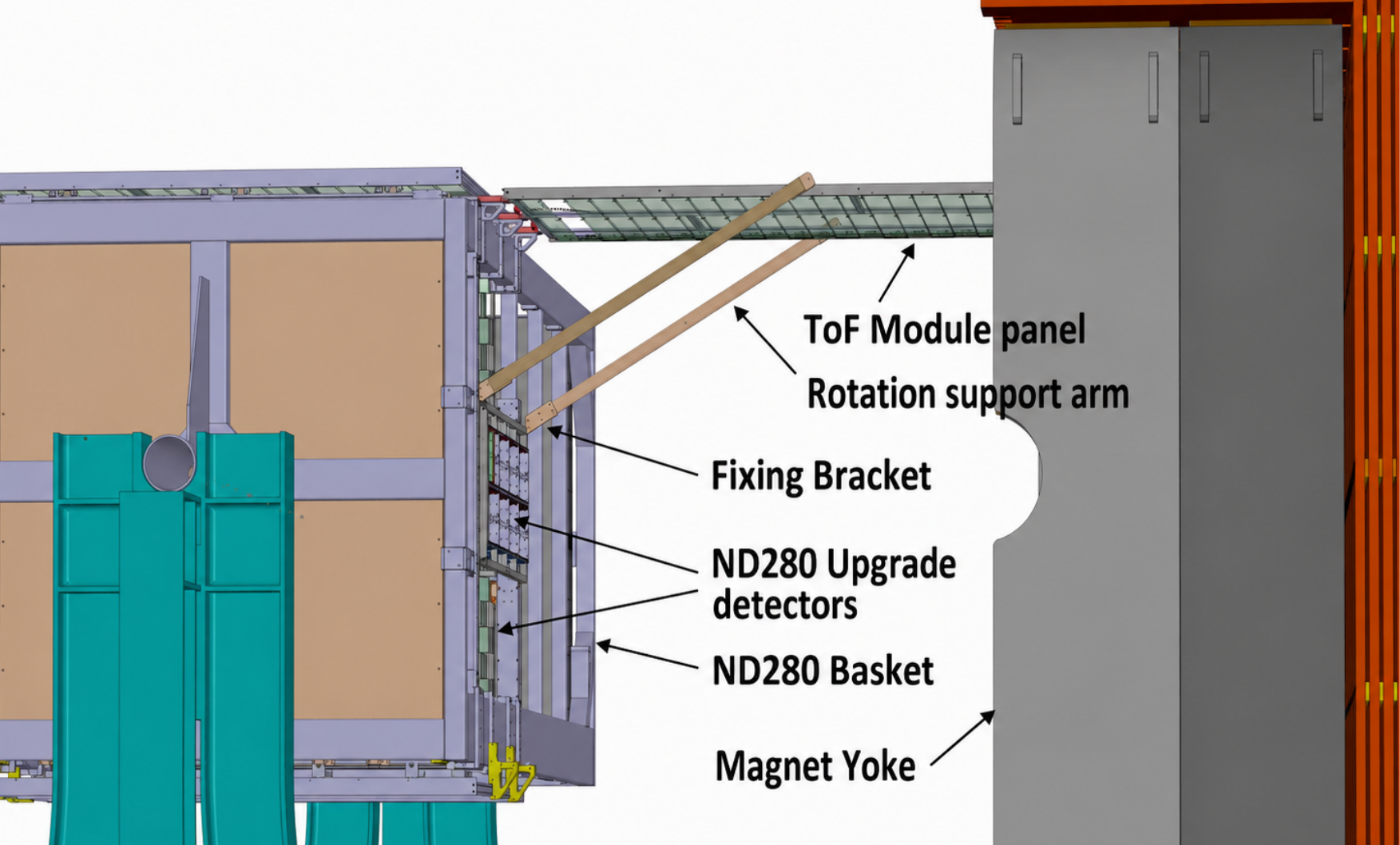}

    \caption{Schematic of the North/South ToF panel rotation system in the open configuration. The panel is rotated by approximately $90^\circ$ and supported by side braces, providing access to the inner detectors while maintaining mechanical stability and preserving all cabling connections.}
    \label{fig:tof_rotation}
\end{figure}

The ToF detector panels are mounted on the ND280 support structure using mechanical fixation brackets and reinforcing beams that ensure stable positioning during data taking. The fixation system of the North and South ToF panels was also designed to allow for controlled rotation of these panels, as illustrated in Figure~\ref{fig:tof_rotation}. This capability enables the panels to be opened by approximately~$90^\circ$, providing access to the inner subdetectors, including the SuperFGD, HA-TPC electronics, and part of the upstream ECal electronics, for maintenance interventions.

Rotation can be performed without dismantling the ToF system or disconnecting the high-voltage, low-voltage, signal, or SC cables, preserving detector integrity and reducing the time required to access internal detector components. During normal operation, the panels are securely attached to the ND280 basket through the bottom brackets and reinforcement beams. For the open configuration, the panels are released from the bottom brackets and rotated using dedicated lifting points and the crane system. Once in position, they are supported by removable side braces attached to interface brackets on the ND280 basket. These brace structures stabilise the panels in the rotated configuration, allowing the crane slings to be released while maintaining a safe and mechanically stable setup. The full rotation procedure was rehearsed on both side ToF panels, confirming the stability of the opened configuration.
\subsection{Installation and Integration}

The initial assembly and testing of the ToF detector panels was carried out at CERN under the responsibility of the University of Geneva. The scintillator bars, together with their SiPM readout assemblies and front-end electronics, were mounted into mechanical frames to form the six detector panels. The mechanical construction of the full detector system was completed in 2020.
\begin{figure}[!htbp]
    \centering

    \resizebox{\textwidth}{!}{%
        \includegraphics[height=0.22\textheight,keepaspectratio]{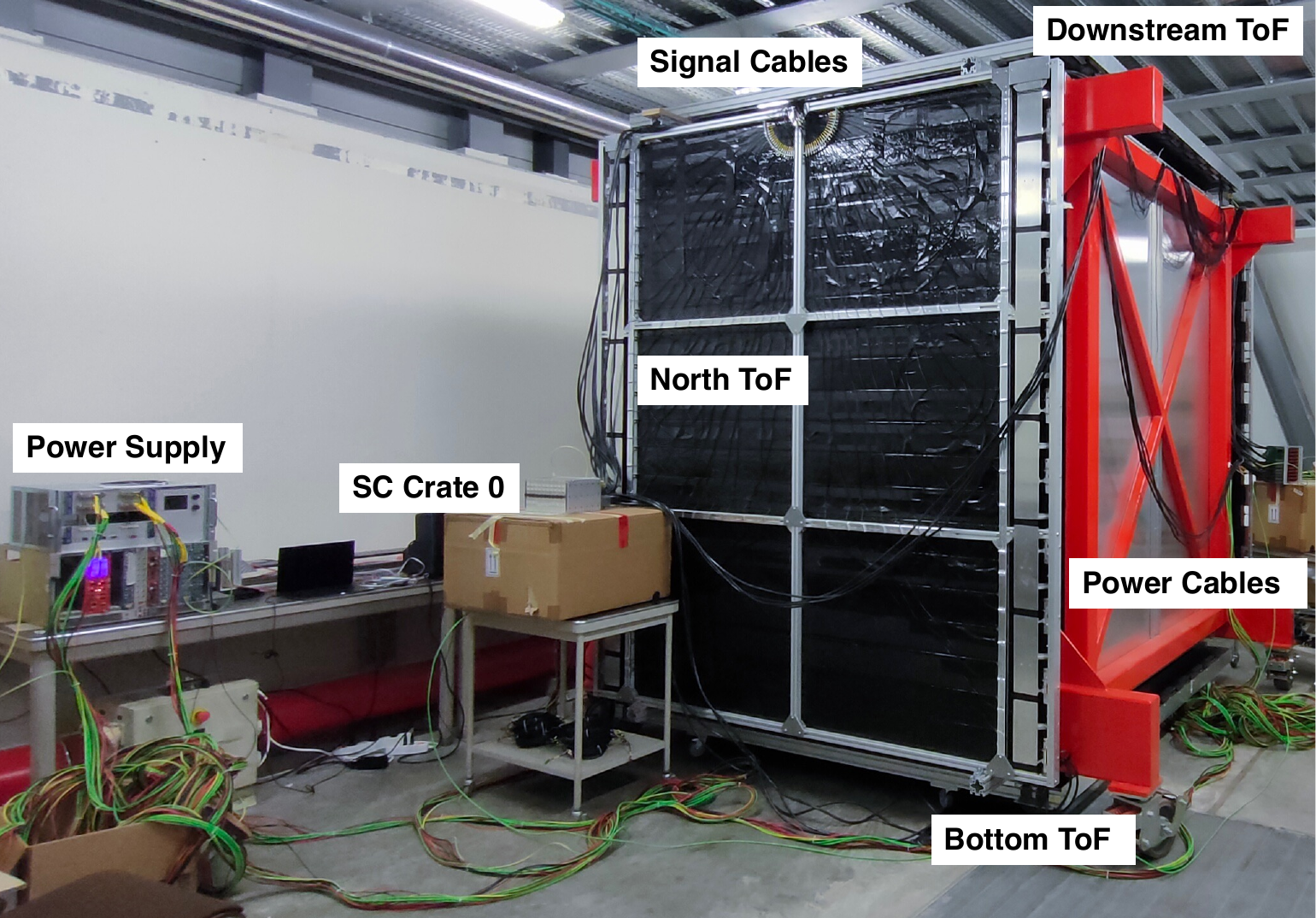}
        \hspace{0.03\textwidth}
        \includegraphics[height=0.22\textheight,keepaspectratio]{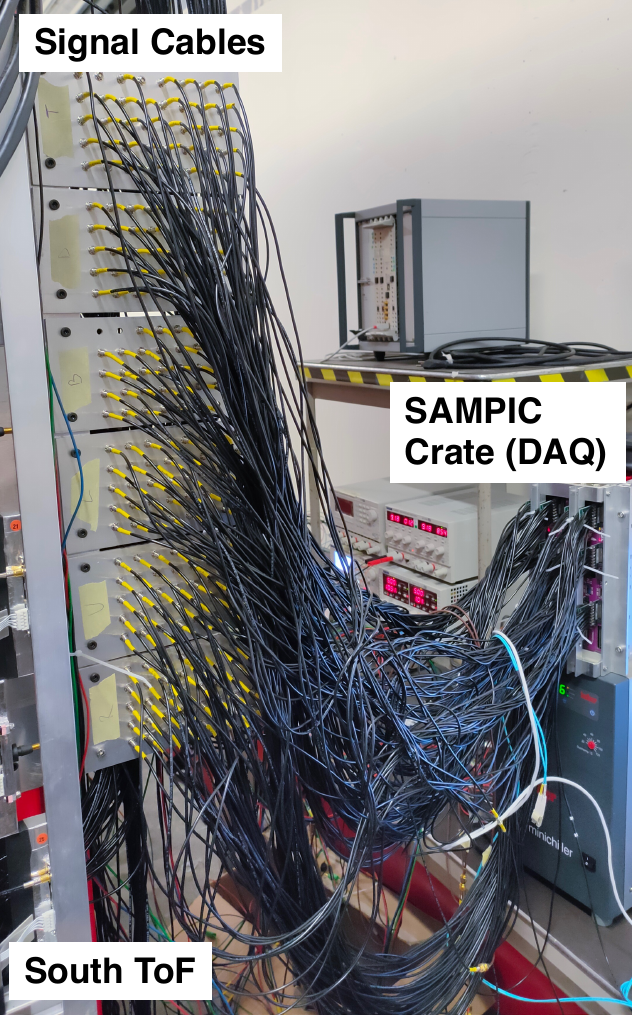}
    }
    \caption{Fully assembled ToF detector during pre-shipment commissioning at the CERN Neutrino Platform. \textit{Left}: detector and slow-control area. \textit{Right}: signal cabling and DAQ system.}
    \label{fig:commissioning_cern}
\end{figure}

\begin{figure}[!htbp]
    \centering

    \begin{minipage}[c]{0.49\textwidth}
        \centering
        \includegraphics[
            width=\linewidth,
            trim=0 1.5 0 1.5,
            clip
        ]{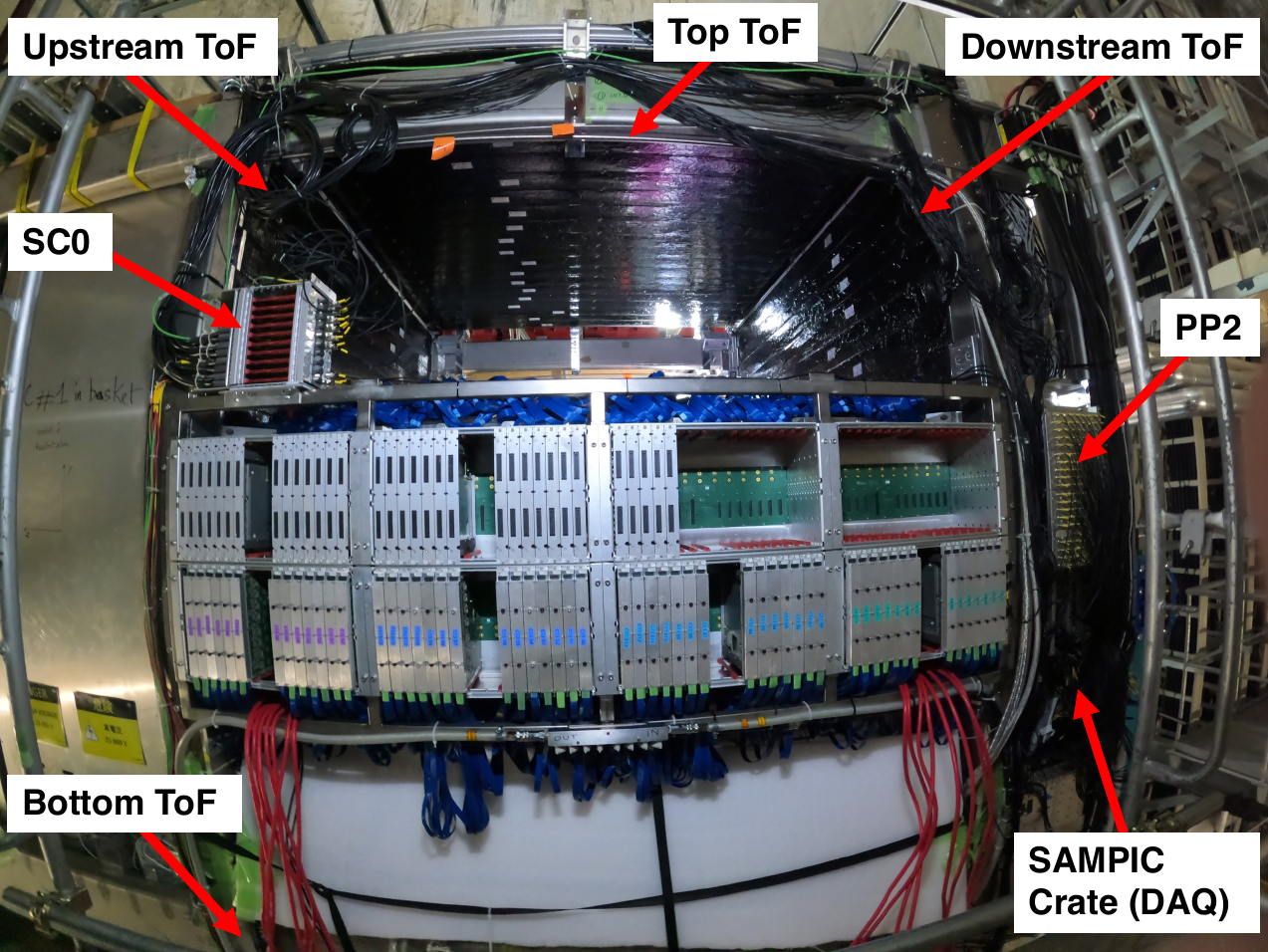}
    \end{minipage}
    \hfill
    \begin{minipage}[c]{0.49\textwidth}
        \centering
        \includegraphics[
            width=\linewidth,
            trim=0 0 0 0,
            clip
        ]{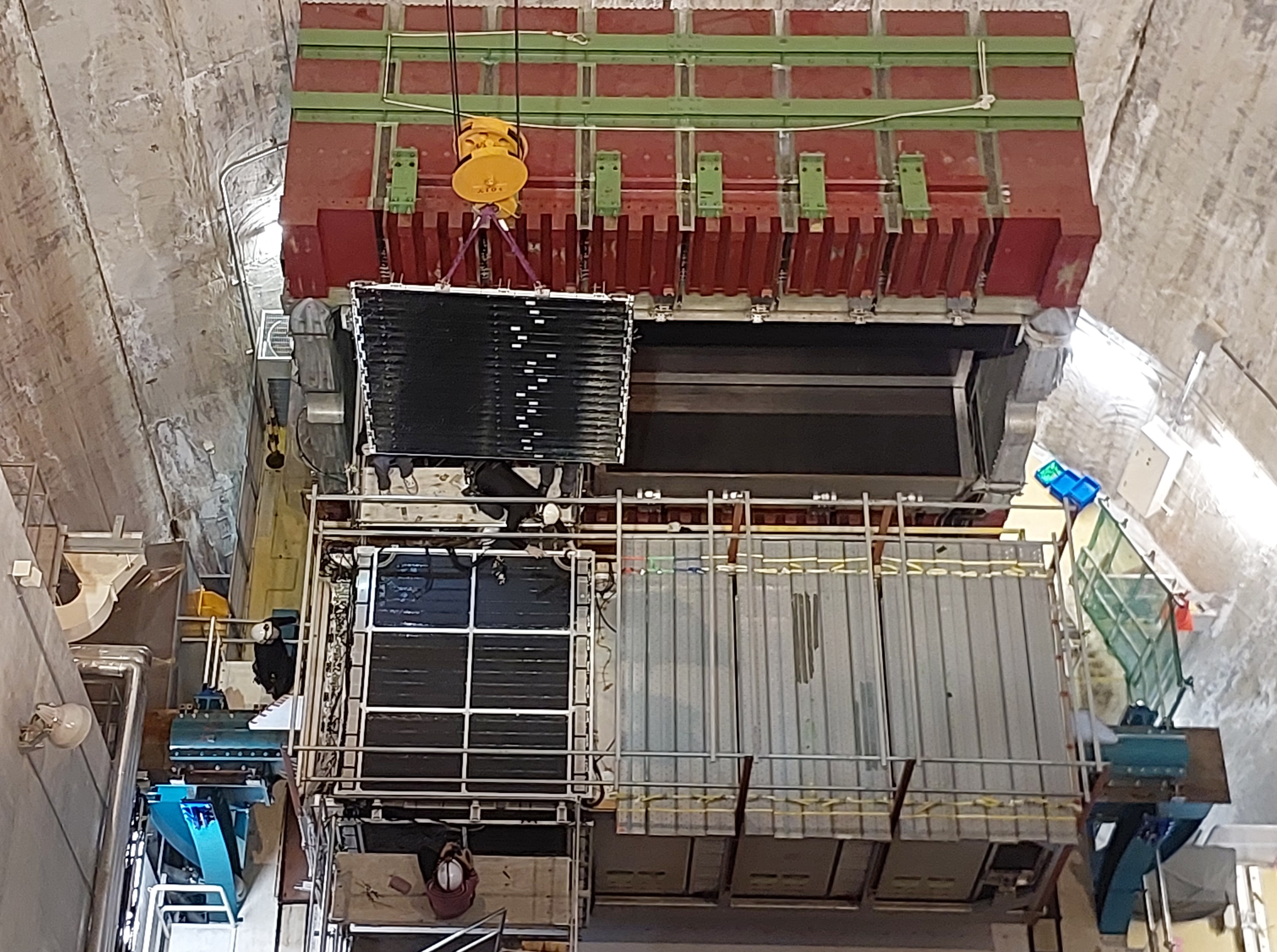}
    \end{minipage}

    \vspace{1mm}

    \begin{minipage}[c]{0.49\textwidth}
        \centering
        \includegraphics[
            width=\linewidth,
            trim=0 0 0 0,
            clip
        ]{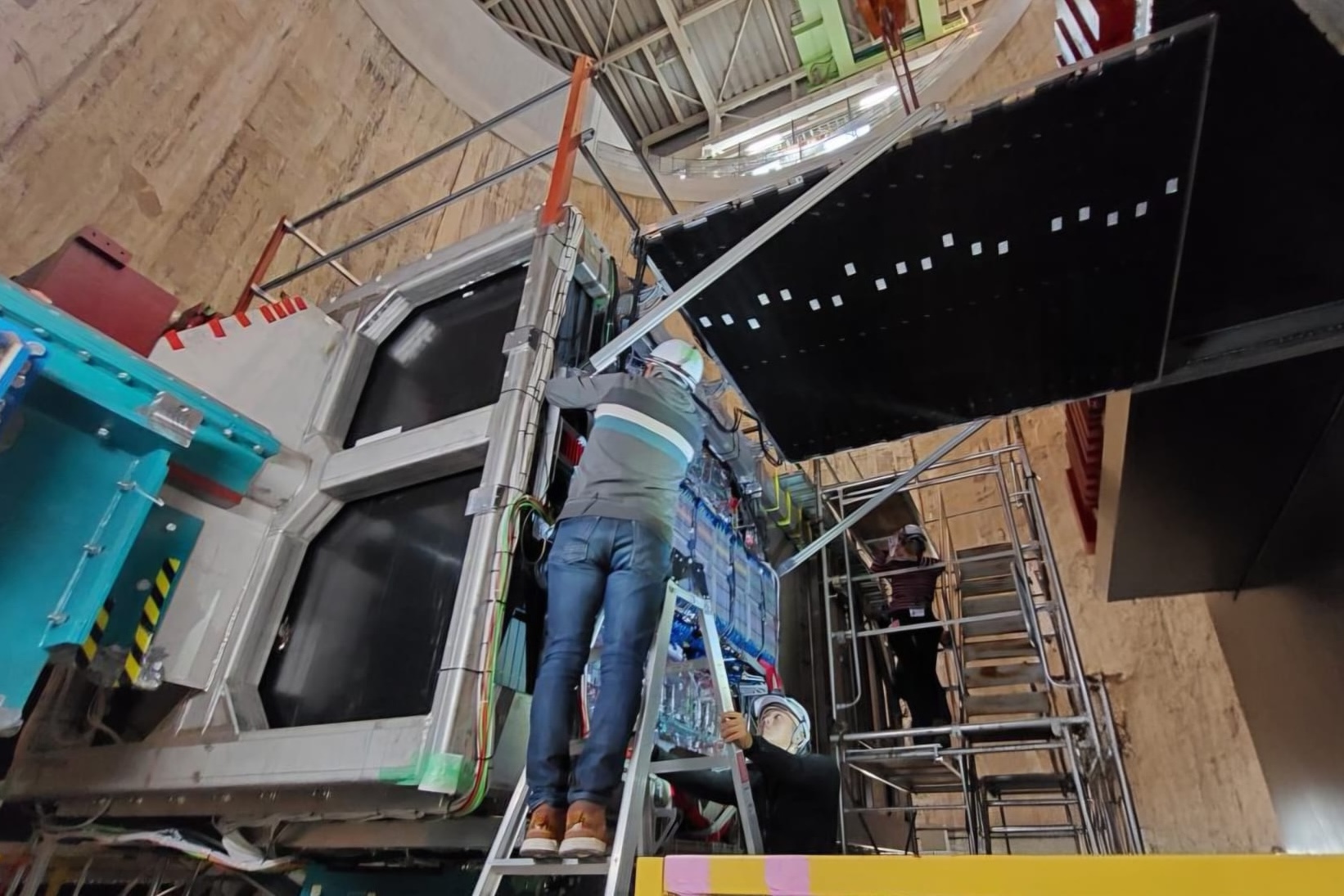}
    \end{minipage}
    \hfill
    \begin{minipage}[c]{0.49\textwidth}
        \centering
        \includegraphics[
            width=\linewidth,
            trim=0 9 0 9,
            clip
        ]{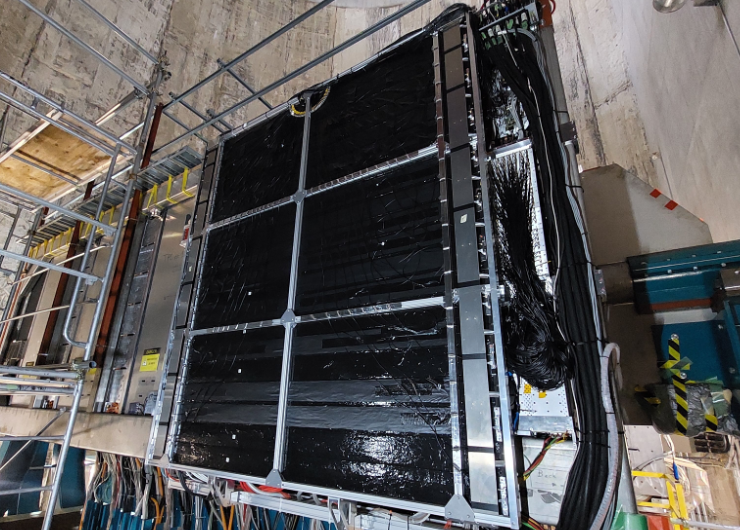}
    \end{minipage}
    \caption{Selected stages of the ToF installation in ND280. \textit{Top left}: four ToF panels, the SuperFGD, and the bottom HA-TPC installed in ND280. \textit{Top right}: installation of the South ToF panel in the ND280 pit. \textit{Bottom left}: North ToF panel in the rotated open configuration during installation. \textit{Bottom right}: completed ND280 Upgrade at J-PARC, with a lateral ToF panel visible.}
    \label{Fig:ToFinstallation}
\end{figure}
Following the assembly of the individual panels, the detector was integrated and commissioned at the CERN Neutrino Platform. Three panels were initially mounted on the ND280 mock-up basket support structure in September 2021, with the remaining panels installed in February 2022. This setup allowed the full detector geometry and cabling scheme to be validated under conditions representative of the final installation. During this phase, the detector was connected to the SC system, power supplies, and DAQ electronics, allowing full-system operation and data-taking tests. The fully assembled detector at the CERN Neutrino Platform is shown in Figure~\ref{fig:commissioning_cern}.

The commissioning campaign included validation of the signal readout chain, verification of the detector response, and optimisation of the cabling layout and integration procedures. The tests also provided an opportunity to rehearse installation procedures, including the handling and positioning of the detector panels using a crane, as well as the routing and connection of signal and control cables in the constrained space around the detector.

The detector was shipped to J-PARC in spring 2023, where installation in ND280 was carried out in several stages between July 2023 and May 2024, coordinated with the installation of the SuperFGD and the HA-TPCs. Integration of the ToF into the global ND280 DAQ system was completed in November 2023, with subsequent installation activities continuing through May 2024. Photographs of selected stages of the installation process are shown in
Figure~\ref{Fig:ToFinstallation}.

\section{Signal Reconstruction and Single-Bar Detector Performance}
\label{sec:reconstruction}
This section describes the reconstruction of ToF signals, including waveform processing, time extraction, and the determination of the particle crossing position and time using the two-ended bar readout. The intrinsic single-bar performance is then evaluated with cosmic-ray data, providing reference timing and position resolutions for the commissioning studies.

\subsection{Waveform Processing and Time Extraction}
\label{sec:waveform_processing}
Signals from the ToF scintillator bars are digitised by the SAMPIC system described in Section~\ref{Sec:DAQ}. Each waveform contains 64 samples recorded at 3.2~GS/s, corresponding to a sampling interval of 312.5~ps and a total acquisition window of 20~ns. The full pulse can extend beyond this window because of the SiPM and front-end electronics response. However, the rising edge, which contains the information used for timing reconstruction, is recorded within the acquisition window.

The first five samples of each waveform are used to estimate the baseline and electronic noise level before the signal time is reconstructed. The hit time is then extracted using a constant-fraction discriminator (CFD), with the threshold set to 10\% of the baseline-subtracted peak amplitude. The signal time is defined as the point at which the rising edge crosses this threshold, reducing the dependence of the reconstructed time on variations in pulse height. A more detailed description of the waveform parametrisation and reconstruction algorithm developed for the ToF detector is given in Ref.~\cite{Alt:2025msx}.

Before the two-ended bar information is combined to reconstruct the particle crossing position and time, the extracted signal times are corrected for channel-dependent offsets arising from differences in cable lengths, electronics delays, and SiPM response variations. The timing calibration procedure used to determine these corrections is described in Section~\ref{sec:timecalibration}.

{
\setlength{\abovedisplayskip}{6pt}
\setlength{\abovedisplayshortskip}{6pt}
\setlength{\belowdisplayskip}{6pt}
\setlength{\belowdisplayshortskip}{6pt}
\subsection{Hit Position and Time Reconstruction}

Each scintillator bar is read out at both ends by SiPM arrays, allowing the particle crossing position along the bar to be reconstructed from the difference in signal arrival times at the two ends. Denoting the measured times at the left and right ends as $t_1$ and $t_2$, respectively, the arrival times can be expressed as
\begin{equation}
t_1 = t_0 + \frac{x}{v} + \delta(x),
\qquad
t_2 = t_0 + \frac{L-x}{v} + \delta(L-x),
\end{equation}
where $t_0$ is the particle crossing time, $x$ is the longitudinal crossing position measured from the first end of the bar, $L$ is the bar length, and $v$ is the effective propagation velocity of scintillation light in the plastic scintillator. The terms $\delta(x)$ and $\delta(L-x)$ account for small position-dependent effects, including light attenuation and internal reflections, which can modify the pulse shape and reconstructed signal time. These corrections are small and are neglected in the present reconstruction, with their leading effect absorbed into the effective propagation velocity. Dedicated measurements determined this velocity to be $v \approx 16$~\si{\centi\metre\per\nano\second}~\cite{Alt:2025msx}.

Neglecting these position-dependent corrections, the particle crossing position and time are reconstructed as
\begin{equation}
x = \frac{v}{2}\left(t_1-t_2+\frac{L}{v}\right),
\qquad
t_0 = \frac{1}{2}\left(t_1+t_2-\frac{L}{v}\right).
\label{Eq:TimePositionReconstruction}
\end{equation}

Assuming independent timing uncertainties at the two readout ends, the corresponding uncertainties on the reconstructed position and crossing time are
\begin{equation}
\sigma_x = \frac{v}{2}\sqrt{\sigma_1^2+\sigma_2^2},
\qquad
\sigma_{t_0} = \frac{1}{2}\sqrt{\sigma_1^2+\sigma_2^2},
\label{Eq:ErrorSimple}
\end{equation}
where $\sigma_1$ and $\sigma_2$ are the timing resolutions of the individual readout ends.

A more precise determination of the particle crossing time is possible when the crossing position along the ToF bar, \(x_{\rm trk}\), is constrained by the other upgraded ND280 detectors. In a combined reconstruction, this constraint can be obtained iteratively. An initial estimate of \(t_0\), derived from the fast timing information of the ToF and SuperFGD, can be used to determine the absolute drift coordinate of a track reconstructed in the HA-TPC. The resulting HA-TPC trajectory provides a more precise estimate of the position at which the track intersects the ToF bar, allowing the propagation-time correction and hence \(t_0\) to be refined. Assuming that the uncertainty on \(x_{\rm trk}\) is negligible compared with the timing uncertainties at the two readout ends, the crossing time can be estimated as a weighted average:
\begin{equation}
t_0 =
\frac{
\sigma_2^2 \left(t_1-\frac{x_{\rm trk}}{v}\right)
+
\sigma_1^2 \left(t_2-\frac{L-x_{\rm trk}}{v}\right)
}{
\sigma_1^2+\sigma_2^2
}.
\label{Eq:TimeWeighted}
\end{equation}
The corresponding optimal time resolution is
\begin{equation}
\sigma_w =
\sqrt{
\frac{\sigma_1^2\sigma_2^2}
{\sigma_1^2+\sigma_2^2}
}.
\label{Eq:ErrorWeighted}
\end{equation}

For the full-detector commissioning studies presented below, an independent measurement of the crossing position is unavailable, so the ToF-only reconstruction is used. The dedicated single-bar study in Section~\ref{sec:singlebar} is an exception, where external trigger counters provide the reference position.
}
\subsection{Single-Bar Detector Characterisation}
\label{sec:singlebar}

\begin{figure}[htbp]
    \centering
    \includegraphics[
        width=0.63\textwidth
    ]{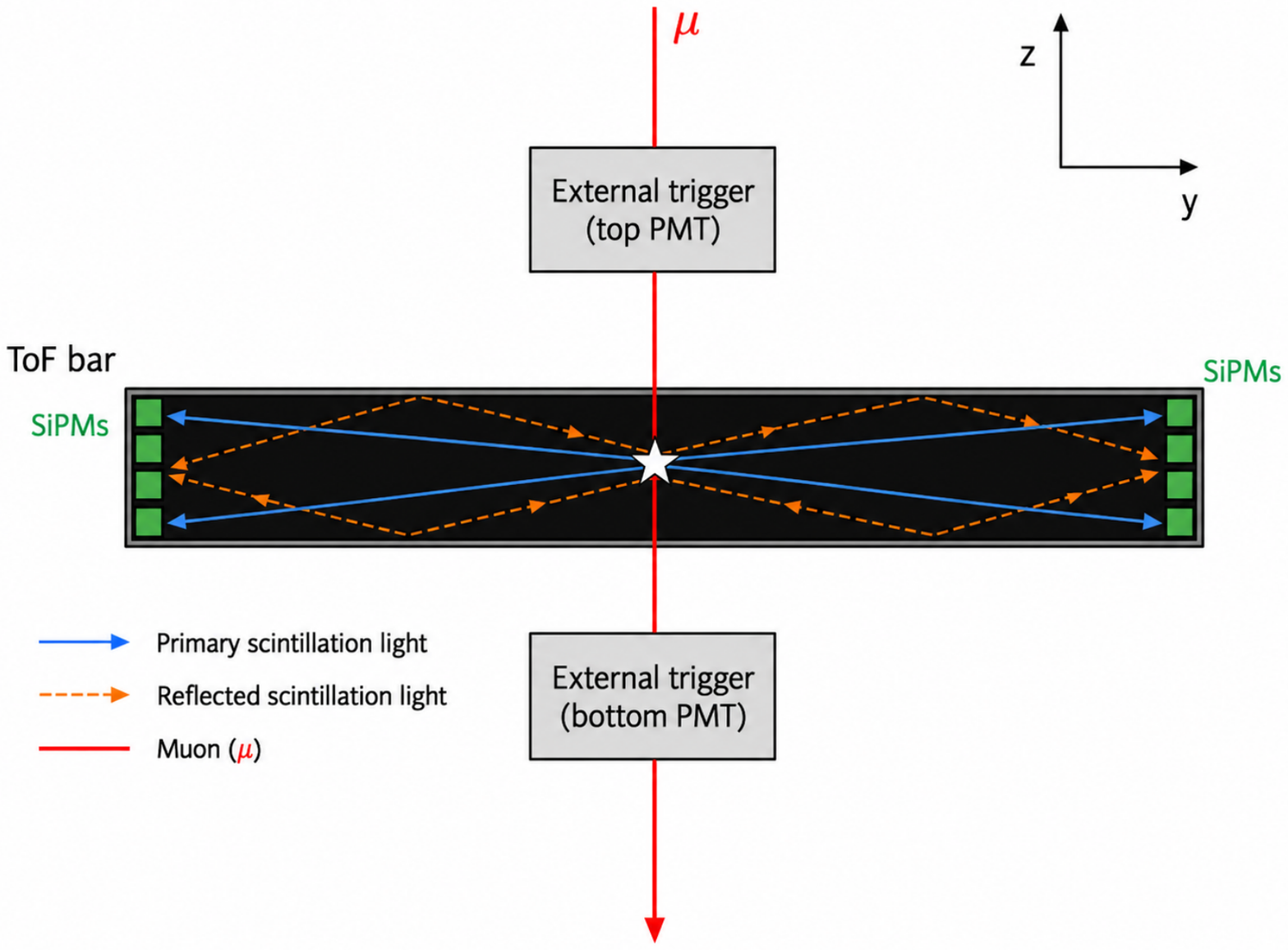}

    \caption{Schematic of the single-bar coincidence configuration. External trigger counters positioned above and below the ToF bar select approximately vertical cosmic-ray tracks and define the reference crossing position along the bar.}
    \label{fig:singlebar_schematic}
\end{figure}

\begin{figure}[htbp]
    \centering

    \makebox[\textwidth][c]{%
        \includegraphics[
            height=0.3\textheight,
            keepaspectratio
        ]{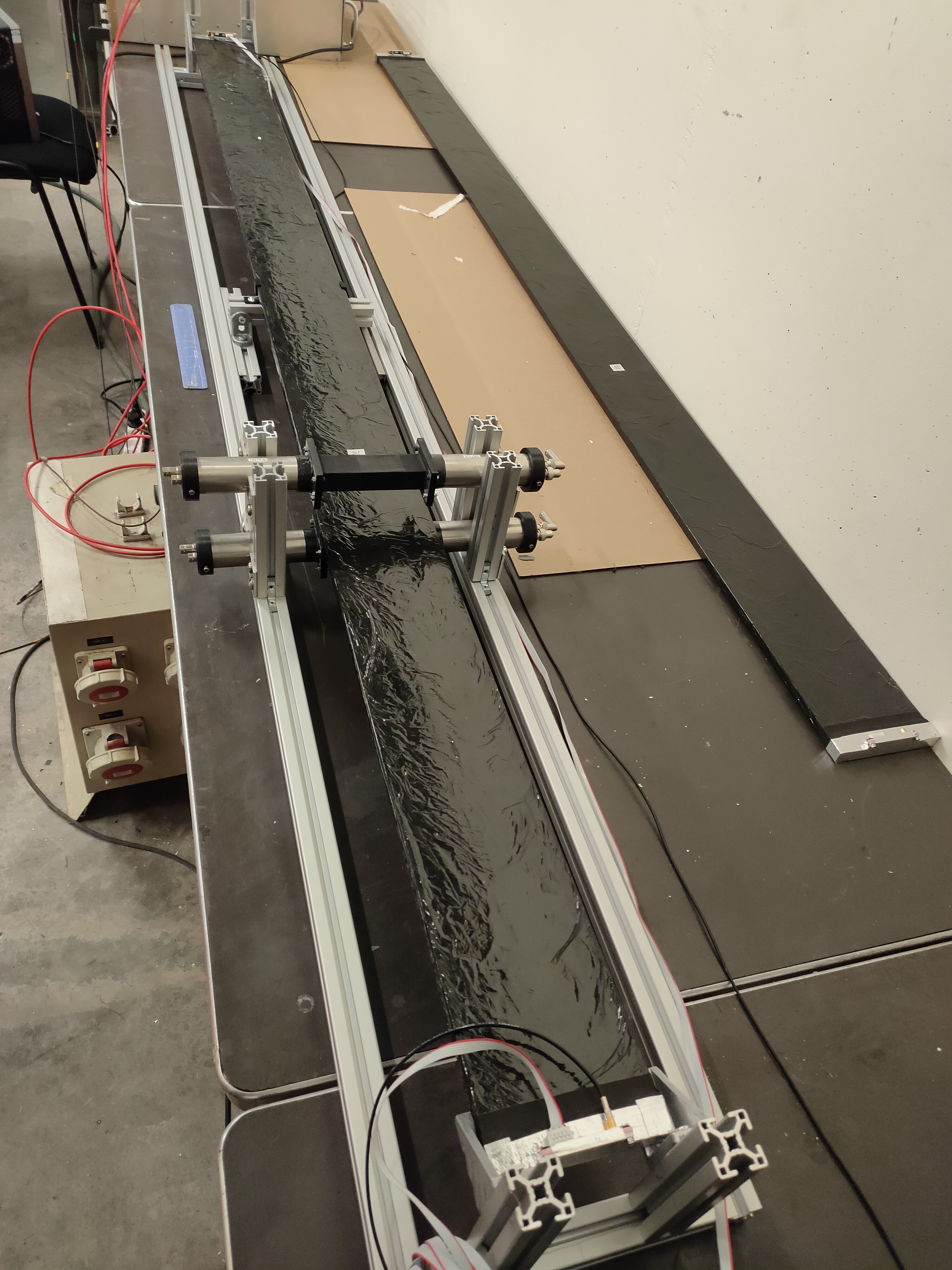}
        \hspace{0.04\textwidth}
        \includegraphics[
            height=0.3\textheight,
            keepaspectratio
        ]{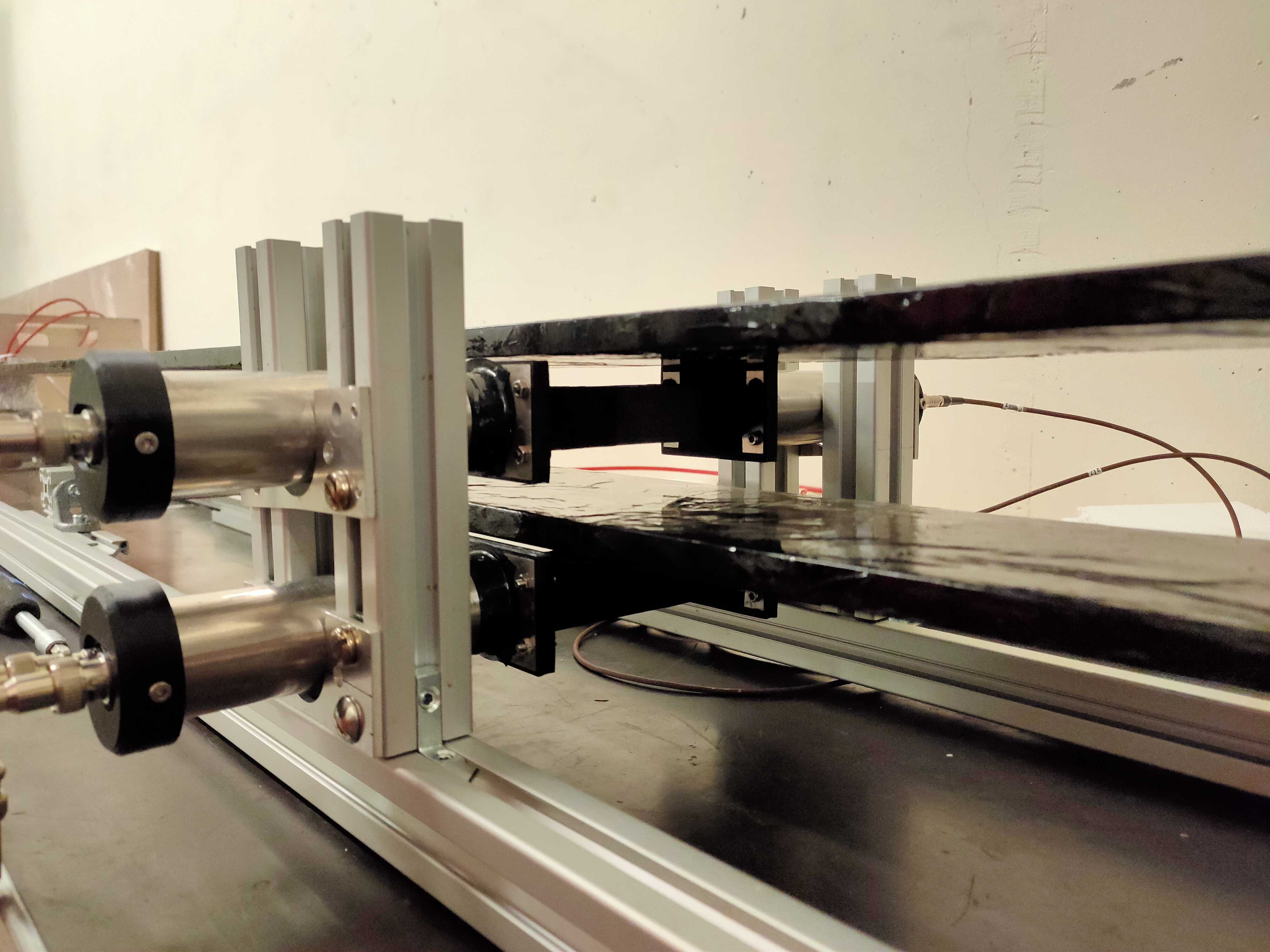}
    }

    \caption{Photographs of the single-bar cosmic-ray test bench. \textit{Left}: instrumented ToF scintillator bar. \textit{Right}: movable external trigger-counter setup used to select approximately vertical cosmic-ray tracks at different reference positions along the bar.}
    \label{fig:singlebar_setup}
\end{figure}

The intrinsic single-bar performance of the ToF detector was studied using a dedicated cosmic-ray test bench developed during the detector R\&D phase~\cite{Villa:2025thesis}. In this setup, a spare ToF scintillator bar instrumented with SiPM arrays at both ends was placed between two $2 \times 2 \times 2~\mathrm{cm}^3$ scintillator trigger counters positioned above and below the bar. The coincidence of these counters selected cosmic-ray tracks crossing the bar approximately vertically. The trigger-counter assembly could be moved along the bar to sample different crossing positions, allowing the detector response to be characterised as a function of distance from the readout ends. The coincidence configuration is illustrated schematically in Figure~\ref{fig:singlebar_schematic}, while photographs of the corresponding test-bench arrangement are shown in Figure~\ref{fig:singlebar_setup}.

For each measurement position, the waveform timing from the SiPM signals was reconstructed using the CFD algorithm described in Section~\ref{sec:waveform_processing}. The reconstructed position along the bar was obtained from the time difference between the two ends according to Equation~\ref{Eq:TimePositionReconstruction}.

\begin{figure}[b]
\centering
\includegraphics[width=0.95\textwidth,trim=10 5 10 5,clip]{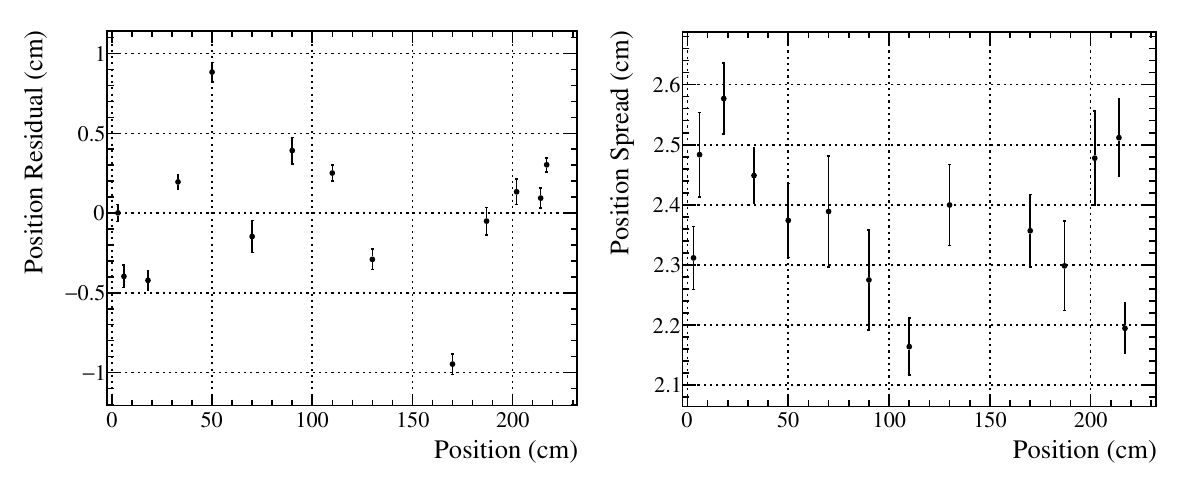}
\caption{Single-bar position reconstruction performance. \textit{Left}: residual between the reconstructed and reference positions along the bar. \textit{Right}: Gaussian width of the reconstructed position distribution at each reference position.}
\label{fig:singlebar_position}
\end{figure}

Figure~\ref{fig:singlebar_position} shows the deviation of the reconstructed position from the reference position across all measurement points. The maximum absolute bias is approximately 1~cm, while the average Gaussian width of the reconstructed-position distributions is 2.4~cm. Combining these contributions in quadrature gives an overall position resolution of
\begin{equation}
    \sigma_{\text{position}}
    =
    \sqrt{\sigma_{\text{bias}}^2 + \sigma_{\text{width}}^2}
    =
    \sqrt{1.0^2 + 2.4^2}~\mathrm{cm}
    \approx 2.6~\mathrm{cm}.
\end{equation}

The reconstructed position depends on the effective light-propagation velocity in Equation~\ref{Eq:TimePositionReconstruction} and is also affected by position-dependent variations in the signal waveform. In particular, the signal rise time varies with the distance from the photosensor, as discussed in Ref.~\cite{Alt:2025msx}. One contribution to this behaviour arises from photons reflected near the SiPM end of the bar, which can propagate back along the bar and modify the waveform observed at the opposite readout end.

\begin{figure}[t]
\centering
\includegraphics[width=\textwidth]{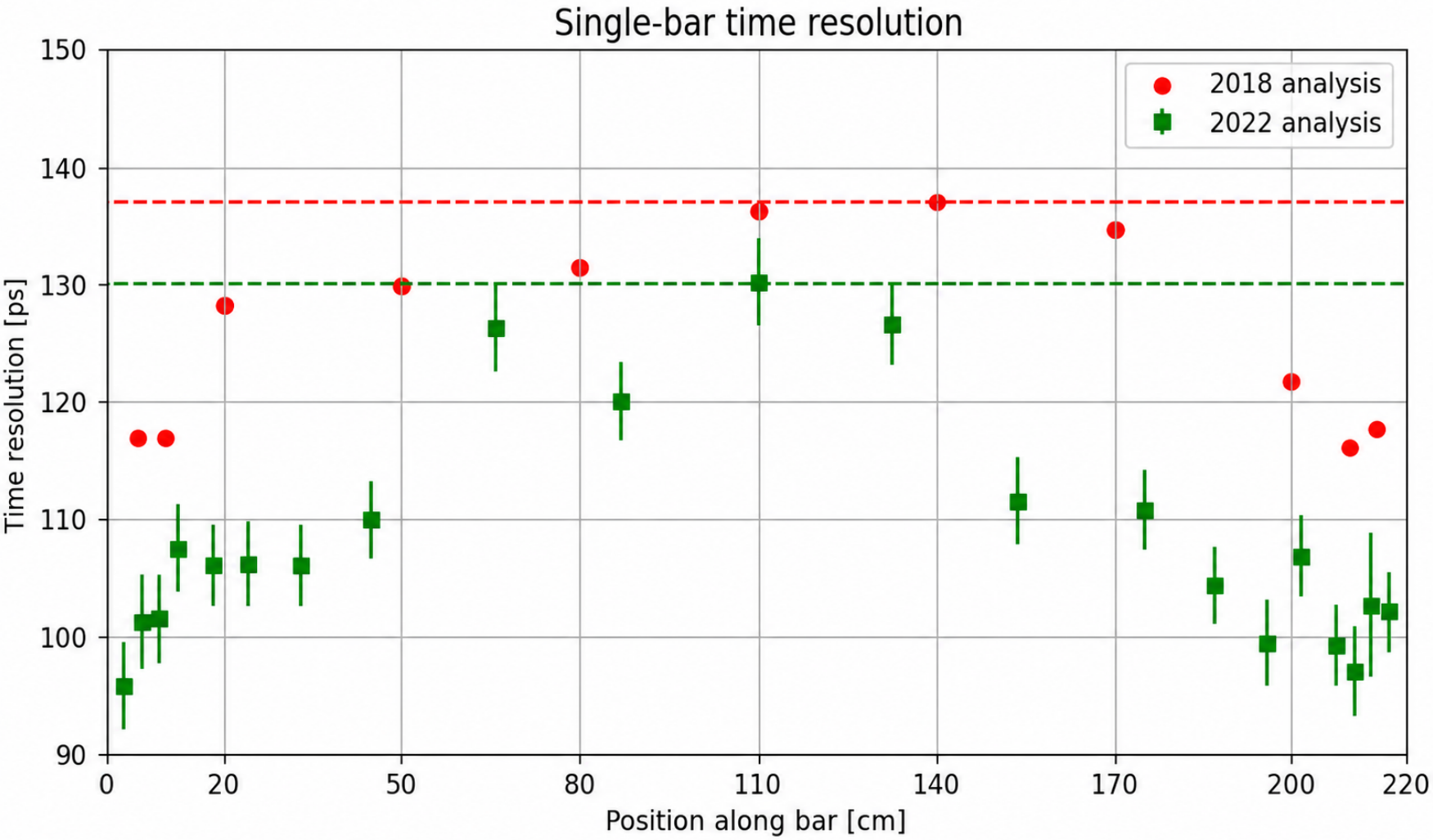}
\caption{Single-bar time resolution as a function of the reference crossing position along the bar. The resolution is obtained from the inverse-variance weighted combination of the two SiPM readouts. Red circles show the 2018 analysis from Ref.~\cite{Korzenev:2021mny}, while green squares show the updated 2022 reconstruction. The dashed lines indicate the corresponding representative resolutions of 137~ps and 130~ps.}
\label{fig:singlebar_timeres}
\end{figure}

To quantify the timing performance, the time resolution is evaluated from the distribution of relative times between the SiPM signals and the external trigger system, which provides an independent reference crossing time. The contribution from the trigger resolution is subtracted in quadrature to obtain the intrinsic time resolution of the bar readout. Figure~\ref{fig:singlebar_timeres} compares the earlier analysis~\cite{Korzenev:2021mny} with the updated reconstruction, for which the improved waveform treatment yields a small improvement in the timing resolution. The comparison also shows that the resolution varies systematically with the reference position along the bar. Combining the timing measurements from both ends yields resolutions ranging from approximately 100~ps near the bar ends to about 130~ps in the central region.

The position dependence of the timing resolution is associated with changes in light propagation, photon statistics, and waveform shape along the bar. When the crossing point is close to one readout end, reflected photons can traverse nearly twice the bar length before reaching the opposite end, corresponding to a delay of approximately $2L/v \approx 27$~ns. The reflected component is then sufficiently separated from the primary signal that it has little effect on the rising edge used for the timing reconstruction. As the crossing point moves farther along the bar, the separation between the primary and reflected components decreases, causing the waveforms to overlap and deforming the rising edge. Near the opposite end, the two components can overlap more strongly, increasing the amount of light detected within a short time interval and improving the timing resolution. In addition, proximity to either SiPM array provides higher photon statistics and a sharper rising edge. Since the combined time estimate is inverse-variance weighted, it is dominated near the bar ends by the more precise of the two readout-end measurements. Conversely, towards the centre of the bar, the resolution degrades because of lower photon statistics and greater waveform distortion from the overlap of primary and reflected light components. A value of 130~ps is therefore adopted as a conservative estimate of the timing performance across the full bar length.

These measurements demonstrate that the ToF detector meets the timing requirements of the ND280 Upgrade and can provide precise time-of-flight information for particle identification and background rejection. The reconstruction methods described here are used in the commissioning and detector-performance studies presented in the following sections.
\section{Detector Commissioning}
\label{sec:commissioning}
The commissioning of the detector was carried out using both dedicated ToF-only runs and global ND280 runs. The ToF-only runs, in which the detector was operated independently of the ND280 DAQ system, were used to validate the performance of individual panels and subsystems. Global ND280 runs were then used to verify the correct integration of the ToF detector within the experiment-wide DAQ and trigger framework.

\subsection{Online Performance Monitoring}
\label{sec:om}

An online monitoring system provides a prompt overview of the ToF detector during operation. It processes a fraction of the acquired events in real time and combines detector-readout information with operational quantities from the SC system. This allows the detector response and running conditions to be tracked continuously and enables the rapid identification of issues affecting data quality, including noisy or inactive channels, missing signals, abnormal occupancies, and timing instabilities.

A dedicated monitoring application produces diagnostic quantities such as hit rates, channel occupancies, signal amplitudes, and timing observables. Hit-rate and occupancy distributions provide a direct check of the detector mapping and identify missing or excessively noisy channels. The baseline and peak amplitude monitor the stability of the front-end electronics and SiPM response, while timing quantities are used to verify synchronisation and stable operation throughout data taking. The system is also sensitive to transient DAQ-related problems. Communication errors can occasionally occur following global ND280 DAQ start or stop transitions, appearing as missing channels or abnormal occupancies. These issues are generally resolved by restarting the affected acquisition processes, whereas persistent anomalies may indicate a hardware problem requiring further intervention.

\begin{figure}[htbp]
    \centering
    \includegraphics[
        width=0.495\textwidth,
        trim=0 3 0 3,
        clip
    ]{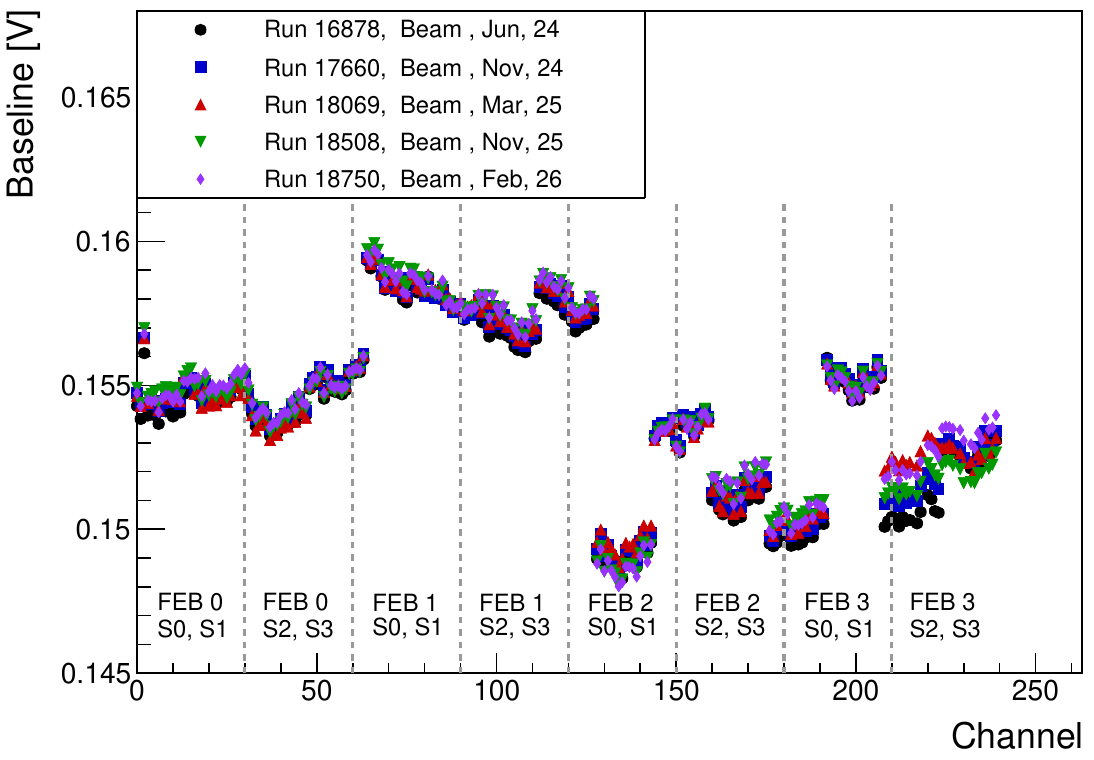}
    \hfill
    \includegraphics[
        width=0.495\textwidth,
        trim=0 3 0 3,
        clip
    ]{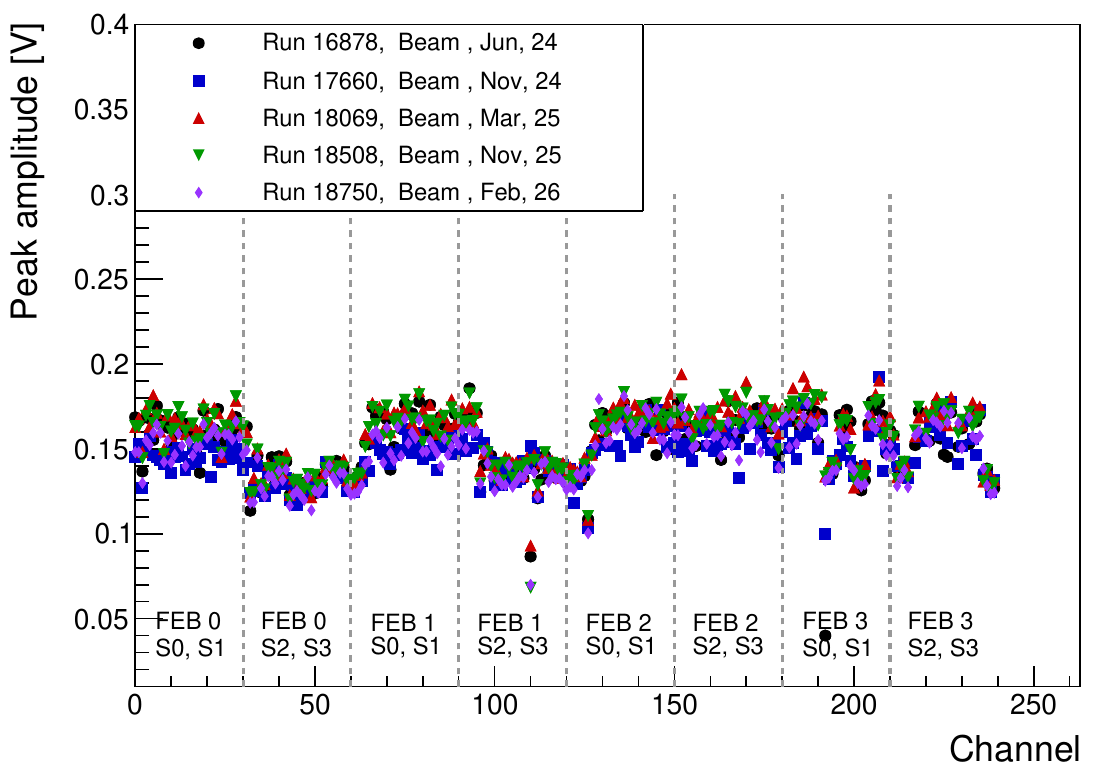}

    \caption{Representative online-monitoring observables for the ToF detector. \textit{Left}: average waveform baseline per channel. \textit{Right}: average peak amplitude per channel. These quantities are monitored throughout data taking to identify changes in the front-end electronics or SiPM response.}
    \label{fig:monitoring}
\end{figure}

Representative monitoring outputs are shown in Figure~\ref{fig:monitoring} for several beam periods. The baseline values remain stable across all channels and front-end boards from June~2024 to February~2026. The peak-amplitude distributions are also broadly consistent between run periods, with the observed channel-to-channel variations reflecting the detector geometry and particle-crossing distributions. These measurements provide an important check of the detector response following maintenance interventions. The North and South ToF panels were removed and reinstalled in October~2024 and during subsequent maintenance in 2025. No significant change in the baseline or peak-amplitude response is observed following these interventions, indicating stable operation of the detector and front-end electronics.

\begin{figure}[t]
    \centering
    \includegraphics[
        width=0.8\textwidth
    ]{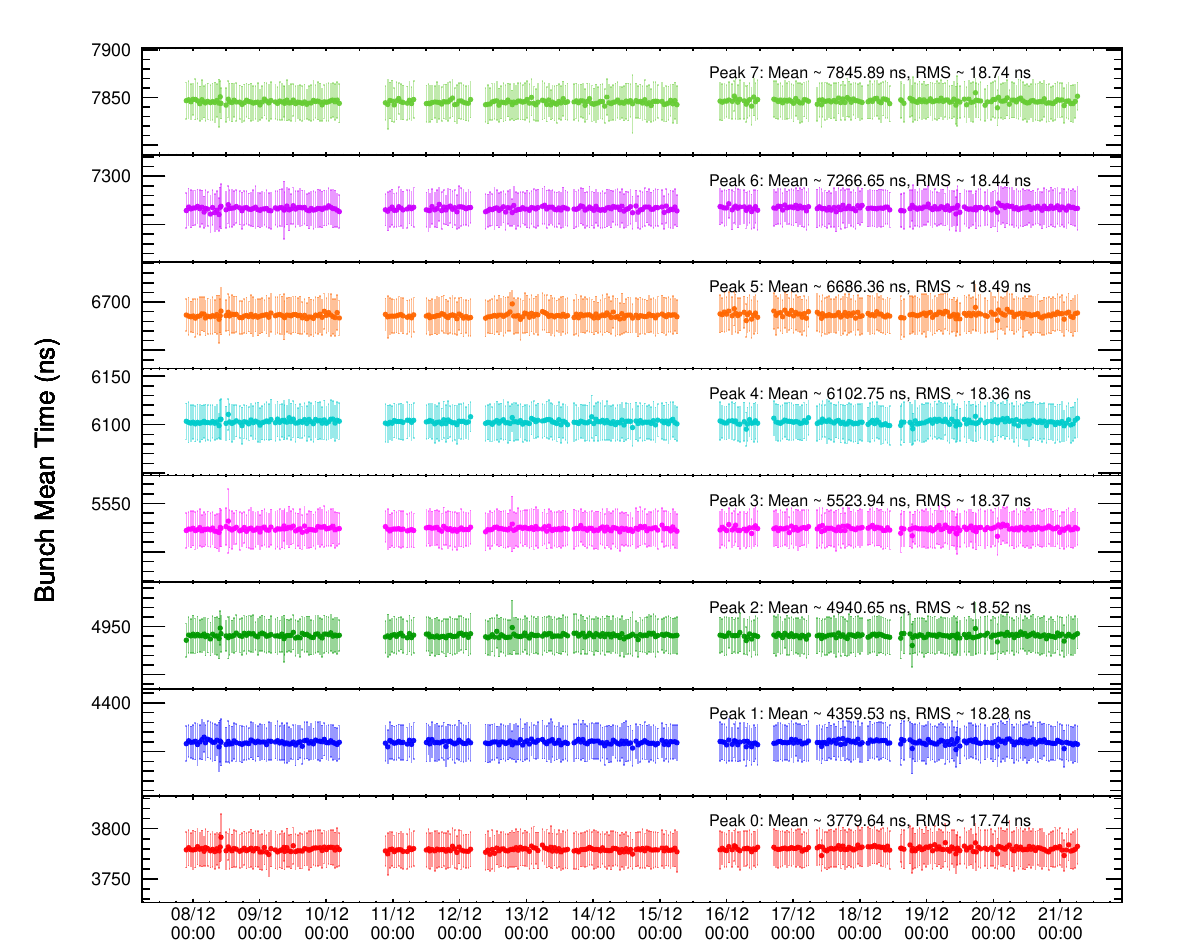}
    \caption{Evolution of the reconstructed ToF beam-bunch timing between 8 and 22 December 2025. The positions and RMS widths of the eight bunch peaks remain stable throughout the beam period.}
    \label{fig:bunchtime}
\end{figure}

Figure~\ref{fig:bunchtime} shows the reconstructed ToF bunch timing during beam data taking from 8 to 22 December 2025. The eight T2K beam bunches are clearly resolved, and their peak positions remain stable throughout the run period. The corresponding RMS widths are approximately 18~ns, demonstrating stable timing reconstruction and synchronisation with the global ND280 trigger. Entries outside the beam-bunch bands primarily correspond to non-beam running, including cosmic-ray runs and periods in which the ToF detector was not included in the global run configuration.

\subsection{Data Quality}
The ToF detector data quality (DQ) is evaluated using a dedicated assessment based on the low-level monitoring observables described in Section~\ref{sec:om}, including waveform, occupancy, and timing quantities. This procedure identifies periods affected by detector or readout issues, such as anomalous channel occupancies, FEB communication failures, and timing reconstruction problems. A discrete DQ flag is assigned to each sub-run according to the following categories:

\begin{itemize}[itemsep=0pt]

\item \textbf{Good data (flag 0):} no significant channel-, FEB-, or timing-related issues are identified by the DQ checks.

\item \textbf{Bad data (flag 1):} significant channel-level occupancy problems. Channels with no recorded events, or with event counts differing from the average by more than $5\sigma$, are identified as problematic. A sub-run is assigned flag~1 when more than ten channels satisfy these criteria.

\item \textbf{FEB error (flag 2):} board-level communication or readout failures that can affect multiple channels connected to the same FEB.

\item \textbf{Position shift (flag 3):} waveform distortions, such as negative baselines or saturation, that bias the hit-time determination at 10\% of the waveform maximum and consequently shift the reconstructed position along the bar.

\item \textbf{Undefined (flag -1):} insufficient monitoring information to determine the DQ reliably.
\end{itemize}

Only sub-runs assigned flag~0 are classified as good-quality data and retained for analysis, whereas all the other flag values are excluded from the good-data sample.
\begin{figure}[htbp]
  \centering

  \resizebox{\textwidth}{!}{%
    \includegraphics[height=0.295\textwidth, keepaspectratio, trim=1 1 1 1, clip]{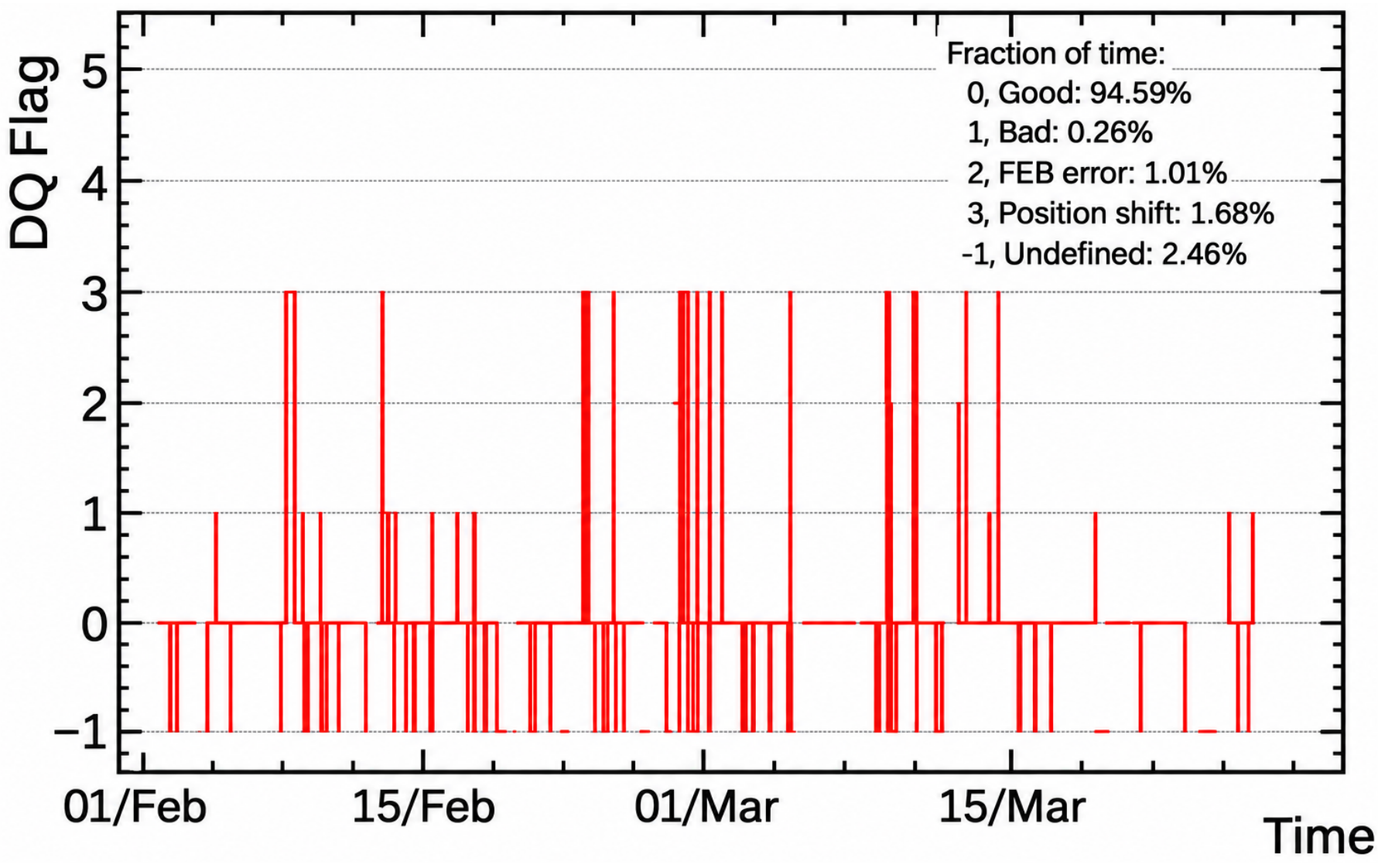}
    \hspace{0.025\textwidth}
    \includegraphics[height=0.3\textwidth, keepaspectratio, trim=2 2 2 2, clip]{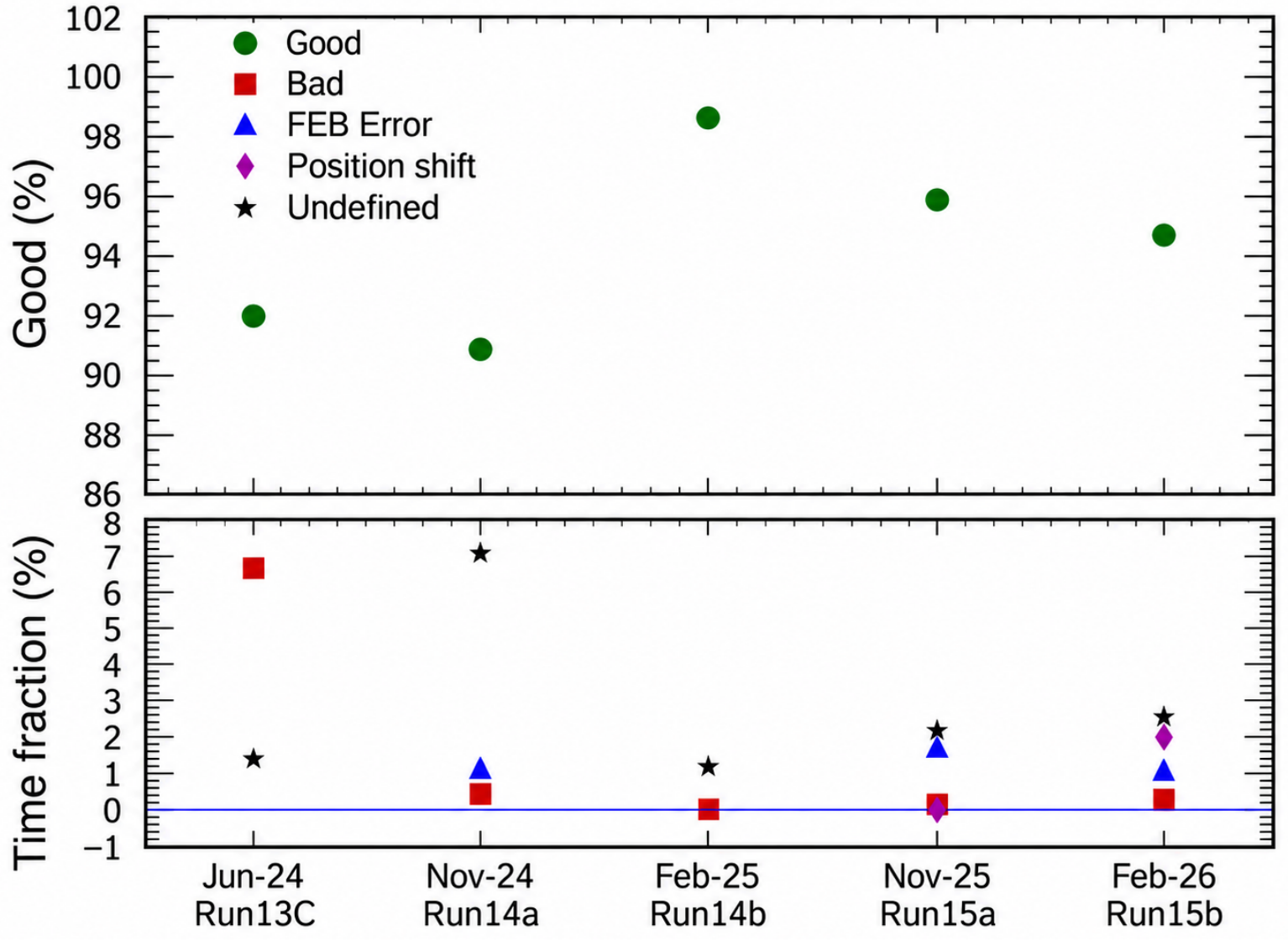}
  }

  \caption{ToF data quality monitoring. \textit{Left}: DQ flag assigned as a function of time during beam data taking from February to March 2026, with the inset showing the fraction of time in each category. \textit{Right}: fraction of time assigned to each DQ category for different beam run periods.}
  \label{fig:DQ_summary}
\end{figure}

The left panel of Figure~\ref{fig:DQ_summary} shows the DQ flag assigned as a function of time for beam data taken from February to March 2026. The detector operated stably for the majority of this period, with 94.6\% of the time classified as good data quality. The remaining 5.4\% is mainly associated with undefined intervals corresponding to data gaps (2.5\%) and position-shift effects (1.7\%), while only a small fraction is attributed to FEB-related errors or bad-data intervals.

The long-term stability of the ToF detector was evaluated by comparing the fraction of time assigned to each DQ category across different beam run periods. The right panel of Figure~\ref{fig:DQ_summary} shows that the detector maintains a high good-data fraction, typically above 95\%, with only small contributions from FEB-related errors and position-shift effects. These results demonstrate stable operation of the ToF detector across multiple beam periods, including periods before and after ToF removal and reinstallation interventions.

\subsection{Detector Response to Cosmic-Ray Data}
Cosmic-ray data provide a high-statistics, beam-independent sample of through-going tracks crossing multiple ToF panels. They are used to validate the detector response and reconstructed track topology, assess panel-to-panel timing consistency, and monitor long-term stability. These data also support the geometrical and timing alignment of the detector and provide a reference sample for comparisons with simulation.

Since the detector was installed in 2023, several ToF panels have been removed and reinstalled to provide maintenance access to the SuperFGD and HA-TPCs. Cosmic-ray measurements therefore provide an important check that the detector response remains stable following these interventions and is unaffected by handling or ageing effects.

\begin{figure}[htbp]
    \centering

    \includegraphics[
        width=0.495\textwidth,
        trim=0 0 0 0,
        clip
    ]{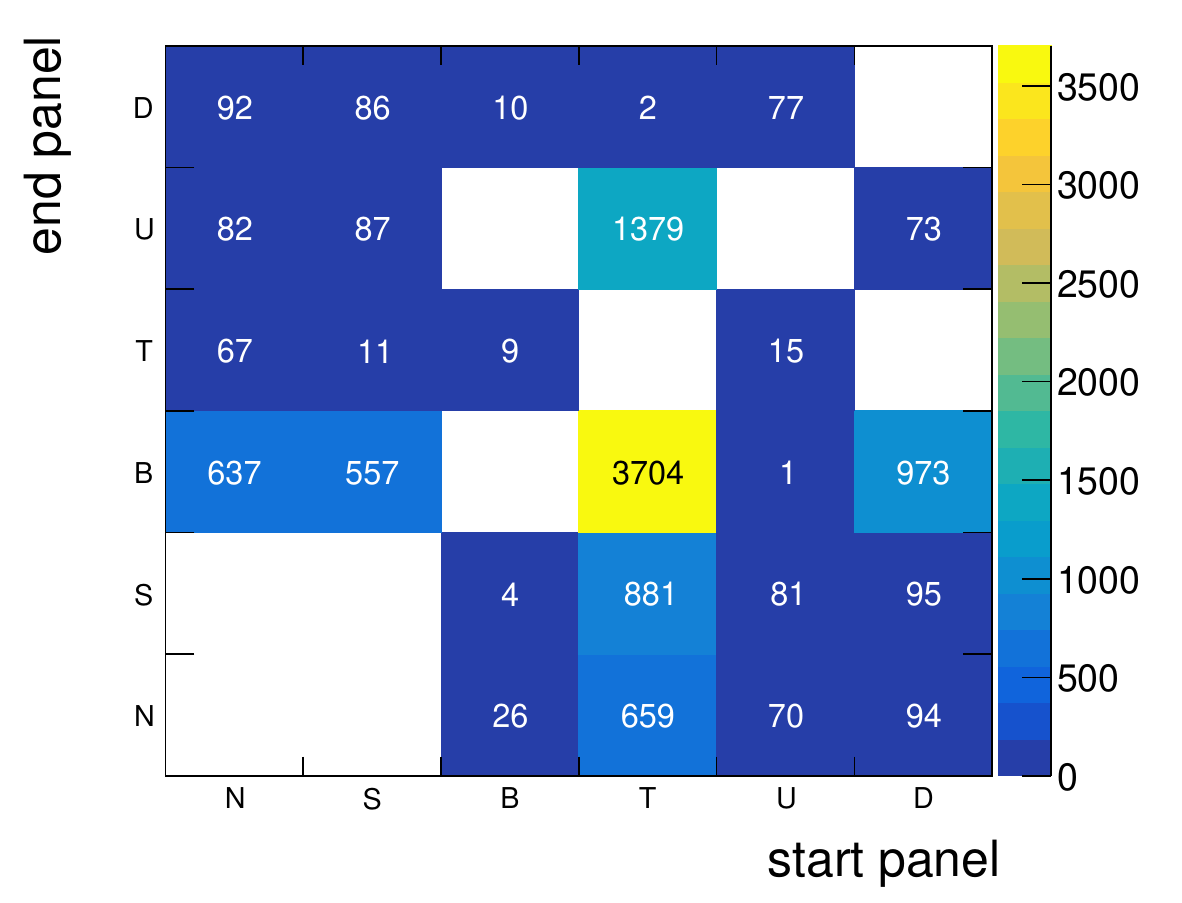}
    \hfill
    \includegraphics[
        width=0.495\textwidth,
        trim=0 0 0 0,
        clip
    ]{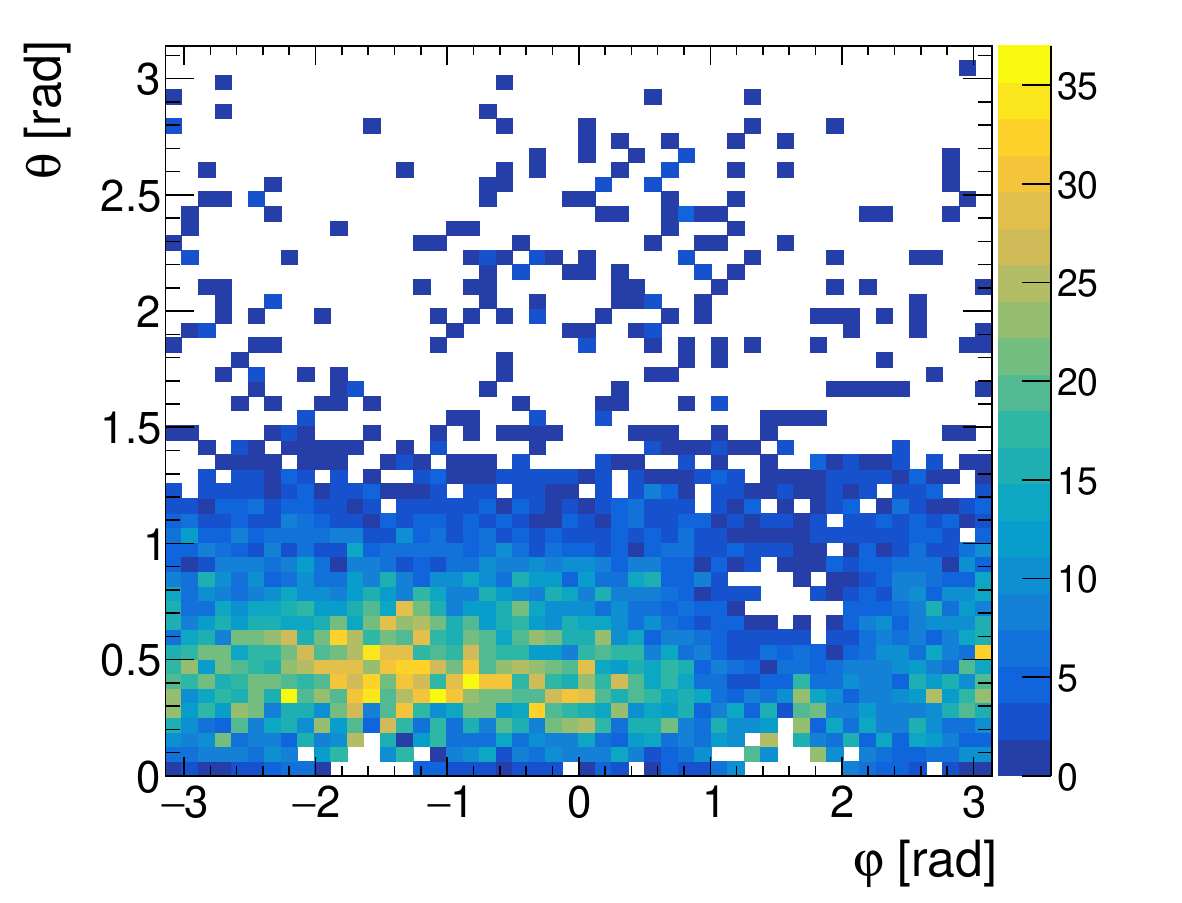}

    \caption{Reconstructed cosmic-ray track distributions from ToF-only runs acquired with the HLT3 trigger in 2026. \textit{Left}: occupancy for each combination of entry and exit panels, with N, S, B, T, U, and D denoting the North, South, Bottom, Top, Upstream, and Downstream panels, respectively. \textit{Right}: two-dimensional distribution of the zenith angle $\theta$ and azimuthal angle $\varphi$.}
    \label{fig:cosmic_angular}
\end{figure}

The reconstructed track topology was first studied using ToF-only cosmic-ray runs acquired with the HLT3 trigger configuration in 2026. The left panel of Figure~\ref{fig:cosmic_angular} shows the start--end panel occupancy and the right panel shows the angular distributions of the selected tracks. The entry and exit panels are identified using calibrated hit times, as described in Section~\ref{sec:timecalibration}, while the track direction is reconstructed from the corresponding hit positions. The zenith angle $\theta$ is defined such that a vertically downward-going track crossing the Top and Bottom panels has $\theta=0$, while the corresponding upward-going track has $\theta=\pi$.

The start--end panel occupancy is dominated by tracks entering through the Top panel and exiting through the Bottom panel. Correspondingly, the angular distribution shows a strong concentration at small values of $\theta$, confirming that most selected cosmic-ray tracks are downward-going and approximately vertical. The distribution is non-uniform in the azimuthal angle $\varphi$, with a reduced event rate around $\varphi \approx \pi/2$ and a larger event rate towards $\varphi \approx -\pi/2$ at small $\theta$. This azimuthal asymmetry is consistent with the geometry of the ND280 pit and the position of the ToF detector relative to the surrounding structures, which can lead to direction-dependent acceptance and shadowing of cosmic-ray tracks. A separate depletion is also visible around $\varphi \approx -\pi$, which arises from the acceptance of the HLT3 trigger configuration.

\begin{figure}[t]
    \centering
    \includegraphics[
        width=0.66\textwidth
    ]{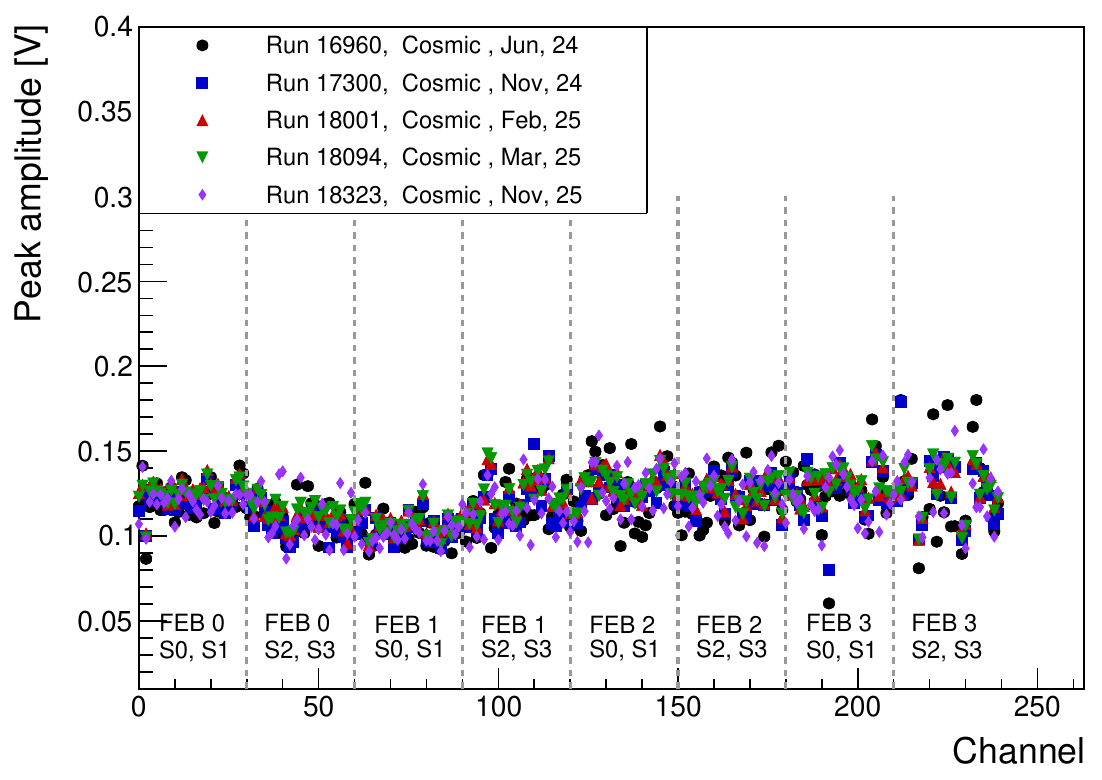}
    \caption{Average maximum peak amplitude per readout channel for five cosmic-ray runs taken during different detector operating periods. The channel-dependent pattern remains stable across the selected runs, demonstrating the long-term stability of the ToF signal response.}
    \label{fig:peakstability}
\end{figure}

The track topology and angular distributions also provide context for the
channel-by-channel signal-amplitude response. Figure~\ref{fig:peakstability}
shows the average maximum peak amplitude per readout channel for five representative
cosmic-ray runs taken during different operating periods. The distributions remain
consistent over time, demonstrating stable SiPM and front-end electronics response
across the selected runs.

The systematic variation in peak amplitude across the readout channels is consistent with differences in the typical track incidence on the corresponding ToF panels. The Top and Bottom panels, read out primarily through FEB~0 and FEB~1 as shown in Figure~\ref{fig:tof_mapping}, are predominantly crossed by near-vertical tracks, which generally produce sharper and higher-amplitude signals. Tracks crossing the lateral panels tend to have larger incidence angles, giving a broader photon-arrival distribution at the bar ends and slightly lower peak amplitudes. The persistence of this channel-dependent signal pattern across the selected runs indicates that the detector response remained stable following the removal and reinstallation of the ToF panels.

\subsection{Time Calibration}
\label{sec:timecalibration}
The segmented geometry of the ToF detector requires a common timing reference across all bars and panels in order to identify crossing tracks with good precision. Several effects can shift the measured hit times between channels, including differences in cable length, small geometrical variations, the measured effective light propagation velocity in the scintillator bars, and residual electronics effects such as PCB trace-length differences and SAMPIC-channel delays. Some of these contributions can be corrected directly from measured detector parameters, while the remaining residuals require a data-driven estimation and calibration.

\begin{table}[htbp]
    \centering
    \small
    \sisetup{
        table-number-alignment = center,
        detect-weight = true,
        detect-family = true
    }

    \begin{tabular*}{\textwidth}{
        @{\extracolsep{\fill}}
        l
        S[table-format=3.0]
        S[table-format=3.0]
        S[table-format=3.0]
        S[table-format=3.0]
        S[table-format=2.2]
        @{}
    }
        \toprule
        \textbf{Panel} &
        {\textbf{SiPM--PP1}} &
        {\textbf{PP1--PP2}} &
        {\textbf{PP2--SAMPIC}} &
        {\textbf{Total length}} &
        {\textbf{Total delay}} \\
        &
        {[\si{\centi\metre}]} &
        {[\si{\centi\metre}]} &
        {[\si{\centi\metre}]} &
        {[\si{\centi\metre}]} &
        {[\si{\nano\second}]} \\
        \midrule
        North      & 260 & 620 & 115 & 995 & 50.25 \\
        South      & 260 & 350 &  90 & 700 & 35.35 \\
        Bottom     & 260 & 620 &  90 & 970 & 48.99 \\
        Top        & 260 & 390 & 140 & 790 & 39.90 \\
        Upstream   & 260 & 350 & 140 & 750 & 37.88 \\
        Downstream & 260 & 620 & 115 & 995 & 50.25 \\
        \bottomrule
    \end{tabular*}

    \caption{Cable lengths and corresponding signal delays for each ToF panel. The total delays are calculated using a signal-propagation speed of \(19.8~\mathrm{cm/ns}\).}
    \label{tab:cable_lengths_delays}
\end{table}

The first correction accounts for signal propagation through the cables connecting the SiPMs to the SAMPIC readout. The cable lengths were measured during production with a precision of about \(1~\mathrm{cm}\), corresponding to a timing uncertainty below \(50~\mathrm{ps}\). The signal propagation velocity in the cables is taken to be \(v_{\text{cable}} = 19.8~\mathrm{cm/ns}\), a reference value for the coaxial cables used in the detector. The resulting cable delays are listed in Table~\ref{tab:cable_lengths_delays}. The cable correction is therefore
applied directly to the \textit{Cell0Time} timestamp:
\begin{equation}
t_{\mathrm{Cell0}}^{\mathrm{corr}}
=
t_{\mathrm{Cell0}}
-
\frac{L_{\mathrm{cable}}}{v_{\mathrm{cable}}}.
\end{equation}
After this first-order correction, residual channel-dependent timing shifts remain. These include small contributions from the readout electronics and from imperfect knowledge of the detector geometry. To correct these residual shifts, a Markov-chain-based iterative calibration method developed jointly for the ToF and SuperFGD is applied~\cite{abeMCcalibration2025}. The method uses pairs of hits that are highly correlated in time and determines a set of fixed offsets that minimises the discrepancy between the measured and expected time differences. For a cosmic-ray track crossing two ToF bars, the relevant residual is defined as
{
\setlength{\abovedisplayskip}{4pt}
\setlength{\abovedisplayshortskip}{4pt}
\begin{equation}
    \Delta t
    =
    t_1 - t_2 - \frac{d}{c},
    \label{eq:tof_time_residual}
\end{equation}
}where \(t_1\) and \(t_2\) are the reconstructed bar times, \(d\) is the distance between the two reconstructed hit positions, and \(c\) is the speed of the crossing particle, approximated by the speed of light. The calibration algorithm iteratively updates the offsets so that the average value of \(\Delta t\) is reduced across all selected hit pairs.

For each scintillator bar \(i\), separate residual timing offsets can be associated with the left and right SiPM readout ends, denoted \(T_{L_i}^{(0)}\) and \(T_{R_i}^{(0)}\), respectively. These are combined into a single bar-level offset,
{
\setlength{\abovedisplayskip}{4pt}
\setlength{\abovedisplayshortskip}{4pt}
\begin{equation}
    T_i^{(0)}
    =
    \frac{T_{L_i}^{(0)} + T_{R_i}^{(0)}}{2}.
    \label{eq:bar_level_offset}
\end{equation}
}

The bar-level treatment assumes that any residual left--right timing difference within a bar is negligible for this calibration step. A separate procedure would be required to determine the relative SiPM-side offset within each bar.

The calibration sample is built from cosmic-ray tracks recorded in a local run. Since ND280 is located underground, the selected sample is dominated by downward-going muons and therefore contains a large fraction of tracks involving the Top panel. To reduce geometrical ambiguities, only tracks crossing two different ToF panels are used. In this calibration, the hit position along the bar is reconstructed from the ToF timing information itself, which introduces the assumption that the residual correction can be treated as a constant offset for each bar.

\begin{figure}[htbp]
    \centering
    \includegraphics[
        width=\textwidth
    ]{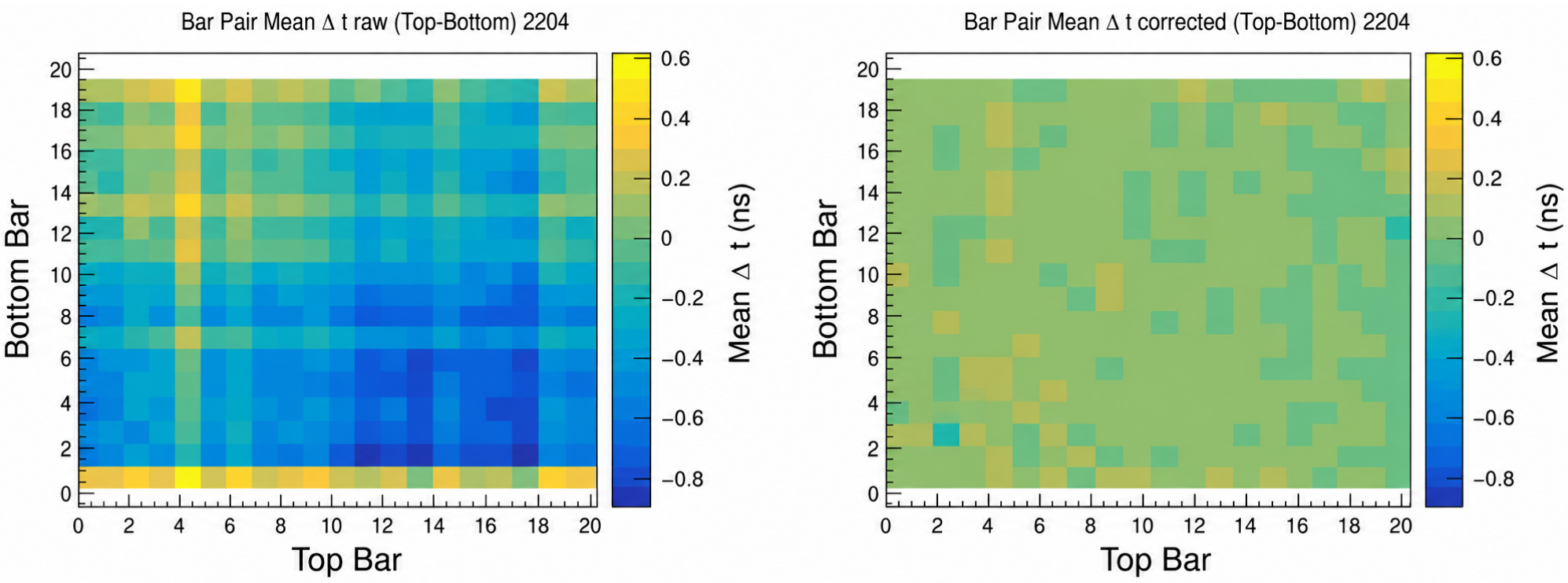}
    \caption{Mean time residual, \(\Delta t\), for each Top--Bottom bar pair. \textit{Left}: before timing calibration. \textit{Right}: after timing calibration. Both histograms use the same colour scale, highlighting the reduction of bar-dependent timing offsets after calibration.}
    \label{fig:bartobarvariation}
\end{figure}

Despite these approximations, the procedure significantly reduces panel-to-panel timing differences, as shown in Figure~\ref{fig:bartobarvariation}. The mean \(\Delta t\) between different bar combinations becomes more uniform after calibration. It is important to note the effect of geometry on both the timing offsets themselves and the calibration procedure. The calibration assumes that the reference geometry is well constrained. Any movement, such as the panel opening described in Section~\ref{ssc:mechanicalSupport}, causes a small difference between the reference and actual geometric positions of the bars. Significant geometric shifts therefore require the calibration to be reprocessed using a fresh dataset.

A more precise calibration of the individual SiPM-side offsets requires an external measurement of the hit position along the ToF bar. This would remove the need to rely only on the ToF timing information for the position reconstruction and would allow the left and right readout sides of each bar to be calibrated separately.

For the ToF detector, such external position information can be provided by the HA-TPCs or the SuperFGD, both located inside the ToF volume. These detectors provide better position resolution than the ToF-only reconstruction, which is limited by the bar timing information. Using an external position reference would also make it possible to select a larger and cleaner calibration sample. This refinement is beyond the scope of this work, but represents a natural next step towards a more detailed SiPM-level timing calibration.

\section{Detector Performance}
\label{sec:performance}
This section presents the main performance studies of the ToF detector after installation in ND280. Beam-triggered data are used to validate the timing synchronisation, study out-of-bunch activity, and characterise DAQ-related efficiency effects, while cosmic-ray data are used to investigate residual reconstruction effects and evaluate the detector-level time resolution.

\subsection{Beam Timing Structure}
\label{sec:beam_timing_structure}
The time distribution of signals recorded by the ToF detector in beam-triggered events provides a validation of its timing with respect to the ND280 beam trigger. Since the ToF readout is synchronised to the common ND280 trigger, the reconstructed hit times are expected to reproduce the bunch structure of the T2K beam spill. Observing this structure in data therefore demonstrates that the trigger distribution, event matching, and ToF timing reconstruction are correctly integrated during global ND280 operation.

\begin{figure}[htbp]
    \centering
    \begin{minipage}[t]{0.504\textwidth}
        \centering
        \includegraphics[width=\linewidth]{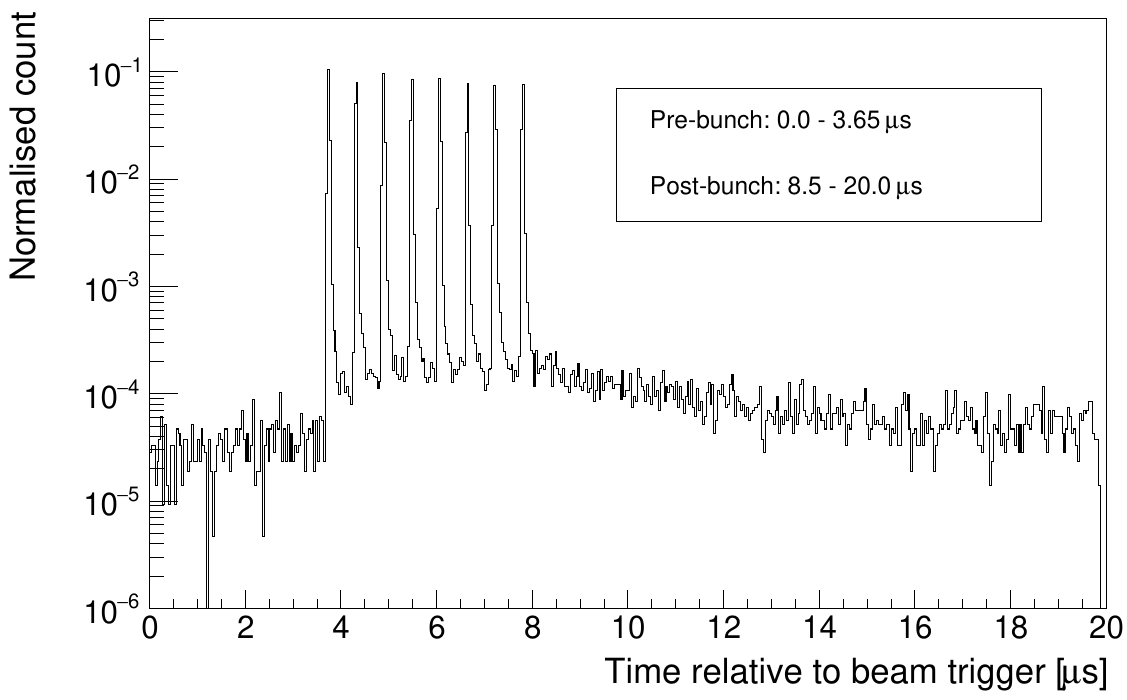}
    \end{minipage}
    \hfill
    \begin{minipage}[t]{0.481\textwidth}
        \centering
        \includegraphics[width=\linewidth]{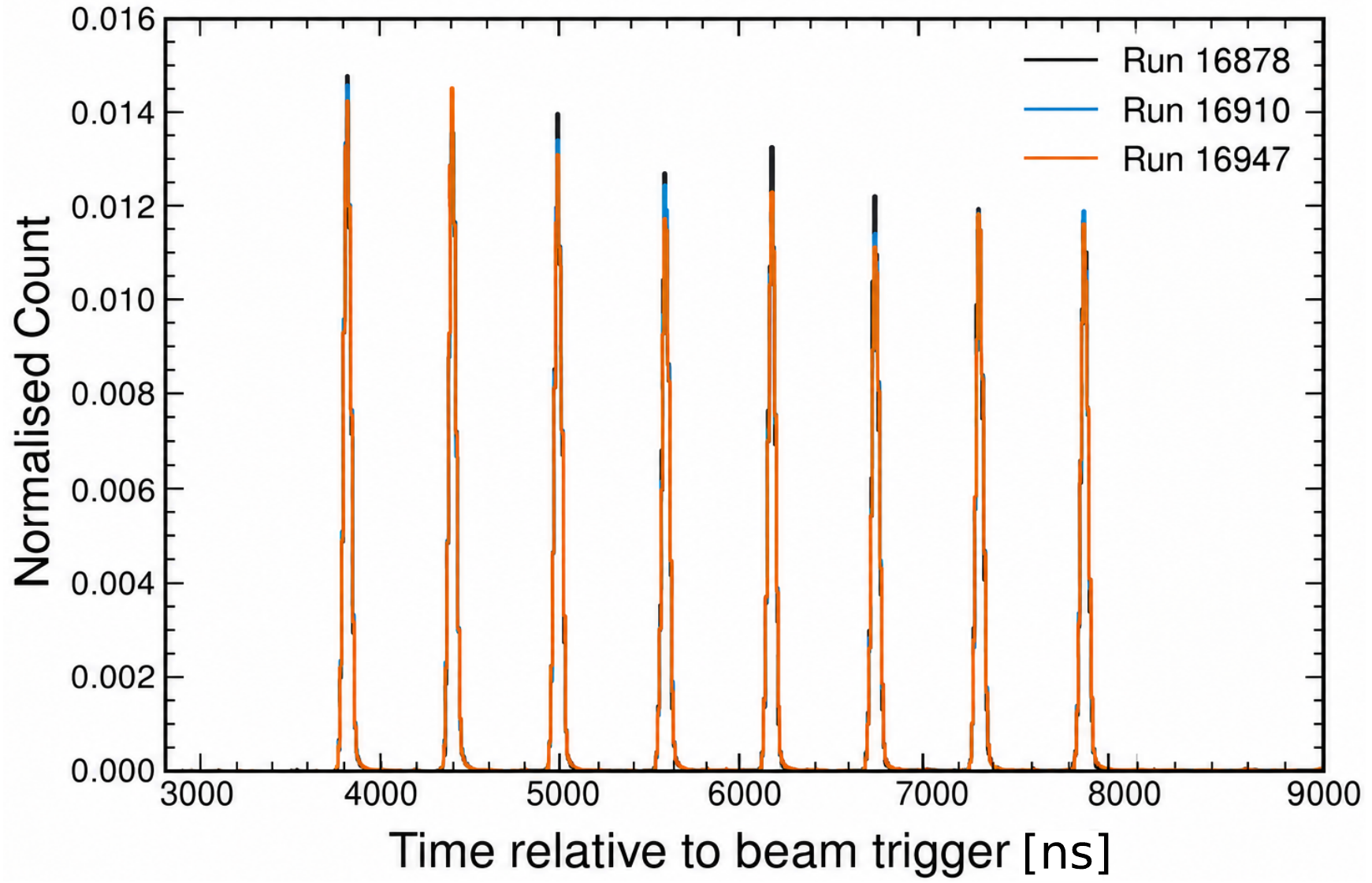}
    \end{minipage}

    \caption{Reconstructed ToF timing in beam-triggered events.
    \textit{Left}: signal time relative to the beam trigger, normalised by the number of gates.
    The pre-bunch region estimates accidental and cosmic-ray background, while the post-bunch region probes delayed activity.
    \textit{Right}: T2K bunch structure for three summer-2024 runs.
    The bunches have an RMS width of approximately 18~ns, are separated by about 580~ns, and show a progressive decrease in occupancy across the spill, discussed in Section~\ref{sec:deadtime}.}
    \label{fig:tof_beam_timing_structure}
\end{figure}

The left panel of Figure~\ref{fig:tof_beam_timing_structure} shows the reconstructed ToF hit time relative to the beginning of the beam-trigger gate for events passing the beam-trigger selection and standard event-quality requirements. The distribution is normalised by the number of selected beam-trigger gates, allowing the activity in different timing regions to be compared consistently, as required for the delayed-activity study presented in Section~\ref{sec:neutron}. The 20~\si{\micro\second} range corresponds to the readout window opened by the beam trigger for the ND280 detectors. Within this window, the T2K bunch structure is clearly visible, with eight narrow peaks separated by approximately 580~ns. The peak positions demonstrate the alignment of the reconstructed ToF times with the expected beam timing structure, while the bunches have an RMS width of approximately 18~ns, consistent with the bunch timing stability observed in the online monitoring described in Section~\ref{sec:om}.

\subsection{Delayed Post-Bunch ToF Activity}
\label{sec:neutron}

The full ToF acquisition window provides sensitivity to activity occurring outside the eight beam-bunch intervals shown in Figure~\ref{fig:tof_beam_timing_structure}. The pre-bunch region, from \(0.0\) to \(3.65\)~\si{\micro\second}, is used to estimate the approximately time-independent contribution from accidental signals and cosmic rays. This provides a data-driven background estimate for the later timing regions. Activity in the pre-bunch interval may also introduce front-end dead time and reduce the efficiency for particles produced early in the beam spill.

The post-bunch region, from \(8.5\) to \(20.0\)~\si{\micro\second}, is used to characterise activity delayed relative to the prompt spill. Its lower boundary excludes the beam-bunch structure and immediate post-spill tail, while its upper boundary is set by the end of the acquisition window. The observed delayed component is compatible with beam-induced secondary activity in and around ND280.

Secondary neutrons may contribute to this population. After being produced in beam-related interactions in the surrounding material, neutrons can undergo moderation and multiple scattering before reaching the ToF detector. Subsequent interactions in or near the ToF panels can therefore produce signals extending beyond the prompt beam-bunch structure and contribute to the delayed timing tail.

\subsubsection{Estimate of the Post-Bunch Excess}

To quantify the out-of-bunch activity, the number of reconstructed ToF signals in each timing interval is normalised by both the number of selected beam-trigger gates and the width of the interval. A gate-normalised signal rate is defined as
\begin{equation}
R = \frac{N_{\mathrm{signals}}}{N_{\mathrm{gates}}\,\Delta t},
\end{equation}

where $N_{\mathrm{signals}}$ is the number of reconstructed ToF signals in the selected timing region, $N_{\mathrm{gates}}$ is the number of selected beam-trigger gates, and $\Delta t$ is the width of the region in \si{\micro\second}. The pre-bunch region is used to estimate the approximately flat accidental and cosmic-ray contribution. The corresponding flat-background rate is
\begin{equation}
R_{\mathrm{flat}} =
\frac{N_{\mathrm{pre}}}{N_{\mathrm{gates}}\,\Delta t_{\mathrm{pre}}},
\end{equation}

where $N_{\mathrm{pre}}$ is the number of reconstructed ToF signals in the pre-bunch region and $\Delta t_{\mathrm{pre}}$ is the width of this region. This rate is extrapolated to the post-bunch region to estimate the expected flat-background contribution:
\begin{equation}
\frac{N_{\mathrm{flat,post}}}{N_{\mathrm{gates}}}
=
R_{\mathrm{flat}}\,\Delta t_{\mathrm{post}},
\end{equation}
where $\Delta t_{\mathrm{post}}$ is the width of the post-bunch region in \si{\micro\second} and \(N_{\rm flat,post}\) is the expected number of flat-background signals in that region. The post-bunch excess per selected beam-trigger gate is then defined as
\begin{equation}
\frac{N_{\mathrm{excess}}}{N_{\mathrm{gates}}}
=
\frac{N_{\mathrm{post}}}{N_{\mathrm{gates}}}
-
\frac{N_{\mathrm{flat,post}}}{N_{\mathrm{gates}}},
\end{equation}
where \(N_{\rm post}\) is the observed number of reconstructed ToF signals in the post-bunch region, and \(N_{\rm excess}\) is the number of signals in excess of the extrapolated flat-background expectation.
\begin{table}[htbp]
    \centering
    \small
    \sisetup{
        table-number-alignment = center,
        detect-weight = true,
        detect-family = true
    }

    \begin{tabular*}{\textwidth}{
        @{\extracolsep{\fill}}
        l
        c
        S[table-format=5.0]
        S[table-format=1.5]
        S[table-format=1.5]
        @{}
    }
        \toprule
        \textbf{Region} &
        \textbf{Time window} &
        {\textbf{Signals}} &
        {\textbf{Signals per gate}} &
        {\textbf{Rate}} \\
        &
        {[\si{\micro\second}]} &
        &
        &
        {[\si{\per\gate\per\micro\second}]} \\
        \midrule
        Pre-bunch baseline       & 0.0--3.65  & 1718  & 0.00678 & 0.00186 \\
        Post-bunch observed      & 8.5--20.0  & 10442 & 0.04121 & 0.00358 \\
        \addlinespace[1mm]
        Post-bunch flat background & 8.5--20.0 & {--} & 0.02136 & 0.00186 \\
        Post-bunch excess        & 8.5--20.0  & {--} & 0.01985 & 0.00173 \\
        \bottomrule
    \end{tabular*}

    \caption{Comparison of pre-bunch and post-bunch ToF activity. Rates are expressed as reconstructed signals per selected beam-trigger gate per unit time.}
    \label{tab:tof_out_of_bunch_rates}
\end{table}

Using this data-driven estimate of the flat accidental and cosmic-ray background, the post-bunch region shows a clear excess above the expected out-of-bunch activity. As summarised in Table~\ref{tab:tof_out_of_bunch_rates}, the observed post-bunch activity is approximately \(1.9\) times larger than the expectation obtained by extrapolating the pre-bunch background.

The difference corresponds to an excess of \(0.0198\) reconstructed ToF signals per selected beam-trigger gate, or approximately \(5.0\times10^{3}\) excess reconstructed ToF signals in the analysed sample. This excess cannot be explained by the approximately flat accidental and cosmic-ray component measured before the beam spill. It therefore indicates an additional delayed component following the bunch train, consistent with beam-induced secondary activity in the detector environment.
\subsubsection{Panel Dependence}
The panel dependence is used to examine whether the delayed activity behaves like a uniform accidental background or shows the spatial structure expected from beam-related secondary activity in the detector environment. This was quantified using the panel-specific excess rate, \(R_{\rm excess}^{p}\), defined as the difference between the post-bunch rate and the pre-bunch baseline rate for each ToF panel:
\begin{equation}
R_{\mathrm{excess}}^{p} =
\frac{N_{\mathrm{post}}^{p}}{N_{\mathrm{gates}}\,\Delta t_{\mathrm{post}}}
-
\frac{N_{\mathrm{pre}}^{p}}{N_{\mathrm{gates}}\,\Delta t_{\mathrm{pre}}},
\end{equation}
where \(N_{\mathrm{pre}}^{p}\) and \(N_{\mathrm{post}}^{p}\) are the numbers of reconstructed ToF signals in panel \(p\) before the first bunch and after the final bunch, respectively. Since the Bottom panel has two fewer active bars than the other panels, its rate is corrected to the equivalent rate for a 20-bar panel when comparing the panel dependence.

\begin{figure}[b]
    \centering
    \includegraphics[
        width=0.85\textwidth,
        trim=0 0 0 0,
        clip
    ]{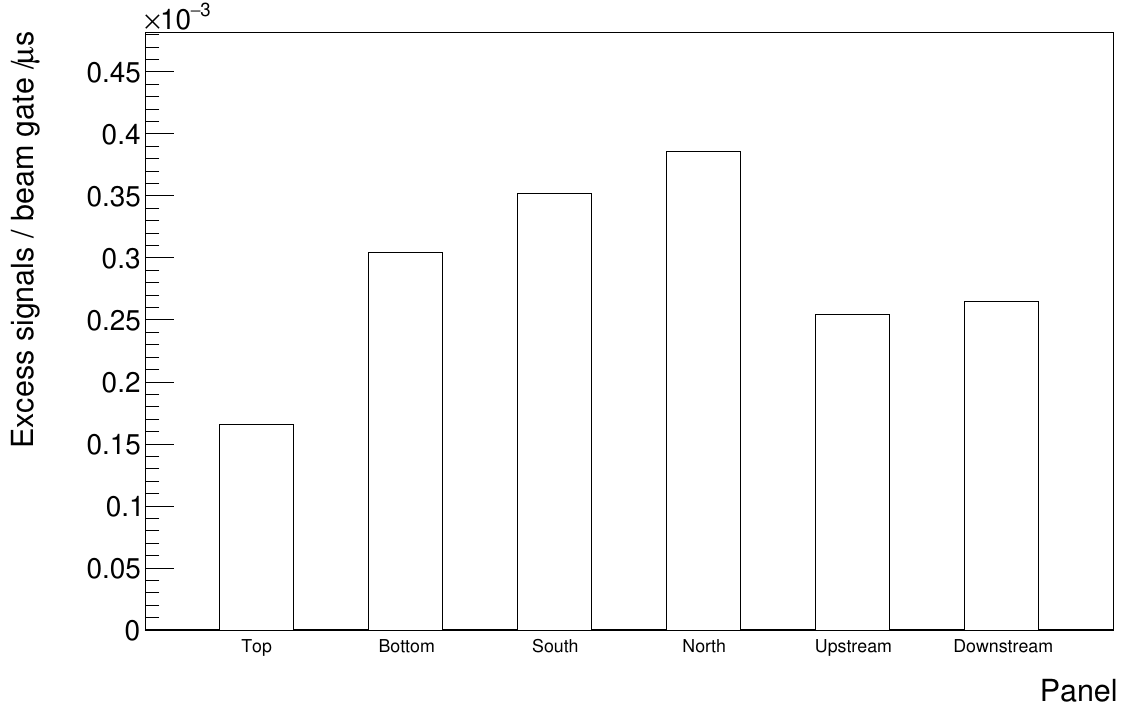}
    \caption{Post-bunch excess rate for each ToF panel, normalised per selected beam-trigger gate and unit time. The Bottom-panel value is scaled to the equivalent rate for a 20-bar panel.}
    \label{fig:tof_module_excess}
\end{figure}

Figure~\ref{fig:tof_module_excess} shows that a positive post-bunch excess is observed in all six ToF panels. The excess is smallest in the Top panel and largest in the side panels, particularly the North and South panels. Since the rates are normalised by the number of selected beam-trigger gates and the timing-window width, and are corrected for the number of active bars, this non-uniformity is not expected from a purely uniform accidental background.

The panel pattern suggests that the delayed activity is influenced by the geometry and material distribution around ND280. Beam-induced secondary neutrons can undergo moderation and multiple scattering before reaching the ToF detector, so their contribution may depend on nearby material and shielding rather than following the prompt beam direction alone. The larger excess observed in the side panels is consistent with additional secondary-particle production or scattering in material close to those panels, including electronics, support structures, and surrounding detector material. A more quantitative interpretation would require a detailed comparison with the simulated neutron component and its dependence on the surrounding detector geometry.

\subsubsection{Time Dependence}

The time profile of the post-bunch excess distinguishes a residual flat background from delayed beam-induced activity. Since a delayed component should decrease after the bunch train, the post-bunch timing distribution was fitted with an exponentially falling component above the flat background level measured before the beam spill. The fit model is
\begin{equation}
N(t) = A\exp\left[-\frac{(t-t_{0})}{\tau}\right] + C_{\mathrm{flat}},
\end{equation}
where \(C_{\mathrm{flat}}\) is fixed to the flat-background estimate obtained from the pre-bunch region, \(t_{0}\) is the start of the fit range, and \(\tau\) is the effective decay timescale. The fit was performed over the full post-bunch interval, from \(8.5\) to \(20.0\)~\si{\micro\second}, using the same beam-gate normalisation as in Section~\ref{sec:beam_timing_structure}.

\begin{figure}[htbp]
    \centering
    \includegraphics[
        width=0.9\textwidth
    ]{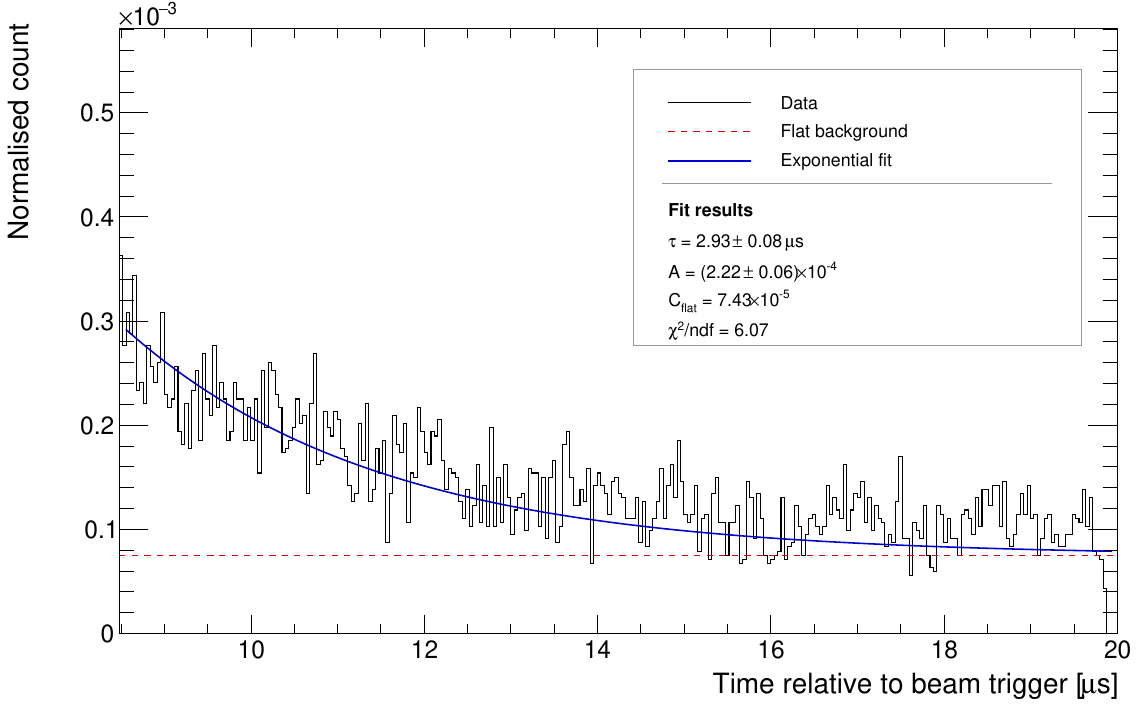}
    \caption{Post-bunch timing distribution fitted over \(8.5\)--\(20.0\)~\si{\micro\second} with an exponential component and a flat background fixed from the pre-bunch region.}
    \label{fig:tof_post_bunch_tail_fit}
\end{figure}

Figure~\ref{fig:tof_post_bunch_tail_fit} shows the post-bunch timing distribution with the flat-background level fixed to the value estimated from the pre-bunch region. The fit captures the broad decrease towards the accidental background level and yields an effective timescale of
\(\tau = 2.93 \pm 0.08\)~\si{\micro\second}.
The fixed background level is
\(C_{\mathrm{flat}} = 7.43\times10^{-5}\)
reconstructed ToF signals per selected beam-trigger gate per \(40\,\text{ns}\) bin. The relatively large $\chi^{2}/\mathrm{ndf}=6.07$ indicates that a single exponential does not fully describe the detailed time structure of the post-bunch distribution. This is expected, since the distribution contains residual features, including small-scale variations associated with electronics dead time, that are not included in the simple fit model. This fitted timescale therefore characterises the effective decay of the delayed post-bunch component, rather than a direct neutron lifetime or a unique time constant.

The important conclusion is that the post-bunch activity is not consistent with a purely flat accidental background. Instead, the timing distribution contains a falling component above the independently estimated pre-bunch background level. Together with the integrated excess and the panel dependence, this supports the interpretation that the post-bunch activity contains a delayed beam-induced component.

\subsection{Readout Dead-Time and Bunch-Dependent Hit Efficiency}
\label{sec:deadtime}
The ToF hit-time distribution in beam-triggered data exhibits the expected T2K spill structure, with eight bunches separated by approximately $580\,\text{ns}$. However, integrating the number of recorded hits in each bunch reveals a progressive decrease in occupancy across the spill, as shown in the right panel of Figure~\ref{fig:tof_beam_timing_structure}. This behaviour is observed consistently across several beam runs, while independent T2K bunch-by-bunch beam measurements show no comparable progressive decrease across the spill~\cite{Ashida:2018,T2K:2015sxa}. The observed trend is therefore attributed to the ToF readout rather than to the beam structure.

This inefficiency arises from dead time in the SAMPIC digitisation and readout chain. Following an accepted trigger, the SAMPIC requires approximately \(1.1\)~\si{\micro\second} to digitise the stored waveform, followed by an additional readout time of approximately \(450\)~ns per active channel on the board~\cite{Villa:2025thesis,Breton:2020kva}. During this sequence, the corresponding board cannot accept another trigger.

The SAMPIC boards are read out in parallel under a common HLT. Although the busy time of each board depends on its local active-channel multiplicity, the detector can only be re-armed once all triggered boards have completed their readout. The effective detector dead time is therefore determined by the slowest board, typically the one with the largest number of active channels.

The standard HLT2 configuration requires coincident signals from the two ends of a scintillator bar, connected to different SAMPIC chips on the same front-end board, as described in Section~\ref{Sec:Trigger}. The minimum HLT2-triggered event therefore contains two active channels, corresponding to a minimum readout dead time of
\begin{equation}
\tau_{\mathrm{min}}
\simeq 1.1~\si{\micro\second}
+ 2 \times 0.45~\si{\micro\second}
= 2.0~\si{\micro\second}.
\end{equation}

Since this minimum readout dead time substantially exceeds the approximately $580\,\text{ns}$ bunch spacing, a trigger associated with an early bunch can suppress the response to several subsequent bunches. The recovery pattern depends on the trigger time, hit multiplicity, hit timing, and distribution of active channels across the readout boards. The observed bunch-dependent occupancy therefore reflects the combined effects of dead time and recovery over many events.

\subsection{Simulation of the Readout Dead-Time}
\begin{figure}[b]
    \centering
    \resizebox{0.79\textwidth}{!}{%
        \includegraphics[
            height=0.30\textheight,
            keepaspectratio,
            trim=5 5 5 5,
            clip
        ]{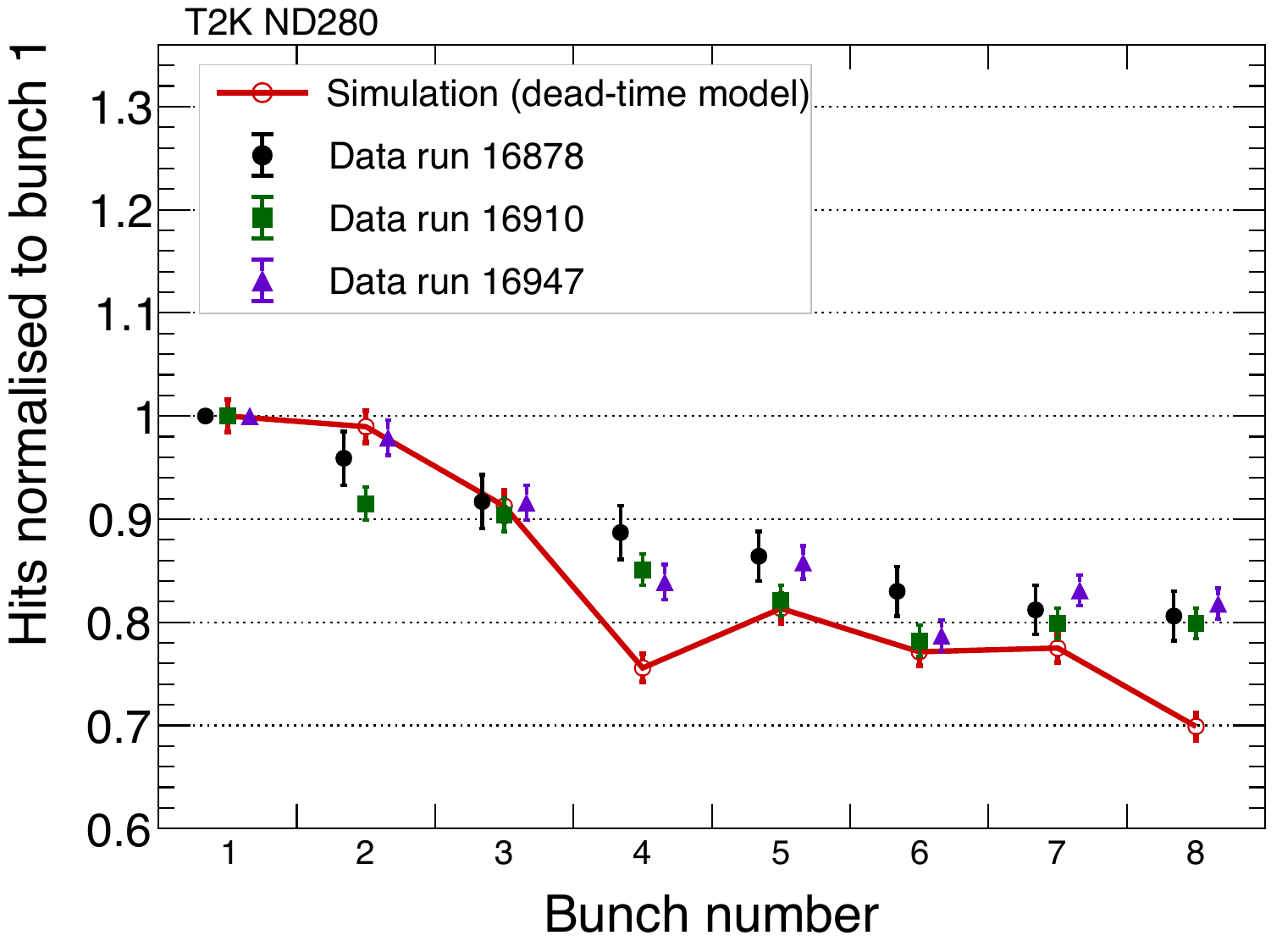}
    }
    \caption{Number of ToF hits per beam bunch, normalised to the first bunch, for selected data runs and simulation with the readout dead-time model applied. The progressive reduction in recorded hits across the spill is reproduced qualitatively by the model; run-dependent beam conditions and out-of-time backgrounds are not fully modelled.}
    \label{fig:efficiencyBeam}
\end{figure}

The ToF detector simulation incorporates the readout inefficiency by modelling the dead time described in Section~\ref{sec:deadtime}. Figure~\ref{fig:efficiencyBeam} compares the bunch-dependent hit distribution observed in selected data runs with the simulation after application of the dead-time model. The model reproduces qualitatively the progressive reduction in recorded hits across successive bunches. A direct quantitative comparison is not expected, since run-dependent beam conditions and out-of-time backgrounds present in data are not fully modelled.

In the simulation, the dead-time efficiency can additionally be evaluated directly as the ratio of accepted hits to the total number of simulated hits. The resulting bunch dependence shows the same overall decreasing trend as the data, with a magnitude closest to Run~16910 for several bunches. However, the simulation predicts a stronger suppression in parts of the spill, most notably around bunch~4 and in the final bunches.

The efficiency of the first bunch can also be affected by activity occurring before the beam spill. Since the acquisition gate opens approximately
\(\tau_{\text{window}} = 3.65\)~\si{\micro\second} before the first bunch, cosmic muons and, to a much lesser extent, accidental SiPM coincidences can produce pre-spill activity in the ToF readout. From the pre-bunch activity measured in Section~\ref{sec:neutron}, the mean pre-bunch occupancy is estimated to be
\(\lambda_{\text{pre}} = 0.00678\) reconstructed ToF signals per selected beam-trigger gate over this interval. Assuming that this activity is uniformly distributed within the pre-bunch window, the mean occupancy within the minimum readout dead-time,
\(\tau_{\text{min}} \approx 2\)~\si{\micro\second}, immediately preceding the first bunch is
\begin{equation}
    \frac{\tau_{\text{min}}}{\tau_{\text{window}}}
    \lambda_{\text{pre}}
    \approx 3.7\times10^{-3}.
\end{equation}
For this small occupancy, this corresponds approximately to a \(0.37\%\) probability that a selected beam-trigger gate contains pre-spill ToF activity sufficiently close to the first bunch to place part of the readout in dead-time. The resulting effect on the overall first-bunch hit efficiency is expected to be small and depends on which readout channels are affected.

The current ND280 simulation models neutron production and transport, but does not include cosmic-ray or environmental-radioactivity backgrounds in beam-neutrino events. Such out-of-time activity can add an approximately flat contribution beneath the bunch structure that is absent from the simulation. Run-dependent bunch intensities and simplifications in the electronics dead-time model may also affect the detailed bunch-by-bunch profile. These effects may account for some of the remaining differences between simulation and data. Modelling the relevant out-of-time backgrounds would improve the quantitative description of the bunch-dependent readout efficiency.

\subsection{Unpaired Hits}
The ToF trigger requires time-coincident signals from the two readout ends of a scintillator bar as described in Section~\ref{Sec:Trigger}. During offline reconstruction, the two signals must also be associated within a separate hit-pairing window. In the absence of reconstruction losses or additional unassociated signals, ToF events are therefore expected to contain an even number of reconstructed hits. However, events with odd hit multiplicity are observed in both beam and cosmic-ray data. Such events contain at least one unpaired hit, for which the signal from the opposite end of the bar is either not reconstructed or not associated within the offline pairing window. 
\subsubsection{Rate and Stability}
\begin{figure}[htbp]
    \centering
    \includegraphics[width=\linewidth]{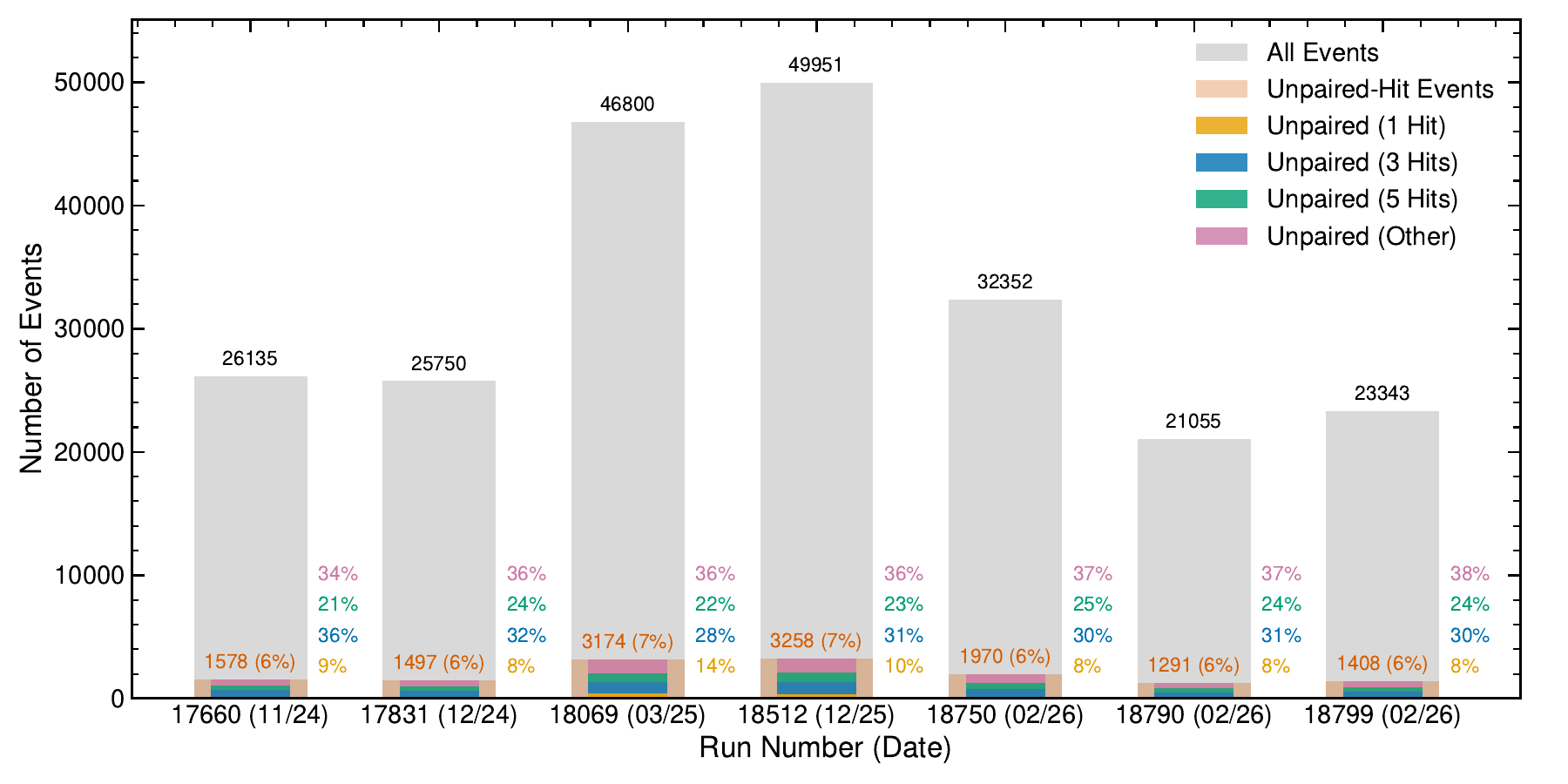}
    \caption{Breakdown of unpaired-hit events by odd hit multiplicity (1-hit, 3-hit, 5-hit, and other), overlaid on the total number of events for each run, grouped by collection date. Runs shown are taken after initial DAQ tuning, and the fractions remain stable across this period.}
    \label{fig:odd_hits}
\end{figure}

Run-to-run variations in the number of unpaired hits were observed during the initial commissioning period, as the detector was progressively instrumented. In the fully instrumented June 2024 runs, spurious associations of out-of-time hits were observed and the overall unpaired-hit fraction was approximately 10--11\%. These observations motivated subsequent improvements to the DAQ buffer clean-up. Following these changes, introduced before run 17660, the unpaired-hit fraction decreased to approximately 6\%.

Figure~\ref{fig:odd_hits} shows several stable beam-triggered runs taken from November 2024 to February 2026, after this initial DAQ tuning. Across this period, the overall unpaired-hit fraction remains stable at approximately 6--7\%. Among the unpaired-hit events, the fraction containing a single reconstructed hit is typically 8--9\%, with higher values of 14\% and 10\% in runs 18069 and 18512, respectively.

The stability of the overall fraction across runs spanning more than a year indicates that the remaining unpaired-hit population persists after the initial DAQ improvements. Understanding the origin of these residual events therefore requires additional information beyond the event multiplicity alone, in particular their timing, amplitude, and association with the detector readout channels. 

\subsubsection{Characterisation of Residual Unpaired Hits}
To investigate the origin of the residual population of unpaired-hit events, three-hit events were studied. These events provide the simplest topology containing one unpaired hit: two hits can be associated with a valid bar coincidence, while the third hit remains unpaired. If the unpaired hit is related to the paired hits, for example through a missed partner hit, charge sharing when a track crosses the overlap between adjacent scintillator bars, or electrical cross-talk, it is expected to be localised within the same region of the readout, typically on the same FEB and, in some cases, on the same SAMPIC chip. This expectation follows from the detector geometry and readout mapping described in Figure~\ref{fig:tof_mapping}. In contrast, random noise is not expected to be correlated with the FEB or SAMPIC of the paired hits and can occur elsewhere in the detector.

In runs 17660 and 17831, about $80\%$ of these unpaired third hits originated from the same FEB as one of the paired hits, consistent with contributions from charge sharing, electrical cross-talk, or missed partner hits. Furthermore, more than $75\%$ were associated with both the same FEB and the same SAMPIC chip as at least one of the paired hits, further supporting a physical association with the paired hits rather than an origin from uncorrelated detector noise. These observations were made following the DAQ buffer clean-up improvements and coincided with a decrease in operating temperature of up to $2^\circ\mathrm{C}$. Since the SiPM dark-count rate is temperature dependent~\cite{Vacheret:2011}, the lower operating temperature is expected to reduce the number of random noise-induced hits above threshold. This may increase the relative fraction of residual unpaired hits that are physically associated with the paired hits, although the available data do not allow this effect to be isolated from other changes in detector or DAQ conditions.

\begin{figure}[t]
    \centering
    \includegraphics[
        width=0.8\linewidth,
        trim=5 10 5 10,
        clip
    ]{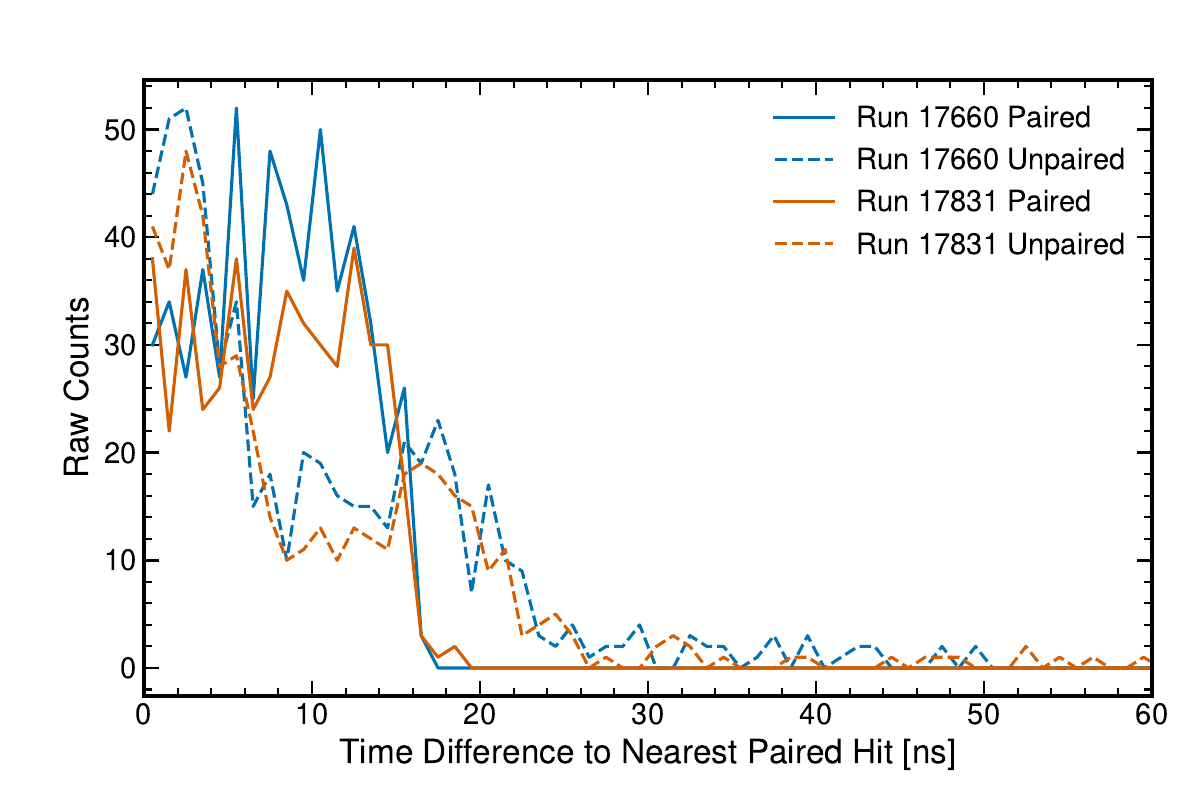}
    \caption{Timing comparison in three-hit events for Run~17660 (blue) and Run~17831 (orange). Solid lines show the time difference for the paired hits, while dashed lines show the time difference between the unpaired hit and the nearest paired hit. The paired-hit distributions are broadly distributed below about \(20~\mathrm{ns}\), whereas the paired--unpaired distributions peak close to zero and show a secondary structure around \(10\)--\(20~\mathrm{ns}\).}
    \label{fig:odd_hits:minCell0Time}
\end{figure}

To further separate cross-talk, missed-hit, and noise contributions, the timing of the unpaired hits was studied in runs 17660 and 17831. The analysis focuses on three-hit events where the unpaired hit is localised on the same FEB and SAMPIC chip as at least one of the paired hits. The timing structure shown in Figure~\ref{fig:odd_hits:minCell0Time} reveals two distinct populations. In about one third of the selected events, the unpaired hit occurs very close in time to one of the paired hits. Within this population, more than 95\% of the time differences lie within approximately 3~ns. These hits generally have low amplitude, with an average below 0.035~V, and preferentially occur in immediately adjacent bars (\(|\Delta\mathrm{bar}|=1\)) on the same FEB. This behaviour is consistent with either electrical cross-talk or charge sharing, in which a track clips the overlap between adjacent bars and deposits a small amount of energy in the neighbouring bar.

The remaining events show a broader timing structure, including a secondary enhancement around 10--20~ns in the time difference between the unpaired hit and the nearest paired hit. These unpaired hits typically have larger amplitudes, with an average above 0.170~V, consistent with a particle traversing the detector. They are therefore interpreted as likely missed-hit candidates, corresponding to cases where the second hit of a genuine bar coincidence was not reconstructed in a one-track or two-track event.

\begin{figure}[htbp]
    \centering
    \includegraphics[
        width=1.\linewidth,
        trim=5 20 5 25,
        clip
    ]{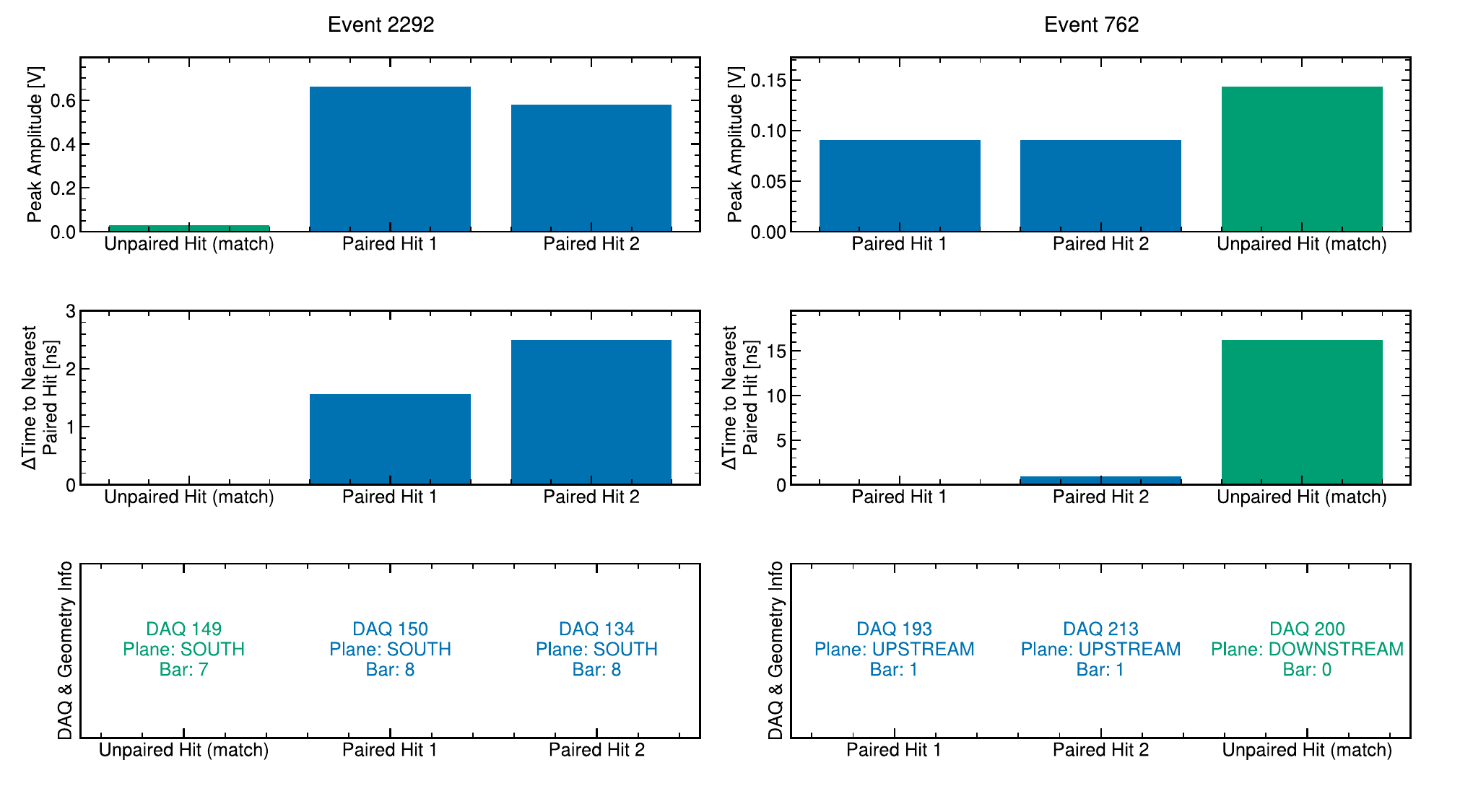}
    \caption{Representative cross-talk/charge-sharing and missed-hit candidates in three-hit events. Each event is shown with its peak amplitude (top), time difference relative to the first hit (middle), and DAQ/geometrical information (bottom). The cross-talk/charge-sharing candidate (left) has a low-amplitude neighbouring hit close in time, while the missed-hit candidate (right) has a higher-amplitude unpaired hit with a larger time separation.}
    \label{fig:odd_hits:cross-talk-and-missed-second-hit}
\end{figure}

Representative examples of these two candidate classes are shown in Figure~\ref{fig:odd_hits:cross-talk-and-missed-second-hit}, together with their timing, amplitude, and DAQ/geometrical information. A smaller population does not exhibit either characteristic behaviour clearly and may include contributions from random noise.

To quantify the possible contribution from charge sharing, adjacent-bar, low-amplitude unpaired hits are selected and their rate is measured relative to the number of reconstructed bar crossings, using only central bars. As summarised in Table~\ref{tab:odd_hits:charge-sharing}, the charge-sharing candidate rate is consistently higher in the B/T/S/N panels than in U/D across all selected runs, with an enhancement of approximately \(1.4\)--\(2.0\). The same ordering is observed in every run, with differences ranging from \(3.4\sigma\) to \(7.1\sigma\), and the rates remain broadly stable across the post-tuning dataset. This pattern is consistent with the detector geometry: beam-like tracks typically intersect the B/T/S/N panels at larger angles to the panel normal than U/D and are therefore more likely to traverse the overlap between adjacent staggered bars. Such tracks can produce a small additional signal in the neighbouring bar, giving rise to a charge-sharing candidate.

\begin{table}[htbp]
  \centering
  \small

  \begin{tabular*}{\textwidth}{
    @{\extracolsep{\fill}}
    r
    c
    c
    c
    c
    @{}
  }
    \toprule
    \textbf{Run} &
    \textbf{U/D} &
    \textbf{B/T/S/N} &
    \textbf{Ratio U/D : B/T/S/N} &
    \textbf{Significance} \\
    &
    \textbf{(\(\perp\) beam)} &
    \textbf{(\(\parallel\) beam)} &
    &
    \\
    \midrule

    17660 & 166 (\(0.687 \pm 0.053\)\%) & 291 (\(0.944 \pm 0.055\)\%) & \(0.73 \pm 0.07\) & \(3.4\sigma\) \\
    17831 & 181 (\(0.625 \pm 0.046\)\%) & 375 (\(0.982 \pm 0.050\)\%) & \(0.64 \pm 0.06\) & \(5.2\sigma\) \\
    18069 & 239 (\(0.636 \pm 0.041\)\%) & 471 (\(1.085 \pm 0.050\)\%) & \(0.59 \pm 0.05\) & \(7.0\sigma\) \\
    18512 & 251 (\(0.596 \pm 0.037\)\%) & 506 (\(1.006 \pm 0.044\)\%) & \(0.59 \pm 0.05\) & \(7.1\sigma\) \\
    18750 & 187 (\(0.621 \pm 0.045\)\%) & 389 (\(1.145 \pm 0.058\)\%) & \(0.54 \pm 0.05\) & \(7.1\sigma\) \\
    18790 & 144 (\(0.702 \pm 0.058\)\%) & 256 (\(1.068 \pm 0.066\)\%) & \(0.66 \pm 0.07\) & \(4.1\sigma\) \\
    18799 & 121 (\(0.552 \pm 0.050\)\%) & 274 (\(1.109 \pm 0.067\)\%) & \(0.50 \pm 0.05\) & \(6.7\sigma\) \\

    \bottomrule
  \end{tabular*}

\caption{Charge-sharing candidates, defined as adjacent-bar, low-amplitude unpaired hits, as a fraction of reconstructed bar crossings in the beam-perpendicular U/D and beam-parallel B/T/S/N panels. Uncertainties are binomial; the final two columns give the rate ratio and significance. Central bars only (FEB3 excluded).}
  \label{tab:odd_hits:charge-sharing}
\end{table}

\subsection{Time Resolution}
The time resolution of the ToF detector was evaluated using cosmic-ray data, which provide a high-statistics sample of tracks crossing multiple ToF panels. The data were taken with the UA1 magnet switched off, allowing the tracks to be approximated as straight lines and avoiding curvature effects in the resolution estimate.

\begin{figure}[b]
    \centering
    \includegraphics[
        width=0.87\textwidth,
        trim=0 0 0 0,
        clip
    ]{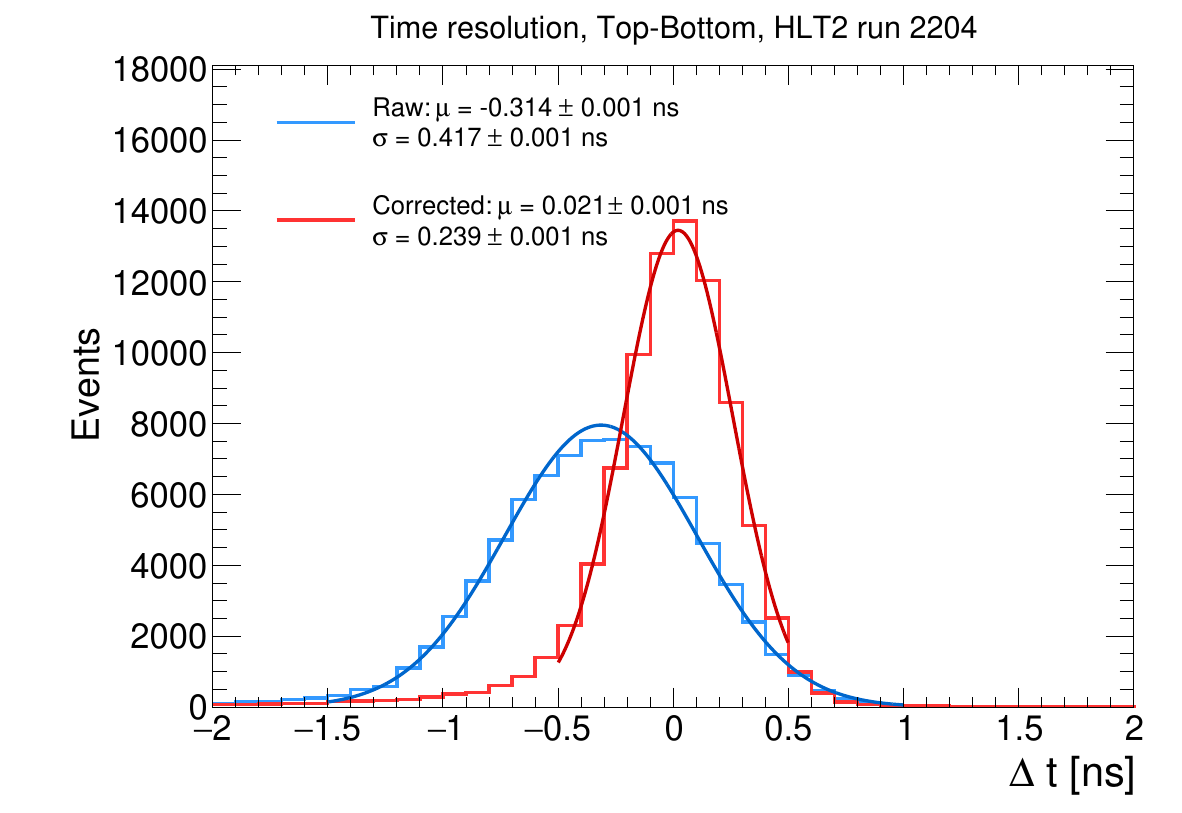}

    \caption{Path-length-corrected time-difference distribution for Top--Bottom cosmic-ray events in HLT2 run 2204, before (blue) and after (red) timing calibration. The solid curves are Gaussian fits, with \(\mu\) and \(\sigma\) denoting the fitted mean and width.}
    \label{fig:Time_resolution_top_bottom}
\end{figure}

The resolution is estimated from the width of the path-length-corrected time residual \(\Delta t\), defined in Equation~\ref{eq:tof_time_residual}. The \(\Delta t\) distributions before (raw) and after (corrected) application of the bar-level timing calibration described in Section~\ref{sec:timecalibration} are fitted with Gaussian functions over the intervals \([-1.5,1.0]\)~ns and \([-0.5,0.5]\)~ns, respectively. Assuming equal and independent timing uncertainties for the two bars, the single-bar time resolution is given by
\begin{equation}
    \sigma_{\mathrm{bar}} = \frac{\sigma_{\Delta t}}{\sqrt{2}},
\end{equation}
where \(\sigma_{\Delta t}\) is the fitted Gaussian width. For HLT2 run 2204, the single-bar resolution improves from \(295 \pm 1\)~ps before calibration to \(169 \pm 1\)~ps after calibration, as shown in Figure~\ref{fig:Time_resolution_top_bottom}. 

As a stability check, the calibration constants derived from the 2024 HLT2 data were applied without refitting to three HLT3 cosmic-ray runs taken in 2026 after the installation of the bracket system. The Top--Bottom resolution improved from \(289 \pm 4\)~ps before calibration to \(172 \pm 4\)~ps after calibration, consistent with the nominal calibrated performance. The larger uncertainty compared with the HLT2 result is due to the lower statistics of the HLT3 sample. This indicates that the calibration is robust to the detector reassembly and associated small geometrical changes.

These in situ resolutions are larger than the approximately \(130\,\text{ps}\) obtained in the dedicated single-bar study described in Section~\ref{sec:singlebar}. There, external trigger counters constrain the particle crossing position, allowing a more precise propagation-time correction and a combined weighted average of the two readout-end times. In the Top--Bottom method used here, the track length \(d\) is instead reconstructed using ToF hit positions, so the finite position resolution contributes in quadrature to the timing uncertainties and broadens the measured \(\Delta t\) distribution.

A more precise determination of the intrinsic timing resolution could be obtained by combining the ToF measurements with tracks reconstructed in the SuperFGD or HA-TPCs. As discussed in Section~\ref{sec:timecalibration}, these subdetectors can provide an independent constraint on the particle trajectory and crossing positions, reducing the uncertainty on the path-length correction \(d/c\).

In addition to this geometrical constraint, the use of ToF timing within the full ND280 detector requires the relative time references of the subdetectors to be aligned. Residual offsets may remain even after correcting for the gate arrival time, owing to differences in cable lengths and intrinsic electronics delays. An inter-detector calibration is therefore applied to align the ToF reference time with the global ND280 clock and the timing references of the other subdetectors.

The calibrated timing resolution is much smaller than the several-nanosecond flight times expected across the metre-scale ToF volume, allowing the particle direction to be determined from the sign of the measured time difference. When combined with the reconstructed path length, the time-of-flight measurement provides an estimate of the particle velocity. Together with the momentum reconstructed in the SuperFGD or HA-TPCs, this contributes to particle identification and provides an additional timing constraint for matching tracks to ToF hits.

\section{Conclusion}

The Time-of-Flight detector of the T2K ND280 Upgrade has been successfully constructed, installed, and commissioned as part of the upgraded near detector. Its design, based on plastic-scintillator bars with dual-ended SiPM readout and SAMPIC waveform digitisation, provides large geometrical coverage around the SuperFGD and HA-TPCs while maintaining a limited number of readout channels. The detector is fully integrated into the global ND280 DAQ and trigger systems, with online monitoring and data-quality procedures supporting stable operation during beam and cosmic-ray data taking.

Dedicated single-bar measurements demonstrate a time resolution of approximately \(130\,\text{ps}\), consistent with prototype studies. After installation, cosmic-ray data were used to validate the detector response, calibrate the timing offsets, and evaluate the full-detector timing performance. For Top--Bottom crossing tracks, the single-bar resolution improved from \(295 \pm 1\)~ps to \(169 \pm 1\)~ps after calibration, demonstrating the effectiveness of the timing-calibration procedure. The calibrated performance satisfies the timing requirements for particle-direction reconstruction and enables the use of ToF information for particle identification and the rejection of backgrounds originating outside the fiducial volume.

Beam data confirm the correct synchronisation of the ToF detector with the ND280 trigger, with the expected T2K bunch structure clearly observed. Studies of out-of-bunch activity reveal a delayed post-bunch excess above the approximately flat accidental background, characterised by an effective timescale of \(\tau = 2.93 \pm 0.08\)~\si{\micro\second} with delayed neutron activity as a possible contribution. Further studies of front-end electronics dead time and residual unpaired-hit events provide important characterisation of the detector response under realistic operating conditions. In particular, the bunch-dependent hit efficiency is understood in terms of SAMPIC readout dead time, while the remaining unpaired-hit population is consistent with contributions from cross-talk or charge sharing between adjacent bars, missed partner hits, and residual noise.

Together, these commissioning and performance studies demonstrate that the ToF detector operates stably and meets the requirements of the ND280 Upgrade. The results establish the ToF as a fully commissioned detector system, providing the timing performance required for reliable event reconstruction, particle identification, and background rejection in the upgraded ND280.

\acknowledgments

This work was partially supported by the Swiss National Science Foundation (SNSF) under grants nos.~\path{20FL20_232671}, \path{20FL20_201477}, \path{20FL21_186178}, \path{200020_182095}, \path{200021_185012}, \path{200020_204609}, and jointly by the SNSF and the Vietnam National Foundation for Science and Technology Development (NAFOSTED) through the Vietnamese--Swiss Joint Research Programme, under grant no.~\path{IZVSZ2_203433}. This work was supported by the European Union’s Horizon 2020 Research and Innovation programme under the H2020 Grant No. RISE-GA644294-JENNIFER 2020.  S.~Samani acknowledges support from the SNSF through a Swiss Postdoctoral Fellowship, under grant no.~\path{TMPFP2_234635}.

The research leading to these results has received funding from the Spanish Ministry of Science and Innovation PID2022-136297NB-I00 /AEI/10.13039/501100011033/ FEDER, UE and through the Spanish State Research Agency, under Severo Ochoa Centres of Excellence Programme 2025-2029 (CEX2024-001441-S) funded by MICIU/AEI/10.13039/501100011033. IFAE is partially funded by the CERCA program of the Generalitat de Catalunya.  We also acknowledge support from the Science and Technology Facilities Council~(STFC) and UK Research and Innovation~(UKRI).

We thank the SAMPIC development and support teams at CEA/IRFU, Saclay, and IJCLab, Orsay, for their assistance with the SAMPIC electronics and readout system. We also thank the engineering teams at the University of Geneva, CERN Neutrino Platform, and J-PARC whose contributions supported the construction, installation, integration, and operation of the ToF detector.

\bibliographystyle{JHEP}
\bibliography{biblio}

@article{T2K:2019bcf,
    author = "Abe, K. and others",
    collaboration = "T2K",
    title = "{Constraint on the matter\textendash{}antimatter symmetry-violating phase in neutrino oscillations}",
    eprint = "1910.03887",
    archivePrefix = "arXiv",
    primaryClass = "hep-ex",
    doi = "10.1038/s41586-020-2177-0",
    journal = "Nature",
    volume = "580",
    number = "7803",
    pages = "339--344",
    year = "2020",
    note = "[Erratum: Nature 583, E16 (2020)]"
}

@techreport{Blondel:2299599,
      author        = "Blondel, A and Yokoyama, M and Zito, M",
      collaboration = "T2K",
      title         = "{The T2K-ND280 upgrade proposal}",
      institution   = "CERN",
      reportNumber  = "CERN-SPSC-2018-001, SPSC-P-357",
      address       = "Geneva",
      year          = "2018",
      url           = "https://cds.cern.ch/record/2299599",
      note          = "This proposal is the follow-up of the Expression of
                       Interest EOI-15 submitted to SPSC in January 2017.",
}

@techreport{T2K:2019bbb,
      author        = "Zito, M and Yokoyama, M and Lux, T",
      collaboration = "T2K",
      title         = "{T2K ND280 Upgrade - Technical Design Report}",
      institution   = "CERN",
      reportNumber  = "CERN-SPSC-2019-001, SPSC-TDR-006",
      address       = "Geneva",
      year          = "2019",
      url           = "https://cds.cern.ch/record/2653463",
}

@article{Blondel:2020hml,
doi = {10.1088/1748-0221/15/12/P12003},
url = {https://doi.org/10.1088/1748-0221/15/12/P12003},
year = {2020},
month = {dec},
publisher = {},
volume = {15},
number = {12},
pages = {P12003},
author = {Blondel, A. and Bogomilov, M. and Bordoni, S. and Cadoux, F. and Douqa, D. and Dugas, K. and Ekelof, T. and Favre, Y. and Fedotov, S. and Fransson, K. and Fujita, R. and Gramstad, E. and Ichikawa, A.K. and Ilieva, S. and Iwamoto, K. and Jesús-Valls, C. and Jung, C.K. and Kasetti, S.P. and Khabibullin, M. and Khotjantsev, A. and Korzenev, A. and Kostin, A. and Kudenko, Y. and Kutter, T. and Lux, T. and Maret, L. and Matsubara, T. and Mefodiev, A. and Minamino, A. and Mineev, O. and Mitev, G. and Nessi, M. and Nicola, L. and Noah, E. and Parsa, S. and Petkov, G. and Sanchez, F. and Sgalaberna, D. and Shorrock, W. and Skwarczynski, K. and Suvorov, S. and Teklu, A. and Tsenov, R. and Uchida, Y. and Vankova-Kirilova, G. and Yershov, N. and Yokoyama, M. and Zalipska, J. and Zou, Y. and Zurek, W.},
title = {The {SuperFGD} Prototype charged particle beam tests},
journal = {Journal of Instrumentation}
}

@article{Attie:2021yeh,
    author = "Atti\'e, D. and others",
    title = "{Characterization of resistive Micromegas detectors for the upgrade of the T2K Near Detector Time Projection Chambers}",
    eprint = "2106.12634",
    archivePrefix = "arXiv",
    primaryClass = "physics.ins-det",
    doi = "10.1016/j.nima.2021.166109",
    journal = "Nucl. Instrum. Meth. A",
    volume = "1025",
    pages = "166109",
    year = "2022"
}

@article{Korzenev:2021mny,
    author = "Korzenev, A. and others",
    title = "{A 4\ensuremath{\pi} time-of-flight detector for the ND280/T2K upgrade}",
    eprint = "2109.03078",
    archivePrefix = "arXiv",
    primaryClass = "physics.ins-det",
    doi = "10.1088/1748-0221/17/01/P01016",
    journal = "JINST",
    volume = "17",
    number = "01",
    pages = "P01016",
    year = "2022"
}

@article{Breton:2020kva,
    author = "Breton, D. and Cheikali, C. and Delagnes, E. and Maalmi, J. and Rusquart, P. and Vallerand, P.",
    editor = "Garibaldi, Franco and Pagano, Angelo",
    title = "{Fast electronics for particle Time-Of-Flight measurement, with focus on the SAMPIC ASIC}",
    doi = "10.1393/ncc/i2020-20007-6",
    journal = "Nuovo Cim. C",
    volume = "43",
    number = "1",
    pages = "7",
    year = "2020"
}

@misc{MIDAS,
    author       = {{PSI and TRIUMF}},
    title        = {Maximum Integrated Data Acquisition System ({MIDAS})},
    howpublished = {\url{https://daq00.triumf.ca/MidasWiki/}},
    year         = {2023}
}

@techreport{SHiP:2015vad,
      author        = "Anelli, M. and Aoki, S. and Arduini, G. and Back, J.J. and
                       Bagulya, A. and Baldini, W. and Baranov, A. and Barker,
                       G.J. and Barsuk, S. and Battistin, M. and Bauche, J. and
                       Bay, A. and Bayliss, V. and Bellagamba, L. and Bencivenni,
                       G. and Bertani, M. and Bezshyyko, O. and Bick, D. and
                       Bingefors, N. and Blondel, A. and Bogomilov, M. and
                       Boyarsky, A. and Bonacorsi, D. and Bondarenko, D. and
                       Bonivento, W. and Borburgh, J. and Bradshaw, T. and
                       Brenner, R. and Breton, D. and Brook, N. and Bruschi, M.
                       and Buonaura, A. and Buontempo, S. and Cadeddu, S. and
                       Calcaterra, A. and Calviani, M. and Campanelli, M. and
                       Capoccia, C. and Cecchetti, A. and Chatterjee, A. and
                       Chauveau, J. and Chepurnov, A. and Chernyavskiy, M. and
                       Ciambrone, P. and Cicalo, C. and Conti, G. and Cornelis, K.
                       and Courthold, M. and Dallavalle, M.G. and D'Ambrosio, N.
                       and De Lellis, G. and De Serio, M. and Dedenko, L. and Di
                       Crescenzo, A. and Di Marco, N. and Dib, C. and Dietrich, J.
                       and Dijkstra, H. and Domenici, D. and Donskov, S. and
                       Druzhkin, D. and Ebert, J. and Egede, U. and Egorov, A. and
                       Egorychev, V. and El Alaoui, M.A. and Enik, T. and Etenko,
                       A. and Fabbri, F. and Fabbri, L. and Fedorova, G. and
                       Felici, G. and Ferro-Luzzi, M. and Fini, R.A. and Franke,
                       M. and Fraser, M. and Galati, G. and Giacobbe, B. and
                       Goddard, B. and Golinka-Bezshyyko, L. and Golubkov, D. and
                       Golutvin, A. and Gorbunov, D. and Graverini, E. and
                       Grenard, J-L and Guler, A.M. and Hagner, C. and Hakobyan,
                       H. and Helo, J.C. and van Herwijnen, E. and Horvath, D. and
                       Iacovacci, M. and Iaselli, G. and Jacobsson, R. and
                       Kadenko, I. and Kamiscioglu, M. and Kamiscioglu, C. and
                       Khaustov, G. and Khotjansev, A. and Kilminster, B. and Kim,
                       V. and Kitagawa, N. and Kodama, K. and Kolesnikov, A. and
                       Kolev, D. and Komatsu, M. and Konovalova, N. and Koretskiy,
                       S. and Korolko, I. and Korzenev, A. and Kovalenko, S. and
                       Kudenko, Y. and Kuznetsova, E. and Lacker, H. and Lai, A.
                       and Lanfranchi, G. and Lauria, A. and Lebbolo, H. and Levy,
                       J.-M. and Lista, L. and Loverre, P. and Lukiashin, A. and
                       Lyubovitskij, V.E. and Malinin, A. and Manfredi, M. and
                       Perillo-Marcone, A. and Marrone, A. and Matev, R. and
                       Messomo, E.N. and Mermod, P. and Mikado, S. and Mikhaylov,
                       Yu. and Miller, J. and Milstead, D. and Mineev, O. and
                       Mingazheva, R. and Mitselmakher, G. and Miyanishi, M. and
                       Monacelli, P. and Montanari, A. and Montesi, M.C. and
                       Morello, G. and Morishima, K. and Movtchan, S. and Murzin,
                       V. and Naganawa, N. and Naka, T. and Nakamura, M. and
                       Nakano, T. and Nurakhov, N. and Obinyakov, B. and Ocalan,
                       K. and Ogawa, S. and Oreshkin, V. and Orlov, A. and
                       Osborne, J. and Pacholek, P. and Panman, J. and Paoloni, A.
                       and Paparella, L. and Pastore, A. and Patel, M. and
                       Petridis, K. and Petrushin, M. and Poli-Lener, M. and
                       Polukhina, N. and Polyakov, V. and Prokudin, M. and Puddu,
                       G. and Pupilli, F. and Rademakers, F. and Rakai, A. and
                       Rawlings, T. and Redi, F. and Ricciardi, S. and Rinaldesi,
                       R. and Roganova, T. and Rogozhnikov, A. and Rokujo, H. and
                       Romaniouk, A. and Rosa, G. and Rostovtseva, I. and Rovelli,
                       T. and Ruchayskiy, O. and Ruf, T. and Saitta, G. and
                       Samoylenko, V. and Samsonov, V. and Sanz Ull, A. and
                       Saputi, A. and Sato, O. and Schmidt-Parzefall, W. and
                       Serra, N. and Sgobba, S. and Shaposhnikov, M. and Shatalov,
                       P. and Shaykhiev, A. and Shchutska, L. and Shevchenko, V.
                       and Shibuya, H. and Shitov, Y. and Silverstein, S. and
                       Simone, S. and Skorokhvatov, M. and Smirnov, S. and
                       Solodko, E. and Sosnovtsev, V. and Spighi, R. and Spinetti,
                       M. and Starkov, N. and Storaci, B. and Strabel, C. and
                       Strolin, P. and Takahashi, S. and Teterin, P. and Tioukov,
                       V. and Tommasini, D. and Treille, D. and Tsenov, R. and
                       Tshchedrina, T. and Ustyuzhanin, A. and Vannucci, F. and
                       Venturi, V. and Villa, M. and Vincke, Heinz and Vincke,
                       Helmut and Vladymyrov, M. and Xella, S. and Yalvac, M. and
                       Yershov, N. and Yilmaz, D. and Yilmazer, A.U. and
                       Vankova-Kirilova, G. and Zaitsev, Y. and Zoccoli, A.",
      collaboration = "SHiP",
      title         = "{A Facility to Search for Hidden Particles (SHiP) at the
                       CERN SPS}",
      institution   = "CERN",
      archivePrefix = "arXiv",
      eprint        = "1504.04956",
      reportNumber = "CERN-SPSC-2015-016, SPSC-P-350",
      address       = "Geneva",
      year          = "2015",
      url           = "https://cds.cern.ch/record/2007512",
      note          = "Technical Proposal",
}

@misc{Hamamatsu2025,
  author       = {{Hamamatsu Photonics}},
  title        = {{MPPC S13360-6050PE}},
  howpublished = {\url{https://www.hamamatsu.com/eu/en/product/optical-sensors/mppc/mppc_mppc-array/S13360-6050PE.html}},
  note         = {Accessed: 21 October 2025},
}

@misc{EljenScintillators,
  author       = {{Eljen Technology}},
  title        = {{EJ-200, EJ-204, EJ-208, EJ-212 - Plastic Scintillators}},
  howpublished = {\url{https://eljentechnology.com/products/plastic-scintillators/ej-200-ej-204-ej-208-ej-212}},
  note         = {Accessed: 2025-10-21}
}

@article{Alt:2025msx,
doi = {10.1088/1748-0221/21/01/P01041},
url = {https://doi.org/10.1088/1748-0221/21/01/P01041},
year = {2026},
month = {jan},
publisher = {IOP Publishing},
volume = {21},
number = {01},
pages = {P01041},
author = {Alt, C. and Blanchet, A. and Bordoni, S. and Collard, P. and Bui, T.H. and Bui, M.H. and Ha, G. and Jesús-Valls, C. and Kasturi, V.S. and Klustová, A. and Korzenev, A. and Le, T.A. and Lux, T. and Nguyen, A.D. and Nguyen, D.T. and Nguyen, H. and Samani, S. and Sanchez, F. and Ta, M. and Thaiduc, T. and Villa, E.},
title = {Modeling scintillation photon transport and reconstruction algorithms for the time-of-flight detector in the {T2K} neutrino experiment},
journal = {Journal of Instrumentation}
}

@article{abeMCcalibration2025,
doi = {10.1088/1748-0221/20/10/P10030},
year = {2025},
month = {oct},
publisher = {IOP Publishing},
volume = {20},
number = {10},
pages = {P10030},
author = "Abe, S. and others",
title = {Introducing a {Markov} chain-based time calibration procedure for multi-channel particle detectors: application to the {SuperFGD} and {ToF} detectors of the {T2K} experiment},
journal = {Journal of Instrumentation}
}

@article{Abe:2026elv,
    author = "Abe, S. and others",
    title = "{The super fine-grained detector for the {T2K} neutrino oscillation experiment}",
    journal = {Nuclear Instruments and Methods in Physics Research Section A:
               Accelerators, Spectrometers, Detectors and Associated Equipment},
    volume = {1092},
    pages = {171882},
    year = {2026},
    doi = {10.1016/j.nima.2026.171882}
}

@article{T2K:2015sxa,
    author = "Abe, K. and others",
    collaboration = "T2K",
    title = "{Upper bound on neutrino mass based on T2K neutrino timing measurements}",
    eprint = "1502.06605",
    archivePrefix = "arXiv",
    primaryClass = "hep-ex",
    doi = "10.1103/PhysRevD.93.012006",
    journal = "Phys. Rev. D",
    volume = "93",
    number = "1",
    pages = "012006",
    year = "2016"
}

@article{Igarashi:2021npv,
    author = "Igarashi, Susumu and others",
    title = "{Accelerator design for 1.3-MW beam power operation of the J-PARC Main Ring}",
    doi = "10.1093/ptep/ptab011",
    journal = "PTEP",
    volume = "2021",
    number = "3",
    pages = "033G01",
    year = "2021"
}

@article{T2K:2011qtm,
    author = "Abe, K. and others",
    collaboration = "T2K",
    title = "{The T2K Experiment}",
    eprint = "1106.1238",
    archivePrefix = "arXiv",
    primaryClass = "physics.ins-det",
    doi = "10.1016/j.nima.2011.06.067",
    journal = "Nucl. Instrum. Meth. A",
    volume = "659",
    pages = "106--135",
    year = "2011"
}

@article{Assylbekov:2011sh,
    author = "Assylbekov, S. and others",
    title = "{The T2K ND280 Off-Axis Pi-Zero Detector}",
    eprint = "1111.5030",
    archivePrefix = "arXiv",
    primaryClass = "physics.ins-det",
    doi = "10.1016/j.nima.2012.05.028",
    journal = "Nucl. Instrum. Meth. A",
    volume = "686",
    pages = "48--63",
    year = "2012"
}

@ARTICLE{Thorpe:2011T2KDAQ,
  author={Thorpe, M. and Angelsen, C. and Barr, G. and Metelko, C. and Nicholls, T. and Pearce, G. and West, N.},
  journal={IEEE Transactions on Nuclear Science}, 
  title={The {T2K} {Near} {Detector} {Data} {Acquisition} {Systems}}, 
  year={2011},
  volume={58},
  number={4},
  pages={1800-1806},
  doi={10.1109/TNS.2011.2141685}}

@article{Vacheret:2011,
    author = "Vacheret, A. and others",
    title = "{Characterization and simulation of the response of Multi-Pixel Photon Counters to low light levels}",
    journal = {Nuclear Instruments and Methods in Physics Research Section A:
               Accelerators, Spectrometers, Detectors and Associated Equipment},
    volume = {656},
    number = {1},
    pages = {69--83},
    year = {2011},
    doi = {10.1016/j.nima.2011.07.022}
}

@article{AMAUDRUZ20121,
title = {The {T2K} fine-grained detectors},
journal = {Nuclear Instruments and Methods in Physics Research Section A: Accelerators, Spectrometers, Detectors and Associated Equipment},
volume = {696},
pages = {1-31},
year = {2012},
issn = {0168-9002},
doi = {10.1016/j.nima.2012.08.020},
author = {P.-A. Amaudruz and M. Barbi and D. Bishop and N. Braam and D.G. Brook-Roberge and S. Giffin and S. Gomi and P. Gumplinger and K. Hamano and N.C. Hastings and S. Hastings and R.L. Helmer and R. Henderson and K. Ieki and B. Jamieson and I. Kato and N. Khan and J. Kim and B. Kirby and P. Kitching and A. Konaka and M. Lenckowski and C. Licciardi and T. Lindner and K. Mahn and E.L. Mathie and C. Metelko and C.A. Miller and A. Minamino and K. Mizouchi and T. Nakaya and K. Nitta and C. Ohlmann and K. Olchanski and S.M. Oser and M. Otani and P. Poffenberger and R. Poutissou and J.-M. Poutissou and W. Qian and F. Retiere and R. Tacik and H.A. Tanaka and P. Vincent and M. Wilking and S. Yen and M. Yokoyama},
}

@phdthesis{Villa:2025thesis,
    author = {Villa, Emanuele},
    title = {Novel Technologies for Next-Generation Neutrino Experiments: Time-of-Flight Detector for T2K ND280, Beam Monitor for ProtoDUNE, and Online Supernova Pointing in DUNE},
    school = {University of Geneva},
    year = {2025},
    doi = {\href{https://doi.org/10.13097/archive-ouverte/unige:192461}{10.13097/archive-ouverte/unige:192461}}
}

@INPROCEEDINGS{Delagnes:2014SAMPIC,
  author={Delagnes, E. and Breton, D. and Grabas, H. and Maalmi, J. and Rusquart, P. and Saimpert, M.},
  booktitle={2014 IEEE Nuclear Science Symposium and Medical Imaging Conference (NSS/MIC)}, 
  title={The {SAMPIC} Waveform and Time to Digital Converter}, 
  year={2014},
  volume={},
  number={},
  pages={1-9},
  doi = {10.1109/NSSMIC.2014.7431231},
  }

@article{D_Allan_2013,
doi = {10.1088/1748-0221/8/10/P10019},
url = {https://doi.org/10.1088/1748-0221/8/10/P10019},
year = {2013},
month = {oct},
publisher = {},
volume = {8},
number = {10},
pages = {P10019},
author = {D Allan and C Andreopoulos and C Angelsen and G J Barker and G Barr and S Bentham and I Bertram and S Boyd and K Briggs and R G Calland and J Carroll and S L Cartwright and A Carver and C Chavez and G Christodoulou and J Coleman and P Cooke and G Davies and C Densham and F Di Lodovico and J Dobson and T Duboyski and T Durkin and D L Evans and A Finch and M Fitton and F C Gannaway and A Grant and N Grant and S Grenwood and P Guzowski and D Hadley and M Haigh and P F Harrison and A Hatzikoutelis and T D J Haycock and A Hyndman and J Ilic and S Ives and A C Kaboth and V Kasey and L Kellet and M Khaleeq and G Kogan and L L Kormos and M Lawe and T B Lawson and C Lister and R P Litchfield and M Lockwood and M Malek and T Maryon and P Masliah and K Mavrokoridis and N McCauley and I Mercer and C Metelko and B Morgan and J Morris and A Muir and M Murdoch and T Nicholls and M Noy and H M O'Keeffe and R A Owen and D Payne and G F Pearce and J D Perkin and E Poplawska and R Preece and W Qian and P Ratoff and T Raufer and M Raymond and M Reeves and D Richards and M Rooney and R Sacco and S Sadler and P Schaack and M Scott and D I Scully and S Short and M Siyad and R Smith and B Still and P Sutcliffe and I J Taylor and R Terri and L F Thompson and A Thorley and M Thorpe and C Timis and C Touramanis and M A Uchida and Y Uchida and A Vacheret and J F Van Schalkwyk and O Veledar and A V Waldron and M A Ward and G P Ward and D Wark and M O Wascko and A Weber and N West and L H Whitehead and C Wilkinson and J R Wilson},
title = {The electromagnetic calorimeter for the {T2K} near detector {ND280}},
journal = {Journal of Instrumentation},
}

@article{AOKI2013135,
title = {The {T2K} {Side} {Muon} {Range} {Detector} {(SMRD)}},
journal = {Nuclear Instruments and Methods in Physics Research Section A: Accelerators, Spectrometers, Detectors and Associated Equipment},
volume = {698},
pages = {135-146},
year = {2013},
issn = {0168-9002},
doi = {10.1016/j.nima.2012.10.001},
url = {https://www.sciencedirect.com/science/article/pii/S0168900212011242},
author = {S. Aoki and G. Barr and M. Batkiewicz and J. Błocki and J.D. Brinson and W. Coleman and A. Dąbrowska and I. Danko and M. Dziewiecki and B. Ellison and L. Golyshkin and R. Gould and T. Hara and J. Haremza and B. Hartfiel and J. Holeczek and A. Izmaylov and M. Khabibullin and A. Khotjantsev and D. Kiełczewska and A. Kilinski and J. Kisiel and Y. Kudenko and N. Kulkarni and R. Kurjata and T. Kutter and J. Łagoda and J. Liu and J. Marzec and W. Metcalf and C. Metelko and P. Mijakowski and O. Mineev and D. Naples and M. Nauman and T.C. Nicholls and D. Northacker and J. Nowak and M. Noy and V. Paolone and G.F. Pearce and O. Perevozchikov and M. Posiadała and P. Przewłocki and W. Qian and M. Raymond and J. Reid and E. Rondio and E. Shabalin and M. Siyad and D. Smith and J. Sobczyk and M. Stodulski and R. Sulej and J. Świerblewski and A.T. Suzuki and T. Szegłowski and M. Szeptycka and M. Thorpe and T. Wąchała and D. Warner and A. Weber and T. Yano and N. Yershov and A. Zalewska and K. Zaremba and M. Ziembicki}
}

@article{ABGRALL201125,
title = {Time projection chambers for the {T2K} near detectors},
journal = {Nuclear Instruments and Methods in Physics Research Section A: Accelerators, Spectrometers, Detectors and Associated Equipment},
volume = {637},
number = {1},
pages = {25-46},
year = {2011},
issn = {0168-9002},
doi = {10.1016/j.nima.2011.02.036},
url = {https://www.sciencedirect.com/science/article/pii/S0168900211003421},
author = {N. Abgrall and B. Andrieu and P. Baron and P. Bene and V. Berardi and J. Beucher and P. Birney and F. Blaszczyk and A. Blondel and C. Bojechko and M. Boyer and F. Cadoux and D. Calvet and M.G. Catanesi and A. Cervera and P. Colas and X. {De La Broise} and E. Delagnes and A. Delbart and M. {Di Marco} and F. Druillole and J. Dumarchez and S. Emery and L. Escudero and W. Faszer and D. Ferrere and A. Ferrero and K. Fransham and A. Gaudin and C. Giganti and I. Giomataris and J. Giraud and M. Goyette and K. Hamano and C. Hearty and R. Henderson and S. Herlant and M. Ieva and B. Jamieson and G. Jover-Mañas and D. Karlen and I. Kato and A. Konaka and K. Laihem and R. Langstaff and M. Laveder and A. {Le Coguie} and O. {Le Dortz} and M. {Le Ross} and M. Lenckowski and T. Lux and M. Macaire and K. Mahn and F. Masciocchi and E. Mazzucato and M. Mezzetto and A. Miller and J.-Ph. Mols and L. Monfregola and E. Monmarthe and J. Myslik and F. Nizery and R. Openshaw and E. Perrin and F. Pierre and D. Pierrepont and P. Poffenberger and B. Popov and E. Radicioni and M. Ravonel and J.-M. Reymond and J.-L. Ritou and M. Roney and S. Roth and F. Sánchez and A. Sarrat and R. Schroeter and A. Stahl and P. Stamoulis and J. Steinmann and D. Terhorst and D. Terront and V. Tvaskis and M. Usseglio and A. Vallereau and G. Vasseur and J. Wendland and G. Wikström and M. Zito}
}

@article{Ashida:2018,
    author = {Ashida, Y and Friend, M and Ichikawa, A K and Ishida, T and Kubo, H and Nakamura, K G and Sakashita, K and Uno, W},
    title = {A new electron-multiplier-tube-based beam monitor for muon monitoring at the {T2K} experiment},
    journal = {Progress of Theoretical and Experimental Physics},
    volume = {2018},
    number = {10},
    pages = {103H01},
    year = {2018},
    month = {10},
    issn = {2050-3911},
    doi = {10.1093/ptep/pty104},
    url = {https://doi.org/10.1093/ptep/pty104},
}

\end{document}